\documentstyle[prb,aps,eqsecnum,multicol,ifthen,epsf]{revtex}
\tighten
\newcommand{\wide}[2]{
\end{multicols}
\widetext
\noindent
\ifthenelse{\equal{#1}{t}}
{}
{
\raisebox{0.1in}[0in][0.02in]{$\rule{3.575in}{0.002in}
\rule{0.002in}{0.08in}$}
}
#2
\ifthenelse{\equal{#1}{b}}
{}
{
{\raisebox{-0.1in}[0in][0.02in]
{\hspace{3.575in}$\rule{0.002in}{0.08in}
\rule[0.08in]{3.575in}{0.002in}$}
}
}
\begin{multicols}{2}
\noindent
}

\begin{document}
\draft
\title{Mesoscopic charge quantization} 
\author{I.L. Aleiner$^{1}$ and
L.I. Glazman$^{2}$} \address{$^{1}$NEC Research Institute, 4
Independence Way, Princeton, NJ 08540\\
 $^{2}$Theoretical Physics
Institute, University of Minnesota, Minneapolis MN 55455} 
\maketitle
\begin{abstract}
We study the Coulomb blockade in a chaotic quantum dot connected to a
lead by a single channel at nearly perfect transmission. We take into
account quantum fluctuations of the dot charge and a finite level
spacing for electron states within the dot. Mesoscopic fluctuations of
thermodynamic and transport properties in the Coulomb blockade regime
exist at any transmission coefficient.  In contrast to the previous
theories, we show that by virtue of these mesoscopic fluctuations, the
Coulomb blockade is not destroyed completely even at perfect
transmission. The oscillatory dependence of all the observable
characteristics on the gate voltage is preserved, its period is still
defined by the charge of a single electron. However, phases of those
oscillations are random; because of the randomness, the Coulomb
blockade shows up not in the averages but in the correlation functions
of the fluctuating observables ({\em e.g.}, capacitance or tunneling
conductance). This phenomenon may be called ``{\em mesoscopic charge
quantization}''.
\end{abstract}

\begin{multicols}{2}

\section{Introduction}
The effect of Coulomb blockade\cite{history,Review,Review2} in chaotic
quantum dots sets up an ideal stage for studying the interplay between
the quantum chaos and interaction phenomena in a many-electron system.
By tuning the connection between the leads and the quantum dot, one
can study a rich variety of non-trivial effects. In the weak tunneling
limit, discrete charging of the dot results in a sequence of sharp
conductance peaks, which carry information about the chaotic motion of
non-interacting electrons confined inside an almost closed
dot\cite{Stone}. In the opposite limit of wide channels, charge
quantization does not occur, and quantum chaos of free electrons in an
open billiard may be studied\cite{Westervelt}. In a broad intermediate
region, the charge quantization is gradually destroyed, and the
chaotic electron motion is affected by fluctuations of charge of the
cavity. The modern experimental technique\cite{Review2,Chang,Marcus}
allows one to continuously traverse between these regimes.

The effect of charging is conventionally described by
Hamiltonian\cite{Review}
\begin{equation}
\hat{H}_C=\frac{E_C}{2}\left(\frac{\hat{Q}}{e}-{\cal N}\right)^2,
\quad E_C=\frac{e^2}{C},
\label{Hc}
\end{equation}
where $C$ is the total capacitance of the dot, the dimensionless
parameter ${\cal N}$ is related to the gate voltage $V_g$ and gate
capacitance $C_g$ by ${\cal N}=V_g/eC_g$, and $\hat{Q}$ is the dot
charge.  Usually, charging energy $E_C$ is much larger than the
one-electron mean level spacing of the dot, $\Delta$. If the
connection of the dot with the leads is weak and temperature $T$ is
small, $T\ll E_C$, the charge is well quantized for almost all $\cal
N$ except narrow vicinities of the charge degeneracy points
(half-integer ${\cal N}$). The behavior of the differential
capacitance of the cavity, $d\langle Q\rangle/dV_g$ and of the
conductance through the cavity is quite different for the system tuned
to the immediate vicinity of charge degeneracy points (Coulomb
blockade peaks), or away from those points (Coulomb blockade
valleys). The statistics of the peaks can be related to the properties
of a single electron energy and wave function\cite{Stone}, so that the
distribution functions for these quantities can be extracted from the
well known Random Matrix theory (RMT) \cite{PorterThomas,Book}. The
transport in the valleys occurs by virtual transitions of an electron
via excited states of the dot\cite{Averin90}. The statistics of the
conductance in this case was recently obtained in
Ref.~\onlinecite{Aleiner96} and was confirmed experimentally in
Ref.~\onlinecite{Sara}.

All the aforementioned results were obtained neglecting quantum
fluctuations of the charge of the cavity. These fluctuations grow with
the increase of the coupling between the dot and
lead\cite{Glazman90}. Then, the difference between the peaks and
valleys becomes less pronounced and eventually instead of the peak
structure, one observes only a weak periodic
modulation\cite{Matveev95}. Clearly, this modulation can be described
neither by the properties of the single-electron wave function nor by
the lowest order virtual transitions via the excited states.

The case of almost perfect transmission of a one-channel point contact
connecting the quantum dot with the lead, see Fig.~\ref{Fig:1.1}, was
analyzed by Matveev\cite{Matveev95} and Flensberg\cite{Flensberg95} in
the framework of an effective one-dimensional Hamiltonian. Employing
the bosonization technique, they showed that the Coulomb blockade
disappears completely if the transmission coefficient of the point
contact is exactly unity $r=0$ and $\Delta = 0$.  The explicit
dependence of the differential capacitance of the system on ${\cal
N}$, and on the small reflection coefficient $0<| r|^2 \ll 1$ was
obtained by Matveev\cite{Matveev95}. It is important to emphasize that
the Coulomb blockade in this situation is non-perturbative in charging
energy effect and it can not be revealed in the standard Hartree-Fock
or Random Phase approximations.

The properties of a quantum dot connected to reservoir by a channel
was analyzed in a series of papers of B\"{u}ttiker and
collaborators\cite{Buttiker1,MelloButtiker}. They have used the Random
Phase approximation to calculate the frequency dependence of the
linear response of the current $I$ through the channel to bias
$Ve^{i\omega t}$ applied to the reservoir: $I=G(\omega)V$. In this
approach, the admittance $C=-{\rm Im}(dG/d\omega)|_{\omega\to 0}$,
coincides by construction with the thermodynamic capacitance of the
non-interacting electrons. The quantum corrections and mesoscopic
fluctuations of this quantity can be then analyzed by using the
distribution of the Wigner delay times of the non-interacting
system\cite{MelloButtiker}. This approach is perfectly valid for a
large number of channels, but for a single channel, there is no any
parameter justifying it. We will see below that the results obtained
by well controlled procedure are significantly different, see
Secs. ~\ref{sec:5}, and \ref{sec:6}.

\narrowtext{
\begin{figure}[h]
\vspace{0.2cm} \epsfxsize=6.0cm
\hspace*{1.0cm} \epsfbox{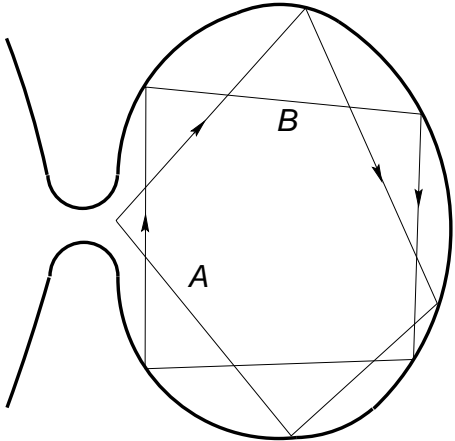}
\vspace{0.6cm}
\caption{Schematic view of a quantum dot connected to a lead.
Periodic orbit ``A'' encounters the entrance to the dot once, $n_A
=1$, and periodic orbit ``B'' does not encounter the entrance, $n_B =
0$.}
\label{Fig:1.1}
\end{figure} 
}

In this paper, we account for both the strong quantum charge
fluctuations, and the chaotic electron motion within the dot. We will
show that backscattering of electrons from the walls of the dot into
the channel connecting it to the lead results in residual Coulomb
blockade oscillations of observables with the gate voltage. In the
limit of perfect channel transmission, the relative magnitude of the
differential capacitance oscillations is $\sqrt{\Delta/E_C}$ and
$\left(\Delta/E_C\right)\ln^2\left(E_C/\Delta\right)$ for the spinless
and for spin $1/2$ cases respectively. If the second lead is attached
to the cavity by a weak tunnel junction with conductance $G_0 \ll
e^2/\hbar$, the two-terminal conductance $G$ can be measured. The
average value of the conductance is suppressed by the Coulomb blockade,
but it remains finite even at zero temperature, $\langle G\rangle
\simeq G_0\Delta/E_C$. Fluctuations of the conductance are of the
order of its average. This resembles the result for the elastic
cotunneling in weak coupling regime\cite{Averin90,Aleiner96}.
However, the dependence of the conductance of the gate voltage is no
longer a sequence of deep valleys and sharp peaks, but rather a weakly
oscillating function.

For a finite reflection coefficient in the channel ($|r|^2\neq 0$), we
found a new contribution, in addition to the averaged differential
capacitance calculated in Ref.~\onlinecite{Matveev95}. This
contribution is fluctuating and provides, in particular, the
dependence of the differential capacitance on the magnetic field.

The paper is organized as follows. In Sec.~\ref{sec:2}, we
qualitatively discuss the mesoscopic fluctuations of the differential
capacitance for spinless electrons.  Section~\ref{sec:3} is devoted to
the formulation of the model and derivation of the effective action
representation.  We will also discuss the conditions of applicability
of the model. Section~\ref{sec:4} describes the bosonization
procedure. Calculation of the ground state energy and differential
capacitance is performed in Sec.~\ref{sec:5}. The tunneling
conductance in a strongly asymmetric setup (one channel is
reflectionless, and the other junction is of conductance $G_0\ll
e^2/\hbar$) is studied in Sec.~\ref{sec:6}. Our findings are
summarized in Conclusion.

\section{Qualitative discussion}
\label{sec:2}
Let us consider first a completely opened channel ($r=0$). In the
limit $\Delta\to 0$, the electron charge of the dot varies with the
gate voltage as $\langle Q\rangle=e{\cal N}$, to assure the minimum of
the electrostatic energy\cite{Matveev95}.  The interaction (\ref{Hc})
depends only on the number of electrons crossing the dot-channel
boundary. Therefore, the properties of the ground state can be
characterized by the asymptotic behavior of the wave-functions far
from the entrance to the dot.  This behavior is described by the
scattering phase, and at low energies can be understood from the
following qualitative argument.

Entrance of an additional electron with energy $\epsilon$ (all the
energies will be measured from the Fermi level) into the dot requires
energy $E_C$. Therefore, the electron may spend in the dot time of the
order of $\hbar/E_C$, and then the extra charge of the dot has to
relax. There are two processes, that lead to the relaxation of the
charge: i) elastic process where the same electron leaves the dot; and
ii) inelastic process where some other electron is emitted from the
dot. At low energies the probability of the inelastic process is small
as $(\epsilon/E_C)^2$, by virtue of the smallness of the phase
volume. (The last statement assumes the Fermi liquid behavior at low
energies and, as we will see later, is valid only for spinless
one-channel case.) Therefore, we may consider only elastic process.
The same consideration is applicable also to an electron leaving the
dot.  Thus we conclude, that the low energy properties of the system
can be mapped onto the dot effectively decoupled from the channel, and
the phase of the scattering amplitude from the entrance of the dot is
given by the Friedel sum rule
\begin{equation}
\delta =\pi\langle Q\rangle/e=\pi{\cal N}.
\label{phaseshift}
\end{equation}
Equation~(\ref{phaseshift}) can be applied to electrons incident from
inside the dot, as well as to electrons incident from the channel.
The outlined description resembles closely the Nozieres description of
the unitary limit in the one-channel Kondo problem\cite{Nozieres}.

The outlined above qualitative picture based on the introduction of
scattering phase $\delta$ is somewhat intuitive; it will be verified
by a calculation in Sec.~\ref{sec:4}. Here instead of rigorous proof,
we demonstrate that this scheme reproduces the result
\begin{equation}
E_g({\cal N})\simeq |r| E_C \cos 2\pi {\cal N}
\label{Matveev}
\end{equation}
obtained by Matveev~\cite{Matveev95} for the ground state energy
$E_g({\cal N})$ of spinless electrons in the limit of zero level
spacing in the dot. Then, we apply the scheme to find the corrections
to the ground state energy arising from a finite $\Delta$. Those
corrections will result in the mesoscopic fluctuations of the ground
state energy.

We start with considering the limit $\Delta=0$. First, we put also
$|r|=0$ and calculate the density of electrons in the channel
$\rho(x)$. Then we take into account the scattering potential $V(x)$
that generates $r\neq 0$, in the first order of perturbation theory,
\begin{equation}
E_g({\cal N}) = \int d x\rho(x)V(x).
\label{pr1}
\end{equation}
As we discussed, the Coulomb interaction leads to the perfect
reflection of electron at low energies; wavefunctions have the form
$\psi_k(x) = \cos \left(k|x| - \delta \right)$, with phase shift
$\delta$ given by Eq.~(\ref{phaseshift}). It leads to the Friedel
oscillation of the electron density $\rho(x)=\sum_{v_F|k-k_F]\lesssim
E_C}\left|\psi_k(x)\right|^2$, where $v_F$ and $k_F$ are the Fermi
velocity and Fermi wavevector respectively.  We obtain
\begin{equation}
\rho(x) =\left\{
\begin{array}{ll}
\displaystyle{\frac{E_C}{v_F} \cos (2k_F|x|- 2\delta)}, \quad & |x| <
v_F/E_C; \\ \displaystyle{\frac{ \sin (2k_F|x|-2\delta)}{|x|}}, \quad
&|x| > v_F/E_C.\\
\end{array}
\right.
\label{pr2}
\end{equation}
Here we omitted the irrelevant constant part of the electron
density. Substituting Eq.~(\ref{pr2}) into Eq.~(\ref{pr1}), assuming
the magnitude of the potential around $x=0$ smaller than $v_F/E_C$,
and using the standard expression $|r| = |\tilde{ V}(2k_F)|/v_F$, we
obtain formula (\ref{Matveev}). Here $\tilde{V}(k)$ is the Fourier
transform of the potential $V(x)$.

Having verified the suggested scheme for the case$\Delta=0$, we
proceed with evaluation of the ground state energy of a finite dot
connected to a reservoir by a perfect channel. According to the above
discussion, the channel is effectively decoupled from the dot due to
the charging effect even though $r=0$. Therefore, we have to relate
the ground state energy of a closed dot to the scattering phase
$\delta$ of Eq.~(\ref{phaseshift}). For a chaotic dot, this problem is
equivalent to finding a variation of the eigenenergies by introduction
of impurity potential $V({\bf r}) =\left(1/\pi\nu\right) \delta({\bf
r}) \tan \delta $, where $\nu$ is the averaged density of states per
unit area.  The relevant contribution to the ground state energy is
given by
\begin{equation}
E_g = \sum_{-E_C \lesssim \xi_i <0} \left[\xi_i(\delta) + \mu \right],
\label{E}
\end{equation}
where $\xi_i$ are the eigenenergies measured from the Fermi level $\mu$.
As soon as the scattering phase changes by $\pi$, one more level
enters under the Fermi level.  Evolution of the energy levels with
changing $\delta$ is shown schematically in Fig.~\ref{Fig:2.1}. The
position of the level $\xi_i(\delta)$ satisfies the gluing condition
\begin{equation}
\xi_i(\delta +\pi) = \xi_{i+1}(\delta ).
\label{gluing}
\end{equation}
From Eqs.~(\ref{E}) and Eq.~(\ref{gluing}) we see that the ground
state energy depends almost periodically on $\delta$:
\begin{equation}
E_g(\delta) = E_g(\delta + \pi ) + {\cal O}(\Delta).
\label{periodic}
\end{equation}
As we will see below, the amplitude of the oscillation of the ground
state energy with $\delta$ is much larger than the mean level spacing
and we will neglect the last term in Eq.~(\ref{periodic}).

In order to estimate the amplitude of the oscillations, we recall that
the correlation function of the level velocities is given
by\cite{AltshulerSimons}
\begin{equation}
\langle \partial _\delta \epsilon_i \rangle =\frac{\Delta}{\pi}, \quad
\langle \partial _\delta \epsilon_i \partial _\delta \epsilon_j\rangle
=\delta_{ij}\frac{2}{\beta}\left(\frac{\Delta}{\pi}\right)^2,
\label{velocities}
\end{equation}
where $\beta = 1,2$ for the orthogonal and unitary ensembles
respectively, and $\langle\dots\rangle$ stands for the ensemble
averaging.  Formula (\ref{velocities}) can be easily understood from
the first order perturbation theory. At $\delta \ll 1$, we have
$\epsilon_i (\delta) \approx \epsilon_i(0) + (\delta/\pi \nu)
|\psi_i^2(0)| $.  For a chaotic system the exact wave functions
$\psi_i$ can be presented in the form
\[
\left|\psi_i(0)\right|^2=\frac{1+ b_i}{\cal A} ,
\]
where the area of the dot ${\cal A}$ appears due to the normalization
condition, and $b_i$ characterizes the fluctuations of the chaotic
wavefunctions. In accordance with the Porter-Thomas
distribution\cite{PorterThomas},
\[
\langle b_i \rangle = 0;\quad \langle b_i b_j\rangle=
\frac{2}{\beta}\delta_{ij}.
\]

\narrowtext{
\begin{figure}[h]
\vspace*{-0.8cm} \epsfxsize=9.5cm
\hspace*{-0.9cm} \epsfbox{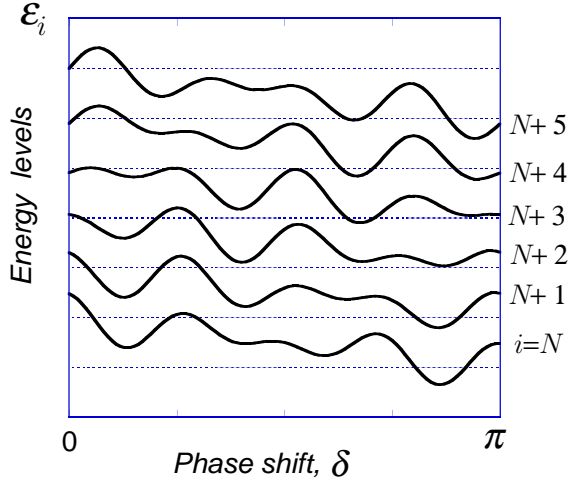}
\vspace*{-4.4cm}
\caption{ Evolution of the energy levels of the quantum dot with the
scattering phase $\delta$. }
\label{Fig:2.1}
\end{figure} 
}

Estimating the mesoscopic fluctuations of the ground state energy
(\ref{E}) with the help of Eq.~(\ref{velocities}), we obtain for
$\delta \ll 1$
\begin{eqnarray}
&&\langle\left[E_g(\delta)-E_g(0)\right]^2\rangle =\label{Egcor} \\
&&\quad \Delta^2\left(\frac{\delta}{\pi}\right)^2\!\!\!  \sum_{-E_C
\lesssim \xi_i,\xi_j <0}\frac{2}{\beta} \delta_{ij} \approx \Delta
E_C{\delta}^2.  \nonumber
\end{eqnarray}
As we have already explained, energy (\ref{E}) is a periodic function
of $\delta$ with period $\pi $. On the other hand, for $\delta\lesssim
1$, Eq.~(\ref{Egcor}) is valid.  Therefore the characteristic
amplitude of the oscillations is of the order of $\sqrt{E_C \Delta}$,
and it is plausible to assume that the correlation function of
energies at two different parameters ${\cal N}_1,\ {\cal N}_2$ takes
the form
\begin{equation}
\langle E_g\left({\cal N}_1\right) E_g\left({\cal N}_2\right)\rangle
\approx \Delta E_C \cos 2\pi \left({\cal N}_1-{\cal N}_2\right),
\label{Egcor1}
\end{equation}
where we use Eq.~(\ref{phaseshift}). It is important to notice that
the variation of the energy of the ground state is much larger than
the mean level spacing $\Delta$. This observation enabled us not to
consider in particular the variation of the chemical potential with
changing $\delta$, because it would generate a correction of the order
of the level spacing only.

In order to explain the functional form of the correlation function
(\ref{Egcor1}) and make our argumentation more precise, it is
instructive to evaluate the shift of the energy levels starting from
the Gutzwiller trace formula\cite{Gutzwiller}.  The energy of the
ground state is given by
\begin{equation}
E_g=-\int_{-\infty}^{0} d \epsilon N(\epsilon) K(\epsilon/E_C),
\label{EgN}
\end{equation}
where $\epsilon $ is measured from the Fermi level, and $N(\epsilon)$
is the integrated density of states,
$N(\epsilon)=\sum_i\theta(\epsilon -\epsilon_i)$.  Here $K(x)$ is some
function which decays rapidly at $x >1$, so only the states which can
be described within the Fermi liquid theory contribute into the
$\delta$-dependent part of the ground state energy
[cf. Eq.~(\ref{E})].

 The integrated density of states can be expressed as a sum over the
classical periodic orbits\cite{Gutzwiller}:
\begin{equation}
N(\epsilon,\delta) ={\rm Re} \sum_j R_j(\epsilon)
\exp\left[\frac{i}{\hbar}{S_j(\epsilon)} +2in_j\delta\right].
\label{Gutzwiller}
\end{equation}
Here $R_j$ is the weight associated with $j$-th orbit, $S_j$ is the
reduced action for this orbit. The last term in the exponent in
Eq.~(\ref{Gutzwiller}) characterizes the reflection from the entrance
of the cavity, and $n_j$ is the number of such reflections for $j$-th
orbit, see Fig.~\ref{Fig:1.1}.  We have omitted the mean value of
$N(\epsilon)$ which is independent of the phase shift
(\ref{phaseshift}).

Integrated density of states (\ref{Gutzwiller}) is a strongly
oscillating function of energy which vanishes upon ensemble (or
energy) averaging.  However, it contributes to the fluctuations of the
density of states:
\begin{eqnarray}
&&\langle N(\epsilon_1,\delta_1) N(\epsilon_2,\delta_2)\rangle =2 {\rm
Re} \sum_j |R_j|^2 \times \label{Ncor} \\ && \ \ \ \ \
\exp\left[\frac{i}{\hbar}\left({S_j(\epsilon_1)}-{S_j(\epsilon_2)}\right)
+2in_j\left(\delta_1 -\delta_2\right)\right], \nonumber
\end{eqnarray}
and we neglected the energy dependence of the pre-exponential factors
$R_j$ because it occurs on the energy scale of the order of the Fermi
energy.  In the double sum over the periodic orbits, arising in
(\ref{Ncor}), one can retain only diagonal terms\cite{Berry85} because
different orbits have different actions; the non-diagonal terms
oscillate strongly and vanish upon the averaging.  Factor of $2$ in
Eq.~(\ref{Ncor}) originates from the fact that, the electron
trajectory, $j$, and the trajectory time reversed to $j$ have the same
action.

In order to calculate correlation function of mesoscopic fluctuations
of ground state energies (\ref{EgN}), we use Eq.~(\ref{Ncor}) and
expand action as $S_j(\epsilon_1) -S_j(\epsilon_2) =
(\epsilon_1-\epsilon_2)t_j$, with $t_j$ being the period of $j$-th
orbit. After integration over energies $\epsilon_{1,2}$, we find
\begin{eqnarray}
&&\langle E_g(\delta_1)E_g(\delta_2)\rangle = \label{EcorG}\\ && \ \ \
\ \ 2 E_C^2 \sum_j |R_j|^2 \tilde{K}\left(\frac{E_Ct_j}{\hbar}\right)
\cos \left[2n_j\left(\delta_1 -\delta_2\right)\right], \nonumber
\end{eqnarray}
where $\tilde{K}\left(y\right)=\left|\int_{-\infty}^0 d x K(x)
e^{ixy}\right|^2 $, is a function decaying at $y>1$.

Coefficients $|R_j|^2$ have a very simple physical meaning, and are
related to the classical probability $P(t)dt$ to find a periodic orbit
with a period within the interval $\left[t; t+dt\right]$:
\begin{equation}
\sum_j\left|R_j\right|^2\dots \to \frac{1}{2\pi^2}\int_0^\infty
\frac{dt}{t}P(t)\dots.
\label{classical}
\end{equation}
In the same fashion, we obtain from Eq.~(\ref{EcorG}):
\begin{eqnarray}
&&\langle E_g(\delta_1)E_g(\delta_2)\rangle = \label{EcorGI} \\ && \ \
\ \ \ \frac{E_C^2}{\pi^2}\!\sum_n\!\int_0^\infty\!\!\frac{dt}{t}
\tilde{K}\left(\!\frac{E_Ct}{\hbar}\!\right)P_n(t)
\cos\!\left[2n\left(\delta_1\! -\!\delta_2\right)\right].  \nonumber
\end{eqnarray}
The classical probability $P_n(t)$ differs from $P(t)$ by satisfying
an additional constraint: the corresponding periodic orbits are
reflected from the dot entrance exactly $n$ times; $\sum_n P_n(t) =
P(t)$.

In general, $P_n(t)$ depends on a particular shape of the
dot. However, if the motion is chaotic, $P_n(t)$ becomes universal,
\begin{equation}
P_n(t) = \frac{\left(t\Delta/\hbar\right)^n}{n!}e^{-t\Delta/\hbar},
\label{Pn}
\end{equation}
for the periods $t$ much larger than the ergodic time $\hbar/E_T$.
Energy $E_T$ associated with the time scale at which the classical
dynamics becomes ergodic is the counterpart of the Thouless energy for
the diffusive system.  Typically $E_T \gtrsim E_C$, therefore we adopt
approximation $E_T \gg E_C\gg\Delta$. According to Eq.~(\ref{EcorG}),
the characteristic period of the semiclassical trajectory is of the
order of $\hbar/E_C$, which is much smaller than the Heisenberg time
$\hbar/\Delta$.  Therefore, when calculating the oscillatory part of
the correlation function of the ground state energies, it suffices to
keep in Eq.~(\ref{EcorGI}) only the contribution of the trajectories
reaching the entrance only once. As the result, we obtain the
expression similar to our estimate (\ref{Egcor1}),
\begin{equation}
\langle E_g\left({\cal N}_1\right) E_g\left({\cal N}_2\right)\rangle
=\alpha \Delta E_C \cos 2\pi \left({\cal N}_1-{\cal N}_2\right),
\label{corGutz}
\end{equation}
where we have used Eq.~(\ref{phaseshift}).  The numerical coefficient
 $\alpha = \frac{1}{\pi^2}\int_0^\infty dx\tilde{K}(x)$ depends on the
 particular form of function $\tilde{K}$, and cannot be found within
 the simple consideration.

Let us now discuss the correlation of the ground state energies as a
function of magnetic field. The magnetic flux threading a periodic
orbit adds a phase $\phi-j=A_jH/\Phi_0$ to the action in each term $j$
in the Gutzwiller formula (\ref{Gutzwiller}), where $A_j$ is the
directed area under the trajectory, $H$ is the applied magnetic field,
and $\Phi_0$ is the flux quantum. Correspondingly, formula
(\ref{EcorG}) is modified to
\begin{eqnarray}
&&\langle E_g(\delta_1,H_1)E_g(\delta_2,H_2)\rangle = 2 E_C^2 \sum_j
|R_j|^2 \tilde{K}\left(\frac{E_Ct_j}{\hbar}\right)\times \nonumber\\
&& \ \ \ \ \ \cos \left(\frac{H_1A_j}{\Phi_0}\right) \cos
\left(\frac{H_2A_j}{\Phi_0}\right) \cos \left[2n_j\left(\delta_1
-\delta_2\right)\right].
\label{EcorGH}
\end{eqnarray}
In analogy with Eq.~(\ref{EcorGI}), we transform Eq.~(\ref{EcorGH}) to
\begin{eqnarray}
&&\langle E_g(\delta_1,H_1)E_g(\delta_2,H_2)\rangle \!=\!
\frac{E_C^2}{\pi^2}\!\sum_n\!\!\int\!
dA\!\!\int_0^\infty\!\!\frac{dt}{t} P_n(t;A)\!  \times \nonumber\\ &&
\tilde{K}\left(\!\frac{E_Ct}{\hbar}\!\right)\cos
\left(\!\frac{H_1A}{\Phi_0}\!\right) \cos
\left(\!\frac{H_2A}{\Phi_0}\!\right) \cos\!\left[2n\left(\delta_1\!
-\!\delta_2\right)\right],
\label{EcorGHI}
\end{eqnarray}
where $P_n(t; A)dA$ differs from probability $P_n(t)$ by one more
constraint: the directed area swept by a trajectory lies within the
interval $\left[A;A+dA\right]$.

In a chaotic system, the probability $P_n(t; A)$ factorizes:
\begin{equation}
P_n(t; A)=P_n(t)p(t;A).
\label{factorization}
\end{equation}
Here $P_n(t)$ is defined by Eq.~(\ref{Pn}), and the distribution
function of the areas is Gaussian:
\begin{eqnarray}
&&p(t;A)=\frac{1}{2\sqrt{\pi \langle
A^2(t)\rangle}}\exp\left\{-\frac{A^2}{4\langle A^2(t)\rangle}\right\},
\nonumber\\ &&\langle A^2(t)\rangle=\frac{E_Tt}{\hbar}{\cal A}^2.
\label{disareas}
\end{eqnarray}
The formula for $\langle A^2(t)\rangle$ shows that the typical area
under the electron trajectory differs from the area of the dot ${\cal
A}$, and grows as $\sqrt t$. This law is applicable at the time scale
exceeding the ergodic time $\hbar/E_T$, and reflects the time
dependence of r.m.s. of the random winding number for the trajectory
of an electron bouncing off the walls of the dot.

As we already discussed, the characteristic time an electron spends in
the dot is $\hbar/E_C$. The characteristic area accumulated during
this time is $\sqrt{E_T/E_C}$. A magnetic field produces an
appreciable effect if a flux penetrating through this area is of the
order of $\Phi_0$.  Thus, the correlation magnetic field is controlled
by the charging energy:
\begin{equation}
H_c=\frac{\Phi_0}{\cal A}\sqrt{\frac{E_C}{2\pi E_T}}.
\label{Bc}\label{eq:2.25}
\end{equation}
Using Eqs.~(\ref{EcorGHI})--(\ref{disareas}), we find the correlation
function:
\begin{equation}
\langle E_g({\cal N}_1,H_1)E_g({\cal N}_2,H_2)\rangle = E_C\Delta\!
\sum_{\gamma=+,-}\!
\Lambda_E\left(\!\frac{H_\gamma^2}{H_c^2}\!\right) \cos 2\pi n,
\label{EHresult}
\end{equation}
where we introduced the short hand notation $H_\pm=H_1\pm H_2$ and
$n={\cal N}_1-{\cal N}_2$.  Calculation of the exact form of
dimensionless function $\Lambda_E(x) = \int_0^\infty dy
e^{-xy}\tilde{K}(y)$, and of the numerical coefficient in
Eq.~(\ref{eq:2.25}) requires more involved treatment which is a
subject of the following sections.

Equations (\ref{corGutz}) and (\ref{EHresult}) constitute the main
qualitative result of this section. We were able to demonstrate the
oscillations of the ground state energy with the applied gate
voltage. The phase of those oscillations is random, so that the
oscillations can be revealed only in the correlation
functions. Unfortunately, these simple qualitative arguments are not
sufficient for finding the precise form of the correlation functions.
Moreover, the assumption of the Fermi liquid behavior is valid only
for the spinless electrons. It is known that the low-energy behavior
of the $s=1/2$ electrons is equivalent\cite{Matveev95} to that of the
two-channel Kondo problem in its strong coupling fixed point
displaying a non-Fermi liquid behavior. Quantitative study of the
system in this case will be presented later, see Sec:~\ref{sec:5b}.

\section{The model}
\label{sec:3}
The main difficulty of the problem is in the non-perturbative nature
of the Coulomb blockade effect. Derivation of an effective
one-dimensional model is our first step in overcoming this
difficulty. The interaction energy (\ref{Hc}) depends only on the
total number of electrons in the dot. The change of this number is
associated with electrons propagating through the channel. Because the
dynamics of the channel is one-dimensional, the charging effects of
the system can be considered on the basis of a one-dimensional
Hamiltonian\cite{Matveev95,Flensberg95}. However, original problem was
at least two-dimensional, so backscattering of the electrons by the
walls of the dot can not be accounted for by the one-dimensional
Hamiltonian. Instead, of an effective Hamiltonian, we were able to
find an effective action which depends only on the electron variables
of the one-dimensional channel.  If there were no interaction, such an
approach would have no advantage; however, in the presence of
interaction it becomes very powerful. The interaction will be exactly
accounted for by means of bosonization, see Section~\ref{sec:4}.

Electrons are backscattered into the channel by the walls of the dot
at random times, therefore the action we derive in
Subsection~\ref{sec:3a} has a non-local in time, random term. This
term, however, can be treated perturbatively by virtue of the small
parameter, $\Delta/E_c \ll 1$.  With the help of the action,
calculation of the correlation functions of energies and differential
capacitances can be performed by the standard diagrammatic
methods\cite{AltshulerSimons}. At energies less than $E_T$ the
averages become universal. In this regime, it is also possible to
formulate the model starting from the Random Matrix
Hamiltonian\cite{Brouwer1}, see Subsection~\ref{sec:3b}.

The applicability of description of the interaction by Eq.~(\ref{Hc}),
{\em i.e.}, of the constant-interaction model, is discussed in
Subsection~\ref{sec:3c}. We will show that the corrections to this
description are of the order of $1/g$, where $g\gg 1$ is the
dimensionless conductance of the dot.

\subsection{"Conventional" formulation.}
\label{sec:3a}
We start with the Hamiltonian of the system,
\begin{equation}
\hat{H}=\hat{H}_F + \hat{H}_C,
\label{Hamiltonian}
\end{equation} 
where $\hat{H}_F$ is the Hamiltonian of non-interacting electrons,
\begin{equation}
\hat{H}_{F}=\int d{\bf
r}\left[\frac{1}{2m}\nabla\psi^\dagger\nabla\psi + \left(-\mu+U({\bf
r})\right)\psi^\dagger\psi\right].
\label{HF}
\end{equation}
The potential $U({\bf r})$ describes the confinement of electrons to
the dot and channel.

The interaction Hamiltonian $\hat{H}_C$ is given by Eq.~(\ref{Hc}),
and the charge of the dot is
\begin{equation}
\frac{\hat{Q}}{e}= \int_{\mbox{dot}} d{\bf r}\psi^\dagger\psi,
\label{Q}
\end{equation}
where the integration is performed within the dot. Of course, the
boundary separating the dot from the lead is not defined.  However,
this ambiguity can be absorbed into the definition of dimensionless
gate voltage ${\cal N}$.

For the purpose of the evaluation it is more convenient, however, to
change the definition of the charge. Noticing that the total number of
particles in the system is an integer number which can be added to the
parameter ${\cal N}$ without affecting any periodic in ${\cal N}$
observables, we write
\begin{equation}
\frac{\hat{Q}}{e}= - \int_{\mbox{channel}} d{\bf r}\psi^\dagger\psi.
\label{Q1}
\end{equation}

To calculate the ground state energy, we start with the thermodynamic
potential,
\begin{equation}
\Omega=-\frac{1}{\beta}\ln\left({\rm Tr}e^{-\beta\hat{H}}\right),
\label{eq:3.5}
\label{Omega}
\end{equation}
where temperature $T=1/\beta$.

We evaluate the trace in two steps, ${\rm Tr}\dots= {\rm Tr}_{1}{\rm
Tr}_2 \dots$, where $1$ and $2$ indicate the fermionic operators
belonging to the channel and dot respectively.  Because all the
interaction is attributed to the channel, see Eq.~(\ref{Q1}), the
charging energy operator is not affected by the summation in the dot,
and can be omitted in the intermediate formulas. The non-interacting
Hamiltonian (\ref{HF}) can be presented as
$\hat{H}_F=\hat{H}_1+\hat{H}_2+\hat{H}_{12}$, where $\hat{H}_1$ and
$\hat{H}_2$ are the noninteracting Hamiltonians of the channel and of
the dot respectively, and $\hat{H}_{12}$ connects the dot with the
channel. Thus, we write
\begin{eqnarray}
&{\rm Tr}_2e^{-\beta \hat{H}_F}= {\rm Tr}_2e^{-\beta
\left(\hat{H}_1+\hat{H}_2 +\hat{H}_{12}\right)}=&
\label{transformation}\\
&e^{-\beta \hat{H}_1 }{\rm Tr}_2 \left[ e^{-\beta \hat{H}_2} T_\tau
e^{-\int_0^\beta d\tau \hat{H}_{12}(\tau)} \right]=&\nonumber\\
&e^{-\beta \Omega_2}e^{-\beta \hat{H}_1 } T_\tau e^{
\frac{1}{2}\int_0^\beta \!\! d\tau _1\int_0^\beta\!\! d\tau _2
\langle\hat{H}_{12}(\tau_1)\hat{H}_{12}(\tau_2)\rangle_2}&.  \nonumber
\end{eqnarray}
Here $\hat{H}_{12}(\tau ) = e^{\tau \left(\hat{H}_1 +\hat{H}_2\right)}
\hat{H}_{12} e^{-\tau \left(\hat{H}_1 +\hat{H}_2\right)}$ is the
interaction representation of the Hamiltonian connecting the dot and
lead, $T_\tau$ stands for the chronological ordering, and $\Omega_2 =
-T \ln {\rm Tr_2} e^{-\beta \hat{H}_2}$ is the thermodynamic potential
of non-interacting electrons in the dot. Averaging $\langle \dots
\rangle_2$ over the electronic variables of the dot is defined by the
relation: $\langle \dots \rangle_2 = e^{\beta\Omega_2}{\rm Tr_2
}\left(e^{-\beta \hat{H}_2}\dots\right)$. Thermodynamic potential
$\Omega_2$ does not depend on ${\cal N}$, and it will be omitted.

The operator Eq.~(\ref{transformation}) depends only on the electron
variables of the channel. The evaluation of the last factor in
Eq.~(\ref{transformation}) is performed in Appendix~\ref{ap:1}. This
yields
\begin{eqnarray}
&&\frac{1}{2}\langle\hat{H}_{12}(\tau_1)\hat{H}_{12}(\tau_2)\rangle_2=
-\bar{\psi}(\tau_1; 0) {\psi}(\tau_2; 0) \times \label{A1}\\
&&\frac{1}{4m^2}\!
\int\! dydy^\prime
\phi({y})\phi({y}^\prime)
\partial^2_{xx^\prime}
{\cal G}(\tau_1 -\tau_2; {\bf r}, {\bf r}^\prime),
\nonumber
\end{eqnarray}
where $\psi(\tau; x)=e^{\tau \hat{H}_1}\psi( x)e^{-\tau \hat{H}_1}$
are the one dimensional fermionic operators of the channel in the
interaction representation, $\bar{\psi}(\tau)=\psi^\dagger (-\tau )$,
and ${\cal G}$ is the exact Matsubara Green function of the closed dot
subjected to the zero boundary condition. The wave function $\phi(y)$
describes the transverse motion in the single-mode channel, and the
coordinates $x$, $x'$ in the derivative of the Green function ${\cal
G}$ are set to $+0$.

The detailed behavior of the Green function ${\cal G}$ depends on the
particular shape of the dot. It is convenient to separate the
fluctuating part of the Green function, ${\cal G}={\bar{\cal G}}
+{\tilde{\cal G}}$, and to combine the proportional to ${\bar{\cal
G}}$ sample-independent part of
$\langle\hat{H}_{12}(\tau_1)\hat{H}_{12}(\tau_2)\rangle_2$ with the
Hamiltonian $H_1$ in Eq.~(\ref{transformation}).

We rewrite Eq.~(\ref{A1}) in the form
\begin{eqnarray}
\langle\hat{H}_{12}(\tau_1)\hat{H}_{12}(\tau_2)\rangle_2&=&
\overline{\langle\hat{H}_{12}(\tau_1)\hat{H}_{12}(\tau_2)\rangle_2}
-\nonumber\\
&&2\bar{\psi}(\tau_1; 0) {\psi}(\tau_2; 0) L(\tau_1-\tau_2), 
\label{A2}
\end{eqnarray}
where kernel $L$ is given by
\begin{equation}
L(\tau)=\!
\frac{1}{4m^2}\!
\int\! dydy^\prime
\phi({y})\phi({y}^\prime)
\partial^2_{xx^\prime}
\tilde{{\cal G}}(\tau; {\bf r}, {\bf r}^\prime).
\label{kernel}
\end{equation}
One can check by a direct calculation that
\begin{equation}
e^{-\beta \hat{H}_1 } 
T_\tau e^{ \frac{1}{2}\int_0^\beta \!\! d\tau _1\int_0^\beta\!\! d\tau _2
\overline{\langle\hat{H}_{12}(\tau_1)\hat{H}_{12}(\tau_2)\rangle_2}}
\propto {\rm Tr}_> e^{-\beta \hat{H}_{1D} } ,
\label{allaxis}
\end{equation}
where ${\rm Tr}_>$ stands for the trace of the one-dimensional
fermionic operators on the positive half-axis, and
\begin{equation}
\hat{H}_{1D}=\int_{-\infty}^\infty
dx\left[\frac{1}{2m}\nabla\psi^\dagger\nabla\psi -\mu\psi^\dagger\psi\right]
\label{H1D}
\end{equation}
is the one dimensional Hamiltonian defined on the whole real axis. The
proportionality coefficient in Eq.~(\ref{allaxis}) does not contribute
to any observable quantity and we omit it.  We substitute
Eq.~(\ref{A2}) into Eq.~(\ref{transformation}), use
Eq.~(\ref{allaxis}), restore the charging energy, see Eq.~(\ref{Q1}),
and obtain
\begin{equation}
{\rm Tr} e^{-\beta\hat{H}}\propto {\rm Tr}\left(e^{-\beta\hat{H}_0}
T_{\tau}e^{-\hat{S}}\right) .
\label{eq:3.12}
\label{Z}
\end{equation}
One-dimensional effective Hamiltonian is given by
\begin{equation}
\hat{H}_{0}=\hat{H}_{1D} + E_C
\left({\cal N} + \int_{-\infty}^0:\psi^\dagger\psi: dx\right)^2,
\label{H0}
\end{equation}
where $:\dots:$ stands for the normal ordering. The effective action
$\hat{S}$ in Eq.~(\ref{Z}),
\begin{equation}
\hat{S}=\int_0^\beta d\tau_1 d\tau_2
L\left(\tau_1-\tau_2\right)\bar{\psi}(\tau_1;0) {\psi}(\tau_2;0),
\label{action}
\end{equation}
has kernel $L$ defined by Eq.~(\ref{kernel}). If the electrons with
spin are considered, the summation over spin indices is implied in the
above formulas.

As we are interested in the dynamics of the system at energies much
smaller than the Fermi energy, we can linearize the spectrum of one
dimensional fermions.  Writing $\psi(x)=e^{-ik_Fx}\psi_L(x)
+e^{ik_Fx}\psi_R(x) $, where $\psi_L$ and $\psi_R$ are the left- and
right-moving fermions respectively, we obtain from Eqs.~(\ref{H1D}) and
(\ref{H0}):
\begin{eqnarray}
\hat{H}_0 &=& 
iv_F\int_{-\infty}^\infty dx \left\{\psi_L^\dagger\partial_x\psi_L -
\psi_R^\dagger\partial_x\psi_R
\right\} \label{Heff}\label{eq:3.15}\\
&&+ \frac{E_C}{2}\left(\int_{-\infty}^0dx
:\psi_L^\dagger\psi_L +\psi^\dagger_R\psi_R:+{\cal N}\right)^2,
\nonumber
\end{eqnarray}
where $v_F$ is the Fermi velocity in the channel. In Eq.~(\ref{Heff})
the fermionic fields are assumed to be smooth on the scale of the Fermi
wavelength.  The action has the following form in terms of the left- and
right-movers:
\begin{eqnarray}
\hat{S}=&&\int_0^\beta d\tau_1 d\tau_2 L\left(\tau_1-\tau_2\right)\times
\label{lraction}\\
&&\left[\bar{\psi}_L(\tau_1)+
\bar{\psi}_R(\tau_1)\right]\left[\psi_L(\tau_2)+\psi_R(\tau_2)\right].
\nonumber
\end{eqnarray}
Finite reflection in the channel can be taken into account by adding
to the Hamiltonian (\ref{Heff}) one more term:
\begin{equation}
c\hat{H}_{bs} = |r| v_F 
\left(\psi_L^\dagger(0) \psi_R(0) + \psi_R^\dagger(0) \psi_L(0)\right),
\label{reflection}
\end{equation}
where $|r|^2 \ll 1$ is the reflection coefficient.
The thermodynamic potential $\Omega ({\cal N})$ can be found from
\begin{equation}
\Omega ({\cal N}) = -T \ln {\rm Tr} 
\left(e^{-\beta\left(\hat{H}_0 +\hat{H}_{bs} \right) }T_\tau e^{-\hat{S}}\right).
\label{O}
\end{equation}
The differential capacitance $C_{\it diff}\left({\cal N}\right)$ of the system
is then given by
\begin{equation}
C_{\it diff}\left({\cal N}\right)= C\left(1 - 
\frac{1}{E_C}\frac{\partial^2\Omega}{\partial{\cal N}^2}\right).
\label{Cdif}
\end{equation}
Equations (\ref{Heff}) and (\ref{reflection}) were first suggested in
Refs.~\onlinecite{Matveev95,Flensberg95}.  

So far we succeeded in reducing the original problem to the effective
one-dimensional problem, where all the features of the chaotic motion
of electrons in the dot are incorporated into the non-local in time
action. The action fluctuates strongly from sample to sample, and we
should study the statistics of these fluctuations.

\subsubsection{Statistics of $L(\tau)$.}

It is convenient to use the Lehmann representation for the function
$L(\tau)$ of Eq.~(\ref{kernel}):
\begin{equation}
L(\tau)= \int_{-\infty}^{\infty} \frac{dt}{2\pi}\left[L^R(t)-L^A(t)\right]\frac{\pi
T} {\sinh\left[\pi T(t+i\tau)\right]},
\label{Lehman}
\end{equation}
where the retarded and advanced kernels $L^{R,A}$ are given by
Eq.~(\ref{kernel}), with $\tilde{\cal G}$ replaced by the exact
advanced and retarded Green functions $\tilde{\cal G}^{R,A}$
respectively. It is well known that the averaged products of the type
$\overline{\tilde{\cal G}^{R}\tilde{\cal G}^{R}}$ and
$\overline{\tilde{\cal G}^{A}\tilde{\cal G}^{A}}$ vanish, and the
products of the retarded and advanced Green functions can be expressed
in terms of the classical propagators -- diffusons ${\cal P}^D$ and
Cooperons ${\cal P}^C$,
\begin{mathletters}
\label{eq:3.21}
\label{classics}
\begin{eqnarray}
&&\overline{\tilde{\cal G}^{R}_{H_1}(t_1;{\bf r}_1^+, {\bf r}_2^+)
\tilde{\cal G}^{A}_{H_2}(t_2; {\bf r}_2^- , {\bf r}_1^-)}
=2\pi\nu\delta(t_1+t_2) \times \label{diffuson}\\
&&\quad\quad 
{\cal F}({\bf r}_1^+, {\bf r}_1^- )
{\cal F}({\bf r}_2^+, {\bf r}_2^-  )
{\cal P}^D\left(t_1;  {\bf R}_1, {\bf R}_2\right);
\nonumber\\
&&\overline{\tilde{\cal G}^{R}_{H_1}(t_1; {\bf r}_1^+, {\bf r}_2^+)
\tilde{\cal G}^{A}_{H_2}(t_2; {\bf r}_1^-, {\bf r}_2^-)}
=2\pi\nu\delta(t_1+t_2) \times \label{cooperon}\\
&&\quad\quad
 {\cal F}({\bf r}_1^+, {\bf r}_1^-  )
{\cal F}({\bf r}_2^+, {\bf r}_2^- )
{\cal P}^C\left(t_1; {\bf R}_1,{\bf R}_2\right),
\nonumber
\\
&&{\cal F}({\bf r}_1, {\bf r}_2  )\!=\!
 {\rm Im}\frac{\overline{G^A}}{\pi}\! =\!
J_0\left(k_F|{\bf r}_1 -{\bf r}_2 |\right)\! -\!
J_0\left(k_F|{\bf r}_1 -\hat{\cal R}{\bf r}_2 |\right),
\nonumber
\end{eqnarray}
where $J_0(x)$ is the zeroth Bessel function,
${\bf r}_i^\pm = {\bf R}_i\pm {\bf r}_i/2$, and 
$k_F$ is the Fermi wavevector. Point $\hat{\cal R}{\bf r}$ is the
coordinate of the image charge created by the charge in the point ${\bf r}$,
so that the propagators (\ref{eq:3.21}) satisfy proper zero boundary
conditions. Green functions here are taken at
different values of magnetic fields $H_1, H_2$. In Eqs.~(\ref{classics}),
$\nu=m/2\pi$ is the density of states per unit area. Equations (\ref{classics})
are valid if the arguments of the Green functions are close to each other
pairwise: $|r_{1,2}|$ must be much smaller than the elastic mean free
path for a diffusive dot, and much smaller than the dot size for a
ballistic dot. Boundary is assumed to be smooth on the scale of the
Fermi wavelength.  
\end{mathletters}

Retarded classical propagators in the diffusive dot 
 satisfy  diffusion-like
equations\cite{AltshulerKhmelnitskii}
\begin{equation}
\left\{
\frac{\partial}{\partial t}\! -\!
D\left[ \nabla +\frac{ie}{c}
\left(\!\!\!
\begin{array}{ll}
&{\bf A}_1 - {\bf A}_2\\
&{\bf A}_1 + {\bf A}_2
\end{array}
\!\right)
\right]^2
\right\}
\left\{\!
\begin{array}{ll}
{\cal P}^D\\
{\cal P}^C
\end{array}
\!\right\}\! =\!
 \delta\left(\!{\bf R}_1\! -
\! {\bf R}_2\!
\right),
\label{DC}
\end{equation}
where $D$ is the diffusion coefficient, and the vector potentials
${\bf A}_{1,2}$ are defined so that $\nabla \times {\bf A}_{1,2} =
H_{1,2}$. For a ballistic dot, Eqs.~(\ref{DC}) should be substituted
by the corresponding Liouville (or, to be more precise,
Perron-Frobenius) equation, and Eq.~(\ref{eq:3.21}) should be somewhat
changed\cite{AleinerLarkin}.
However, in the universal limit considered
in this paper, there is no difference between the ballistic and
diffusive dots.

The universal limit corresponds to a large time scale at which the
semiclassical electron orbit covers all the available phase space.  At
such a time scale, the classical probabilities no longer depend on the
coordinate and acquire the form:
\begin{equation}
{\cal P}^{D,C} = \frac{1}{{\cal A}}\theta (t) e^{-t/\tau_H^{D,C}},
\label{universal}
\end{equation}
where $\theta(x)$ is the step function, ${\cal A}$ is the area of the
dot, and the decay times associated with the magnetic field are given
by
\begin{equation}
\frac{1}{\tau_H^{D,C} }= E_T\left(\frac{\Phi_1\mp \Phi_2}{\Phi_0}\right)^2.
\label{tauH}
\end{equation}
Here $\Phi_0=e/c\hbar$ is the flux quantum, $\Phi_{1,2} = {\cal A}H_{1,2}$
are the fluxes through the dot corresponding to the fields $H_1$ and $H_2$,
and the Thouless energy $E_T$ is of the order of $\hbar D/{\cal A}$ for a
diffusive dot, and of the order of $\hbar v_F/\sqrt{\cal A}$ for a ballistic dot.

The correlation functions of the retarded and advanced parts of the
kernel $L$ can be expressed, with the help of Eqs.~(\ref{kernel})
and (\ref{classics}), in terms of the diffuson and Cooperon. The kernel
$L$, see Eq.~(\ref{kernel}), depends on the Green function at
coinciding spatial arguments. Therefore, both pairings leading to the
diffuson and Cooperon upon averaging, should be taken into account. In
the universal regime, see Eq.~(\ref{universal}), integrals in the
transverse direction in Eq.~(\ref{kernel}) can be calculated using the
normalization condition $\int dy \phi^2(y) =1$. As the result, we find
\begin{eqnarray}
\langle
L^R_{H_1}(t_1)L^A_{H_2}(t_2)\rangle&=&\frac{v_F^2\Delta}{2\pi}\delta
(t_1+t_2)\theta (t_1) \times\label{LRLA}\label{eq:3.25} \\
&&\left\{e^{-t_1/\tau_H^D}+e^{-t_1/\tau_H^C}\right\}, \nonumber
\end{eqnarray}
where $\Delta = 1/(\nu{\cal A})$ is the mean level spacing
of the dot.  Averages of the type $\overline{L^RL^R}$ vanish. Because
we are interested in times smaller than the Heisenberg time
$\hbar/\Delta$, the higher moments can be decoupled by using the Wick
theorem and the pair correlation functions are defined by
Eq.~(\ref{LRLA}).

\subsection{Random matrix formulation.}
\label{sec:3b}

We start from dividing the entire system into two parts, the leads and the
dot. In general, the Hamiltonian $\hat{H}$ of the system can be
represented as
\begin{equation}
\hat{H} = \hat{H}_L + \hat{H}_D + \hat{H}_{LD}.
\label{eq:3.26}
\end{equation}
The Hamiltonian of the leads is of the form
\begin{equation}
\hat{H}_L = v_F \sum_{k,j} k \psi^\dagger_{k,j} \psi_{k,j} ,
\label{eq:3.27}
\end{equation}
where we linearized the electron spectrum in the leads, and measured all the
energies from the Fermi level. Index $k$ is the longitudinal momentum in a mode
propagating along the channel connected to the dot, and index $j=1,\dots,N$ labels
these modes (summation over spin indices is implied whenever it is necessary).
For the sake of simplicity, we assume the same Fermi velocity  in all the
modes, the general case can be reduced to Eq.~(\ref{eq:3.27}) by the corresponding
rescaling. Hamiltonian
$\hat{H}_D=\hat{H}_n + \hat{H}_C$ of the dot consists of the non-interacting part 
\begin{equation}
\hat{H}_n = \sum_{\alpha,\beta}{\cal H}_{\alpha,\beta}\psi^\dagger_\alpha\psi_\beta,
\label{eq:3.28}
\end{equation}
and the interaction term $\hat{H}_C$, which is described by the Hamiltonian
(\ref{Hc}) with the charge $\hat{Q}$ given by 
\begin{equation}
\frac{\hat{Q}}{e} = \sum_{\alpha}\psi^\dagger_\alpha\psi_\alpha.
\label{eq:3.29}
\end{equation}
(In this subsection, we reserve Greek and Latin letters for labeling the
fermionic states in the dot and in the leads respectively.) 
For definiteness, we restrict the discussion to the case of orthogonal
ensemble, generalization to other cases is straightforward. 
Elements ${\cal H}_{\alpha,\beta}$ in Eq.~(\ref{eq:3.28}) form a real random
 Hermitian
matrix ${{\cal H}}$ of size $M\times M$, ($M\to\infty$),  belonging to the Gaussian
ensemble
\begin{equation}
P({\cal H}) \propto \exp\left( - \frac{\pi^2}{4\Delta^2 M}{\rm Tr}{\cal H}^2\right),
\label{eq:3.30}
\end{equation}
where $\Delta$ is the mean level spacing near the center of the band. 
Finally, Hamiltonian $\hat{H}_{LD}$ in Eq.~(\ref{eq:3.26}) describes the
coupling  of the dot to the leads, and has the form
\begin{equation}
\hat{H}_{LD} = \sum_{k,j,\alpha} \left(W_{\alpha, j}\psi^\dagger_\alpha
\psi_{k,j} + {\rm h.c.}\right),
\label{eq:3.31}
\end{equation}
where the coupling constants $W_{\alpha,j}$ is a real $M\times
N$ matrix $W$.

Let us list the needed averaged quantities corresponding to the Gaussian
ensemble Eq.~(\ref{eq:3.30}). The averaged density of states is given by the
semicircle law
\begin{equation}
\rho(\epsilon) = \overline{{\rm Tr}\delta\left(\epsilon - {\cal
H}\right)} = {\rm Re}\frac{1}{\Delta}
\sqrt{1-\left(\frac{\pi\epsilon}{\Delta M}\right)^2}.
\label{eq:3.32}
\end{equation}
We will need only properties of the system at energies $\epsilon$ much
smaller than the width of the band $\Delta N/\pi$, and we will neglect
the energy dependence of the averaged density of states.
Average of the Green functions ${\cal G}^{R,A}(\epsilon ) =
\left(\epsilon - {\cal H}\pm i 0\right)^{-1}$ have the form
\begin{equation}
\overline{{\cal G}^{R,A}_{\alpha,\beta }} = 
\mp i\pi\delta_{\alpha\beta}\frac{1}{M\Delta}.
\label{eq:3.33}
\end{equation}
Random matrix counterparts of the diffuson and Cooperon propagators
(\ref{classics}) can be written as
\begin{eqnarray}
&&\overline{{\cal G}^{R}_{\alpha,\beta }(\epsilon +\omega )
{\cal G}^{A}_{\gamma,\delta }(\epsilon)
} = \nonumber\\
&&\quad
\frac{2\pi }{M\Delta}
\left( {\cal P}^D(\omega) \delta_{\alpha\delta}\delta_{\beta\gamma} +
 {\cal P}^C(\omega) \delta_{\alpha\gamma}\delta_{\beta\delta}\right),
\nonumber\\
&&{\cal P}^D(\omega)= {\cal P}^C(\omega) =\frac{1}{M}\frac {1}{-i\omega
+ 0},\label{eq:3.34}
\end{eqnarray}
where $\tilde{{\cal G}}={\cal G} -\overline{\cal G}$. Averages of the
type $\overline{\tilde {\cal G}^R\tilde {\cal G}^R}$  and
 $\overline{\tilde {\cal G}^A\tilde {\cal G}^A}$ vanish.
Formulas (\ref{eq:3.32}) - (\ref{eq:3.34}) have the accuracy $\sim
1/M$ and they neglect the oscillatory $\omega$-dependences on the scale of the
order of $\Delta$. This accuracy, however, is sufficient for us because, as we already
discussed, the relevant results are contributed by the energy strip of the
width $E_C \gg \Delta$.

Our purpose now is to derive the effective action theory similar to
Eqs.~(\ref{Heff}) -- (\ref{O}) starting from the random matrix model. 
Before doing so, let us review some useful properties
of the system (\ref{eq:3.26}) in the absence of the interaction,
$E_C=0$. In this case, electron transmission at energy $\epsilon$ is
completely characterized by $N\times N$ scattering matrix
$S(\epsilon)$:
\begin{equation}
S(\epsilon) = 1- 2\pi i \nu W^{\dagger}\left[\epsilon -{\cal H}
+i\pi\nu WW^\dagger\right]^{-1}W,
\label{eq:3.35}
\end{equation}
where $\nu =1/(2\pi v_F)$ is the one-dimensional density of states in
the leads. Coupling matrix $W$ can be represented in the form
\begin{equation}
W = \sqrt{\frac{\Delta M}{\pi^2 \nu }} UO\tilde{W},
\label{eq:3.36}
\end{equation} 
where $U$ is an orthogonal $M \times M$ matrix, $\tilde{W}$ is a real
$N\times N$ matrix and $O$ is $M\times N$ matrix, $O_{\alpha
j}=\delta_\alpha j,\quad 1 \leq\alpha\leq N$. Because the distribution
function (\ref{eq:3.26}) is invariant under rotations ${\cal H} \to
U{\cal H}U^\dagger$, matrix $U$ in Eq.~(\ref{eq:3.36}) can be
omitted. Substituting Eq.~(\ref{eq:3.36}) into Eq.~(\ref{eq:3.35}) and
performing ensemble averaging with the help of Eq.~(\ref{eq:3.33}), we
obtain for the average scattering matrix:
\begin{equation}
\overline{S} = \frac{1- \tilde{W}^{\dagger}\tilde{W}}
{1+ \tilde{W}^{\dagger}\tilde{W}}.
\label{eq:3.37}
\end{equation}
At $\tilde{W}^\dagger \tilde{W}=1$ the average scattering matrix vanishes,
see Eq.~(\ref{eq:3.37}). It indicates that matrix $S$ belongs to the circular
ensemble (corresponding to the regime of ``ideal contacts''). Deviation of
$\tilde{W}^\dagger\tilde{W}$ from a unit matrix can be attributed to the scattering on
the contacts between the leads and the dot. This scattering is described by unitary
symmetric $2N \times 2N$ matrix
\begin{equation}
S_c = \left(\begin{array}{ll}
r\quad & t^\dagger\\
t\quad & r^\prime
\end{array}
\right)
\label{eq:3.38}
\end{equation}
with $N\times N$ matrices $r,\ r^\prime,\ t$ being defined
by 
\begin{eqnarray}
&\displaystyle{r =\frac{1- \tilde{W}^{\dagger}\tilde{W}}{1+
\tilde{W}^{\dagger}\tilde{W}},\quad r^\prime =-\tilde{W}\frac{1-
\tilde{W}^{\dagger}\tilde{W}} {1+ \tilde{W}^{\dagger}\tilde{W}}{\tilde W}^{-1},}&
\nonumber\\ &\displaystyle{t= \tilde{W}\frac{2}{1+\tilde{W}^\dagger\tilde{W}}.}&
\label{eq:3.39}
\end{eqnarray}
The explicit relations between the coupling matrices $W$ and the scattering
matrix in the contacts (\ref{eq:3.38}) were first obtained by
Brouwer\cite{Brouwer95}.

Now we turn to the derivation of the effective action. For technical
reasons, [see discussion above Eq.~(\ref{Q1})], we replace the charge
operator (\ref{eq:3.29}) with
\begin{equation}
\frac{\hat{Q}}{e} = - \sum_{k,j} \psi^\dagger_{k,j}\psi_{k,j}.
\label{eq:3.40}
\end{equation} 
After this replacement, the Hamiltonian of the system becomes
quadratic in fermionic operators of the dot, so that this part of the
system can be integrated out:
\begin{eqnarray}
&&{\rm Tr}_D e^{-\beta\hat{H}}= {\rm Tr}_D
e^{-\beta\left(\hat{H}_L+\hat{H}_C + \hat{H}_n + \hat{H}_{LD}\right)}=
\label{eq:3.41}\\
&&\ \ 
e^{-\beta\left(\hat{H}_L+\hat{H}_C \right)}e^{-\beta\Omega_n}
T_\tau e^{\frac{1}{2}\int_0^\beta d\tau_1d\tau_2 
\langle\hat{H}_{LD}(\tau_1)\hat{H}_{LD}(\tau_2) \rangle_D}.
\nonumber
\end{eqnarray}
Here $\hat{H}_{LD}(\tau)$ is the interaction representation of the
coupling operator $\hat{H}_{LD}$ 
and averaging over Hamiltonian of the dot is defined
as $\langle \dots\rangle_D = e^{\beta\Omega_D}{\rm Tr}_D\left(
e^{-\beta\hat{H}_n}\dots\right)$. Thermodynamic potential $\Omega_D =
T\ln {\rm Tr} e^{-\beta \hat{H}_n}$ of the closed dot is independent
of the gate voltage ${\cal N}$ and it will be omitted.

The average in the last factor in Eq.~(\ref{eq:3.41}) is calculated
with the help of Eq.~(\ref{eq:3.31}) and of the definition of the
Matsubara Green function for the closed dot:
\begin{equation}
{\cal G}_{\alpha\beta}(\tau) = - \langle T_\tau \psi_\alpha
(\tau)\bar{\psi}_\beta (0)\rangle_D = \sum_{\omega_n}
e^{i\omega_n\tau}\left[\frac{1}{i\omega_n - {\cal
H}}\right]_{\alpha\beta},
\label{eq:3.42}
\end{equation}
where $\psi_\alpha (\tau) = e^{\hat{H}_n\tau}\psi_\alpha 
e^{-\hat{H}_n\tau}, \ \bar{\psi}(\tau)=\psi^\dagger(-\tau)$, 
and $\omega_n = \pi T (2n+1)$ is the fermionic Matsubara frequency.
The result is
\begin{eqnarray}
&&\langle\hat{H}_{LD}(\tau_1)\hat{H}_{LD}(\tau_2)\rangle_2=\\\label{eq:3.43}
&&\ \ \ 
-\sum_{k_1,k_2,j_1,j_2}\bar{\psi}_{k_1,j_1}(\tau_1) 
\left[W^\dagger {\cal G}(\tau_1 - \tau_2)
W\right]_{j_1j_2}{\psi}_{k_2,j_2}(\tau_2). \nonumber
\end{eqnarray}
We separate the averaged part of the Green function ${\cal G}=
\overline{\cal G} + \tilde{\cal G}$, use Eqs.~(\ref{eq:3.33}) and
(\ref{eq:3.36}), and obtain the Fourier transform of the kernel in
Eq.~(\ref{eq:3.43}),
\begin{equation}
W{\cal G}(i\omega_n)W^{\dagger}=-i\frac{ {\rm sgn}\,
\omega_n}{\pi\nu}\tilde{W}^\dagger\tilde{W} + \frac{\Delta
M}{\pi_2\nu}\tilde{W}^\dagger \tilde{\cal G}(i\omega_n)\tilde{W}.
\label{eq:3.44}
\end{equation}
The first term in the RHS of Eq.~(\ref{eq:3.44}) does not contain any
information about the dot, and its frequency dependence is the same as
of the Green function of free chiral fermions. It is, therefore,
possible (and very convenient) to transform this part of the action to
the Hamiltonian form by introducing fictitious fermionic fields
$b_{k,j},\ j=1,...N$.  Then, Eq. (\ref{eq:3.41}) acquires the form
\begin{equation}
{\rm Tr}_D e^{-\beta\hat{H}} \propto
{\rm Tr}_b \left( e^{-\beta \hat{H}_{eff}}T_\tau e^{-\hat{S}}\right);
\label{eq:3.45}
\end{equation}
the omitted ${\cal N}$- independent proportionality coefficient is
irrelevant. The effective Hamiltonian $\hat{H}_{eff}$ is given by
\begin{eqnarray}
\hat{H}_{eff}& =&v_F \sum_{j,k}k
\left(\psi^\dagger_{k,j}\psi_{k,j} +b^\dagger_{k,j}b_{k,j}\right)
+ \nonumber\\ &&\frac{1}{\pi\nu} \sum_{k_1,k_2;j_1,j_2}
\left[
b_{k_1,j_1}^\dagger
w_{j_1j_2}\psi_{k_2,j_2}
 + h.c.\right] +
\nonumber\\ 
&&E_C \left(\sum_{k,j}\psi^\dagger_{k,j}\psi_{k,j} +{\cal N}\right)^2,
\label{eq:3.46}
\end{eqnarray}
where $w_{j_1j_2}$ are the elements of the Hermitian matrix $w$
defined as
\begin{equation}
w =\left(\tilde{W}^\dagger {W}\right)^{1/2}.
\label{eq:3.47}
\end{equation}
Action ${\cal S}$ has the form 
\begin{equation}
{\cal S}=4\!\!\sum_{k_1,k_2;\ j_1,j_2}\!\!\int_0^\beta\! d\tau_1 d\tau_2 
\bar{\psi}_{k_1,j_1}(\tau_1)
L_{j_1j_2}(\tau_1\!-\!\tau_2)
{\psi}_{k_2,j_2}(\tau_2),
\label{eq:3.48}
\end{equation}
where the kernel $L$ is a $N\times N$ matrix given by
\begin{equation}
L(\tau)=\frac{\Delta M}{4\pi^2\nu}\tilde{W}^\dagger \tilde{\cal
G}(\tau)\tilde{W}.
\label{eq:3.49}
\end{equation}
Equation (\ref{eq:3.45}) can be easily checked by tracing
out fermions $b$ and using the relation
\[
\sum_{k_1,k_2}\langle T_\tau b_{k_1,j_1}(\tau)\bar{b}_{k_2,j_2}(0)\rangle
=i\pi\nu \delta_{j_1j_2}\sum_{\omega_n}e^{i\omega_n\tau}{\rm sgn}\, \omega_n.
\]

Hamiltonian $\hat{H}_{eff}$ can be rewritten in a more familiar
form. Introducing the Fourier transform of the fermionic fields
$\psi_j(x)=\sum_k e^{-ikx} \psi_{j,k}$ and  $b_j(x)=\sum_k e^{-ikx} b_{j,k}$,
we obtain from Eq.~(\ref{eq:3.46})
\begin{eqnarray}
\hat{H}_{eff}& =&iv_F \sum_{j}\int_{-\infty}^{\infty}dx
\left(\psi^\dagger_{j}\partial_x\psi_{j} +b^\dagger_{j}\partial_x
b_{j}\right) + \nonumber\\ &&\frac{1}{\pi\nu} \sum_{j_1,j_2} \left[
b_{j_1}^\dagger(0)w_{j_1j_2}\psi_{j_2}(0) + h.c.\right] + \nonumber\\
&&E_C \left(\sum_{j}\int_{\infty}^\infty dx \psi^\dagger_{j}\psi_{j}
+{\cal N}\right)^2.
\label{eq:3.50}
\end{eqnarray}
We are interested in the case of almost open ideal contacts,
$\parallel w -1\parallel \ll 1$. In this case, it is natural to change
the variables and reveal the small parameter of the perturbation
theory. Introducing left- and right-moving fermions,
\begin{eqnarray*}
\psi_j(x) = \psi_{L,j}(x)\theta(-x) + \psi_{R,j}(-x)\theta(x),\\ 
b_j(x)= i\left[ \psi_{R,j}(-x)\theta(-x) - \psi_{L,j}(x)\theta(x)\right]
\end{eqnarray*}
(the ambiguity of this definition at the origin should be resolved as
$\psi (0)=  \left[\psi(+0) + \psi(-0) \right]/2$), we obtain from
Eq.~(\ref{eq:3.50})
\begin{mathletters}
\label{eq:3.51}
\begin{eqnarray}
&&\hat{H}_{eff} = \hat{H}_0 +\hat{H}_{bs};\label{eq:3.51a}\\
&&\hat{H}_0=iv_F \sum_{j}\int_{-\infty}^{\infty}dx
\left(\psi^\dagger_{L,j}\partial_x\psi_{L,j} -
\psi^\dagger_{R,j}\partial_x \psi_{R,j}\right)+\nonumber\\ && E_C
\left(\sum_{j}\int_{-\infty}\!\! dx
:\psi^\dagger_{L,j}\psi_{L,j}+\psi^\dagger_{R,j}\psi_{R,j}: +{\cal
N}\right)^2; \label{eq:3.51b}\\
&&\hat{H}_{bs}=-i\sum_{ij}r_{ij}\psi_{L,i}^\dagger\psi_{R,j} + h.c.;
\label{eq:3.51c}\\
&&\hat{S}=\sum_{ij}\int_0^\beta d\tau_1 d\tau_2
L_{ij}\left(\tau_1-\tau_2\right)\times
\label{eq:3.51d}\\
&&\ \ \ \ \ \ 
\left[\bar{\psi}_{L,i}(\tau_1;0)+
\bar{\psi}_{R,i}(\tau_1;0)\right]
\left[\psi_{L,j}(\tau_2;0)+\psi_{R,j}(\tau_2;0)\right].
\nonumber
\end{eqnarray}
Here we neglected the terms related to the discontinuities of the
fermionic field at the origin which induces higher order terms in $r$,
and approximated reflection matrix $r \approx 1- w$, as follows from
Eqs.~(\ref{eq:3.47}) and (\ref{eq:3.39}) for $\parallel r\parallel \ll
1$. Within the same approximation we can put $\tilde{W} = 1$ in
Eq.~(\ref{eq:3.49}). Formulas (\ref{eq:3.51}) are analogous to
Eqs.~(\ref{Heff}) -- (\ref{O}) derived in a previous subsection.
\end{mathletters}

Finally, we have to study the statistics of the kernel (\ref{eq:3.49})
which can be expressed in terms of its advanced and retarded
counterparts by Lehmann formula (\ref{Lehman}). Performing time
Fourier transform of Eq.~(\ref{eq:3.34}), and using $\nu=1/(2\pi
v_F)$, we obtain
\begin{eqnarray}
\langle L^R_{ij}(t_1)L^A_{rs}(t_2)\rangle&=&\frac{v_F^2\Delta}{2\pi}\delta
(t_1+t_2)\theta (t_1) \times\label{eq:3.52} \\
&&\left\{\delta_{is}\delta_{jr}+\delta_{ir}\delta_{js} \right\}.
\nonumber
\end{eqnarray}
For the one-mode lead this result agrees  with the zero magnetic field version of
Eq.~(\ref{eq:3.25}).

The statistics of the kernel at different magnetic fields can be
obtained by adding purely imaginary Hermitian matrix to the original
matrix ${\cal H}$ in Eq.~(\ref{eq:3.28})\cite{Book,MehtaPandey}. This
would lead to the result analogous to the exponential decay in
Eq.~(\ref{eq:3.25}).  We will not describe details of such calculation
here, and refer the reader to the extensive literature on parametric
correlations\cite{Efetov,RMT}.

\subsection{Applicability of the model.}
\label{sec:3c}

So far, we were using a very simple model of the interaction (\ref{Hc})
which ascribes all the interaction effects to the variation of the number
of particles in the dot. However, a natural question arises: what is the
accuracy of this approximation? One may even think, that the effects
considered in this paper are completely washed out by remaining
interaction terms which we neglected. 

The purpose of this subsection is to show that the simple model (\ref{Hc}) of 
interactions in the dot leaves out only small, $\propto 1/g$, effects, where
$g=E_T/\Delta$ is the dimensionless conductance of the cavity ($E_T$ is the Thouless
energy).  For a diffusive dot in the metallic regime, and for a ballistic
non-integrable dot the conductance is large, $g\gg 1$.  Mesoscopic charge quantization
is adequately described by the model of interaction (\ref{Hc}), as long as the
$g \gg 1$, and the number of modes propagating to the lead is much smaller than
$g$. In other words, ergodic time of the dot $\hbar/E_T$ should be much larger than
the escape time of electron from the dot.

The electrons in the dot are described by the Hamiltonian
$\hat{H}={\hat H}_n + \hat{H}_{int}$, where the non-interacting part of
the Hamiltonian, $\hat{H}_{n}$, is given by Eq.~(\ref{eq:3.25}).  The
validity of the random matrix theory for $\hat{H}_{n}$ for the
energy scale smaller than Thouless energy was proven in
Ref.~\onlinecite{Berry85} for chaotic systems and in
Ref.~\onlinecite{Altshuler86} for diffusive systems.  

The general form of the interaction Hamiltonian is
\begin{equation}
\hat{H}_{int}= \frac{1}{2}\sum V_{\alpha\beta\gamma\delta}
\psi^\dagger_{\alpha,\sigma_1} \psi^\dagger_{\beta,\sigma_2} 
\psi_{\gamma,\sigma_2}\psi_{\delta,\sigma_1}.
\label{eq:3.53}
\end{equation}
In this subsection, we will write explicitly the spin indices $\sigma$
for  the fermionic operators. 
The interaction Hamiltonian (\ref{Hc}) corresponds to the
approximation of the matrix $V$ by
\begin{equation}
V\approx E_C \delta_{\alpha\delta}\delta_{\beta\gamma}.
\label{eq:3.54}
\end{equation}
Our goal now is to show that all the other matrix elements as well as the
mesoscopic fluctuations of matrix elements (\ref{eq:3.54}) are small.
Some of these calculations already appeared in the
literature\cite{1overg,MirlinBlanter}, however, we will present brief derivation to
make this paper self-contained.

The easiest way to study the statistics of the one electron
wave-function $\phi_\alpha ({\bf r})$ is to relate them to the Green
function and then use Eq.~(\ref{classics}).  By definition of the
retarded and advanced Green functions we have
\begin{equation}
G^A_\epsilon({\bf r}_1,{\bf r}_2 ) - G^R_\epsilon({\bf r}_1,{\bf r}_2)
=2\pi i \sum_\alpha \phi_\alpha ({\bf r}_1)\phi_\alpha({\bf r}_2
)\delta (\epsilon - \epsilon_\alpha),
\label{eq:3.55}
\end{equation}
where we assumed no magnetic field for simplicity. At given energy
$\epsilon$ only one function contributes into the sum in
Eq.~(\ref{eq:3.55}), so that the statistics of the Green functions is
related to that of the wave-functions.  Furthermore, it is known that
there is no correlation between level statistics and wave function in
the lowest order in $1/g$, see {\em e.g.}
Ref.~\onlinecite{Altshuler86}, so we can neglect the level
correlations and average $\delta$-function in Eq.~(\ref{eq:3.55})
independently.  As the result, we can estimate
\begin{equation}
\phi_\alpha ({\bf r}_1)\phi_\alpha({\bf r}_2)\approx \frac{\Delta}{2\pi i}
\left[G^A_{\epsilon_\alpha}({\bf r}_1,{\bf r}_2 ) -
G^R_{\epsilon_\alpha}
({\bf r}_1,{\bf r}_2)\right].
\label{eq:3.56}
\end{equation}
Now we can use Eq.~(\ref{eq:3.55}) and (\ref{classics}) to study the
average of different momenta of the matrix elements
\[
V_{\alpha\beta\gamma\delta}
=\int d{\bf r}_1 d{\bf r}_2 V({\bf r}_1- {\bf r}_2)
\phi_{\alpha}({\bf r}_1)
\phi_{\beta}({\bf r}_1)
\phi_{\gamma}({\bf r}_2)
\phi_{\delta}({\bf r}_2).
\]
Averaging this matrix element itself with the help of Eq.~(\ref{eq:3.56}),
we obtain
\begin{equation}
\overline{V_{\alpha\beta\gamma\delta}}=V_{\alpha\beta\gamma\delta}^{(0)}
+ V_{\alpha\beta\gamma\delta}^{(1/g)}.
\label{eq:3.57}
\end{equation}
First term in Eq.~(\ref{eq:3.57}) comes from the product of the
averaged Green functions
\[
{\rm Im}\overline{G^A({\bf r}_1, {\bf r}_2 )} = \pi\nu\ \langle
e^{i{\bf k}({\bf r}_1-{\bf r}_2)}\rangle_{FS},
\] 
and it is given by
\begin{eqnarray}
&V_{\alpha\beta\gamma\delta}^{(0)}=
E_C\delta_{\alpha\delta}\delta_{\beta\gamma}
+
F\Delta \left(\delta_{\alpha\gamma}\delta_{\beta\delta}+
\delta_{\alpha\beta}\delta_{\gamma\delta}\right),
& 
\label{eq:3.58} \\
&\displaystyle{
E_C=\frac{1}{{\cal A}^2} \int d{\bf r}_1 d{\bf r}_2 V({\bf r}_1 -{\bf
r}_2 ),}& \nonumber \\
 &F= {\nu}\langle\tilde{V}({\bf k})
\rangle_{FS}. & \nonumber
\end{eqnarray}
Here ${\cal A}$ is the area of the dot, $\langle\dots\rangle_{FS}$
stands for the averaging over directions of the wave vector on the
Fermi surface, $\nu$ is the averaged density of states per unit area
and per one spin in the dot, and $\tilde{V}({\bf k})$ is the Fourier
transform of the two-particle interaction $V({\bf r})$. 

Charging energy $E_C$ in Eq.~(\ref{eq:3.58}) is related to the zero
mode of the interaction potential. This mode can not redistribute the
electrons within the dot and that is why it is not screened
(the redistribution of the electrons
between the dot and the leads is taken into account by the
model). As the result, $E_C$ is much larger than the mean level spacing. On
the other hand, coefficient $F$ includes only non-zero modes which are
perfectly screened, $V(k)=V_0(k)/[1+2\nu V_0(k)]$, where $V_0(k)$ is
the bare potential.  (The use of the static screening here is possible
because the screening is established during the characteristic time of
the plasmon propagation through the dot, which is much smaller than
$\hbar/E_T$.)  Therefore, we estimate $F \leq 1/2$, so that the
last two terms introduce a correction only of the order of level
spacing, and may be neglected. (We will not consider here the case of
the attractive interaction when the third term in Eq.~(\ref{eq:3.58})
renormalizes to infinity due to the interaction in the Cooper
channel.)

The second term in Eq.~(\ref{eq:3.57}) originates from the product of the
retarded and advanced Green functions, see Eqs.~(\ref{classics}).  In
the absence of magnetic field, diffuson and Cooperon propagators
coincide, and their spectral expansion for a diffusive system is
\begin{equation}
{\cal P}_\omega ({\bf r}_1,{\bf r}_2 ) = 
\label{eq:3.59} 
\frac{1}{(-i\omega+0 ){\cal A}}
+\sum_{\gamma_\mu \neq 0}
\frac{ f^*_\mu({\bf r}_1)
f_\mu ({\bf r}_2 )}
{-i\omega + \gamma_\mu},
\end{equation}
where $\gamma_\mu$ and $f_\mu({\bf r})$ are the corresponding
eigenvalues and eigenfunctions. For a diffusive system, $\gamma_\mu =D
Q^2_\mu$, where $D$ is the diffusion constant, and wavevectors $Q_\mu$
depend on the shape of the system. For a rectangular diffusive dot
of the size $L_x\times L_y$, one finds $Q^2 =
\pi^2\left(n_x^2/L_x^2+n_y^2/L_y^2\right)$ with $n_x,\ n_y >0$ being
integer numbers. For chaotic systems, $\gamma_\mu$ are the eigenvalues
of the Perron-Frobenius operator.  Zero mode in Eq.~(\ref{eq:3.59})
corresponds to the conservation of the number particles, and all the
other modes describe the relaxation of any initial inhomogeneous
distribution function by the virtue of classical chaotic dynamics,
${\rm Re} \gamma_\mu > 0$.  If the system is integrable, or there are
some additional symmetries of the system, other zero modes appear,
however, we disregard such a possibility and consider only diffusive
or classically chaotic systems.

Using Eqs.~(\ref{classics}), (\ref{eq:3.56}) and (\ref{eq:3.59}), and
taking into account that all the energies are smaller than the
Thouless energy (or, in other words, the lowest non-zero eigenvalue of diffusion or
Perron-Frobenius operator),
$|\epsilon_\alpha |
\ll
\gamma_1$,  we find
\begin{equation}
V_{\alpha\beta\gamma\delta}^{(1/g)} = \frac {\Delta}{g}
\left[ 2F_1\delta_{\alpha\delta}\delta_{\beta\gamma}
+
\left(F_2+F_1\right) \left(\delta_{\alpha\gamma}\delta_{\beta\delta}+
\delta_{\alpha\beta}\delta_{\gamma\delta}\right)\right],
\label{eq:3.60}
\end{equation}
where the dimensionless conductance of the system is defined as
\begin{equation}
g = {\rm Re}\frac{\gamma_1}{\Delta},
\label{eq:3.61}
\end{equation}
and is assumed to be much larger than unity.  Dimensionless
coefficients in Eq.~(\ref{eq:3.60}) are given by
\begin{eqnarray}
\displaystyle{F_i}&=&\displaystyle{ \frac{{\rm Re}\gamma_1}{\pi}
\sum_{\gamma_\mu\neq 0}\frac{F_i^\mu}{\gamma_\mu} , \quad i=1,2;}
\label{eq:3.62}\\
\displaystyle{F_1^\mu}&=&
\displaystyle{{\nu}\langle\tilde{V}({\bf k})\rangle_{FS}
}\nonumber\\
\displaystyle{F_2^\mu}&=&\displaystyle{
\nu \int d{\bf r}_1 d {\bf r}_2 
V\left({\bf r}_1-{\bf r}_2 \right)
f^*_\mu({\bf r}_1) f_\mu({\bf r}_2). }
\nonumber
\end{eqnarray}
For the screened interaction potential, coefficients $F_{1,2}^\mu =1/2 $ and,
therefore,
$F_{1,2}$ are of the order of unity for chaotic systems and of the order of
$(1/4\pi^{2})\ln (L/l)$ for the diffusive dot. Here $L$ is the size of the dot and $l$ is
the transport elastic mean free path. Thus, we have shown that the corrections to
the average matrix element (\ref{eq:3.54}) are parametrically small for the metallic
regime.

Now, we wish to show that the fluctuations of the matrix elements are
small. Indeed, with the help of Eq.~(\ref{classics}), we obtain for
a generic ({\sl i.e.}, with no pair-wise equal indices) matrix element:
\begin{equation}
\overline{\left(V_{\alpha\beta\gamma\delta}\right)^2}= c
\left(\frac{\Delta}{g}\right)^2.
\label{eq:3.63}
\end{equation}
The numerical coefficient $c$ for the diffusive system is given by 
\[
c=\left(\frac{\gamma_1}{\pi}\right)^2
\sum_{\gamma_\mu\neq 0}
\left\{2 \left(F_1^\mu\right)^2+\left(F_2^\mu\right)^2\right\}
\left({\rm Re}\frac{1}{\gamma_\mu}\right)^2,
\]
[where coefficients $F_{1,2}^\mu$ are defined in Eq.~(\ref{eq:3.62})]
and it is of the order of unity, so that the matrix elements are small
at $g\gg 1$. For chaotic systems, the expression for $c$ is more cumbersome, but
still have a similar structure. In the case of ``diagonal'' [in the sense of
Eq.~(\ref{eq:3.60})] matrix elements, the average in Eq.~(\ref{eq:3.63}) should be
replaced by the corresponding variance.
 
The main conclusion of this section is contained in Eqs.~(\ref{eq:3.60})
and (\ref{eq:3.63}). These equations clearly show that the Coulomb
blockade type interaction (\ref{Hc}) is a parametrically justified
description for the dynamics of the system at energies smaller than
the Thouless energy.

Closing this subsection, let us mention numerical works that have been
performed recently\cite{Berkowits,Shklovskii}. These papers addressed either
dirty diffusive systems with small number of electrons\cite{Berkowits},
$g\simeq 1$, or classically localized states\cite{Shklovskii} and are
not relevant for the metallic regime $g\gg 1$ we are dealing with. We
believe that the large quantum dots studied in
Refs.~\onlinecite{Westervelt,Chang,Marcus} belong to the metallic
regime.

\section{Bosonization procedure}
\label{sec:4}

Equations~(\ref{Heff}) -- (\ref{O}) reduce the initial system
consisting of a dot and a single-mode channel to the effective
one-dimensional model. To treat interaction in the model (the second
term in Eq.~(\ref{Heff})), we follow Ref.~\onlinecite{Matveev95}, and
use the bosonization technique. In the bosonic variables, the entire
Hamiltonian (\ref{Heff}) becomes quadratic. The price for this
convenience, is a strongly non-linear form that the backscattering
terms acquire (in the language of left- and right movers, those are
the terms $\propto\psi_L^\dagger \psi_R$ in Eqs.~(\ref{lraction}) and
(\ref{reflection})).  Fortunately, the typical value of the
kernel~(\ref{kernel}) is small ($\sim\Delta/E_C$), and this enables us
to use the perturbation theory, which will be presented in
Secs.~\ref{sec:5} and \ref{sec:6}. In this section we present
bosonization procedure in a form most suitable for our purposes.

\subsection{Spinless electrons}
\label{sec:4a}

One-dimensional fermionic fields can be presented in the form\cite{Affleck}
\begin{equation}
\psi_R(x) =\frac{\hat{\eta}}{\sqrt{2\pi\lambda}}e^{i\hat{\varphi}_R(x)};
\quad
\psi_L(x) =\frac{\hat{\eta}}{\sqrt{2\pi\lambda}}e^{-i\hat{\varphi}_L(x)},
\label{eq:4.0}
\end{equation}
where $\lambda$ is the high energy cut-off of the order of the Fermi
wavelength, and $\hat{\eta}=\hat{\eta}^\dagger,\ \hat{\eta}^2=1$ is the Majorana
fermion, its significance will be discussed later.  One-dimensional bosonic fields
$\hat{\varphi}_{L,R}(x)$ satisfy the following commutation relations:
\begin{mathletters}
\label{eq:4.1}
\begin{eqnarray}
\left[\hat{\varphi}_L(x), \hat{\varphi}_L(y)\right] &=& -i\pi\ {\rm
sgn} (x-y);
\label{eq:4.1a}\\
\left[\hat{\varphi}_R(x), \hat{\varphi}_R(y)\right] &=& i\pi\ {\rm
sgn} (x-y);
\label{eq:4.1b}\\
\left[\hat{\varphi}_R(x), \hat{\varphi}_L(y)\right] &=&- i\pi. 
\label{eq:4.1c}
\end{eqnarray}
\end{mathletters}
It is easy to check, using Eqs.~(\ref{eq:4.1}), that the fermionic
fields (\ref{eq:4.0}) obey the standard commutation relations. The
expressions for the densities of left and right movers are
\begin{eqnarray}
:\psi_L^\dagger(x)\psi_L(x):=\frac{1}{2\pi}\partial_x\varphi_L(x);
\nonumber\\
:\psi_R^\dagger(x)\psi_R(x):=\frac{1}{2\pi}\partial_x\varphi_R(x).
\label{eq:4.2}
\end{eqnarray}
With the help of Eq.~(\ref{eq:4.2}), Hamiltonian (\ref{Heff}) can be
bosonized as
\begin{eqnarray}
\hat{H}_0&=&\frac{v_F}{4\pi}\int_{-\infty}^{\infty}dx
\left[
\left(\frac{\partial \hat{\varphi}_L}{\partial x}\right)^2 +
\left(\frac{\partial \hat{\varphi}_R}{\partial x}\right)^2
\right] + \label{eq:4.3}\\
&& \frac{E_C}{8\pi^2}
\left[
\hat{\varphi}_L(0) +
\hat{\varphi}_R(0) 
+ 2\pi{\cal N}
\right]^2.
\nonumber
\end{eqnarray}

The relations~(\ref{eq:4.1a}) and (\ref{eq:4.1b}) ensure the fermionic
commutation relations within the species $\psi_L$ and $\psi_R$.
The commutation relation  (\ref{eq:4.1c}) guarantees the anticommutation relation
of $\psi_L$ with $\psi_R$ and the commutation relations
\[
\left[\int_{-\infty}^0  \psi_R^\dagger \psi_R + \psi_L^\dagger \psi_L dx;
\ \psi^\dagger_{R,L}(y)
\right] = \psi^\dagger_{R,L}(y) \theta(-y).
\]

Backscattering Hamiltonian (\ref{reflection}) takes the form:
\begin{equation}
\hat{H}_{bs} = \frac{|r|v_F}{\pi \lambda} \cos 
\left[\hat{\varphi}_L(0)+\hat{\varphi}_R(0)\right].
\label{eq:4.4}
\end{equation}
Bosonized version of the effective action Eq.~(\ref{lraction}) is:
\begin{eqnarray}
\hat{S}=&&\frac{1}{2\pi\lambda}\int_0^\beta d\tau_1 d\tau_2
L\left(\tau_1-\tau_2\right)\hat{\eta}(\tau_1)\hat{\eta}(\tau_2)\times
\label{eq:4.5}\\
&&\left[e^{i\hat{\varphi}_L(\tau_1)} +
e^{-i\hat{\varphi}_R(\tau_1)}\right]
\left[e^{-i\hat{\varphi}_L(\tau_2)} +
e^{i\hat{\varphi}_R(\tau_2)}\right],
\nonumber
\end{eqnarray}
where bosonic operators are taken at the origin $x=0$. Majorana
fermion $\eta$ does not enter the effective Hamiltonian, and therefore
it is not a dynamical field. Its role in the effective action is to
take care of the difference in the definition of the operation of
chronological ordering for the fermionic and bosonic
operator. Equality
\begin{equation}
\langle T_\tau\hat{\eta}(\tau_1)\hat{\eta}(\tau_2)\rangle = {\rm sgn}
 (\tau_1-\tau_2),
\label{eq:4.6}
\end{equation}
and Wick's theorem preserves the definition of chronological ordering for
fermions in Eq.~(\ref{O}).

\begin{mathletters}
\label{eq:4.7}
It is convenient to separate the part of the bosonic sector non
affected by the Coulomb interaction and introduce new field
$\hat{\varphi}_+,\ \hat{\varphi}_-,\ \hat{\Phi}$ with the commutation
relations
\begin{eqnarray}
\left[\hat{\varphi}_+(x), \hat{\varphi}_+(y)\right] &=& -i\pi\ {\rm
sgn} (x-y);
\label{eq:4.7a}\\
\left[\hat{\varphi}_-(x), \hat{\varphi}_-(y)\right] &=&- i\pi\ {\rm
sgn} (x-y);
\label{eq:4.7b}\\
\left[\hat{\Phi}, \hat{\varphi}_-(x)\right] &=&\left[\hat{\varphi}_+(x),
\hat{\varphi}_-(y)\right] =0;
\label{eq:4.7c}\\
\left[\hat{\Phi}, \hat{\varphi}_+(x)\right] &=&i\pi. 
\label{eq:4.7d}
\end{eqnarray}
We express operators (\ref{eq:4.1}) in terms of new fields
(\ref{eq:4.7}) as
\end{mathletters}

\begin{eqnarray}
\hat{\varphi}_L(x)=
\frac{\hat{\varphi}_+(x) + \hat{\varphi}_-(x) +\hat{\Phi}}
{\sqrt{2}}
 - \pi {\cal N};
\nonumber\\
\hat{\varphi}_R(x)=
\frac{\hat{\varphi}_+(-x) - \hat{\varphi}_-(-x) -\hat{\Phi}}
{\sqrt{2}}
 - \pi {\cal N};
\label{eq:4.8}
\end{eqnarray}
where $c$ -number $\pi {\cal N}$ is incorporated into the definition
of the field.  It is easy to see that the commutation relations
(\ref{eq:4.1}) are preserved.  In the new variables, the Hamiltonian
(\ref{eq:4.3}) is independent on the gate voltage ({\sl i.e.}, on
${\cal N}$):
\begin{equation}
\hat{H}_0=\frac{v_F}{4\pi}\int_{-\infty}^{\infty}dx
\left[
\left(\frac{\partial \hat{\varphi}_+}{\partial x}\right)^2 +
\left(\frac{\partial \hat{\varphi}_-}{\partial x}\right)^2
\right] + \frac{E_C}{4\pi^2}
\hat{\varphi}_+^2(0) .
\label{eq:4.9}
\end{equation}
All the ${\cal N}$--dependence is transferred now to the backscattering term
in the Hamiltonian:
\begin{equation}
\hat{H}_{bs} = \frac{|r|v_F}{\pi \lambda} \cos 
\left[\sqrt{2}\hat{\varphi}_+(0)-2\pi{\cal N}\right],
\label{eq:4.10}
\end{equation}
and to the action $\hat{S}$ which is contributed by the return trajectories of
electrons after multiple scattering within the dot:
\begin{eqnarray}
\hat{S}=&&\frac{1}{2\pi\lambda}\int_0^\beta d\tau_1 d\tau_2
L\left(\tau_1-\tau_2\right)\hat{\eta}(\tau_1)\hat{\eta}(\tau_2)\times
\label{eq:4.11}\\
&&\exp\left[i\frac{\hat{\varphi}_-(\tau_1)-
\hat{\varphi}_-(\tau_2)}{\sqrt{2}}\right]
\exp\left[i\frac{\hat{\Phi}(\tau_1)-\hat{\Phi}(\tau_2)}
{\sqrt{2}}\right]\times
\nonumber\\
&&\cos\left[\frac{\hat{\varphi}_+(\tau_1)}{\sqrt{2}}+\frac{\pi}{4}-\pi{\cal
N}\right]
\cos\left[\frac{\hat{\varphi}_+(\tau_2)}{\sqrt{2}}+\frac{\pi}{4}-\pi{\cal
N}\right].  \nonumber
\end{eqnarray}
If one neglects such trajectories altogether\cite{Matveev95}, then
the Coulomb blockade oscillations  apparently vanish in the limit $r=0$ (no
backscattering in the channel). However, if return trajectories ({\sl i.e.}, a finite value
of $\Delta/E_C$) are taken into account, the Coulomb blockade oscillations exist, even
if $r=0$. The ${\cal N}$--dependence of the action (\ref{eq:4.11}) and Hamiltonian
(\ref{eq:4.10}) clearly shows that the period of the oscillations does not depend on
the details of the system. This periodic dependence is a direct consequence of the
discreteness of the electron charge.

\begin{mathletters}
\label{eq:4.12}
We are going to develop perturbation theory in $\hat{S}$ and
$\hat{H}_{bs}$.  Every order of the perturbation theory is expressed
in terms of the correlators of the bosonic field governed by the
quadratic Hamiltonian (\ref{eq:4.9}). The necessary correlation
functions are
\begin{eqnarray}
&&{\cal D}_-\left(\tau\right) = \langle T_\tau
\hat{\varphi}_-(\tau)\hat{\varphi}_-(0)\rangle;
\label{eq:4.12a}\\
&&{\cal D}_+\left(\tau\right) = \langle T_\tau
\hat{\varphi}_+(\tau)\hat{\varphi}_+(0)\rangle;
\label{eq:4.12b}\\
&&{\cal D}_{\Phi}\left(\tau\right) = \langle T_\tau
\hat{\Phi}(\tau)\hat{\Phi}(0)\rangle;
\label{eq:4.12c}\\
&&{\cal D}_{\Phi +}\left(\tau\right) = \langle T_\tau
\hat{\Phi}(\tau)\hat{\varphi}_+(0)\rangle,
\label{eq:4.12d}
\end{eqnarray}
where averages are calculated with respect to Hamiltonian $\hat{H}_0$
and all the bosonic fields are taken at $x=0$.
\end{mathletters}

Standard calculation presented in Appendix~\ref{ap:2} yields for 
$T \ll E_C$ and $\tau \gtrsim \lambda/v_F \simeq 1/\epsilon_F$:

\wide{m}
{
\begin{mathletters}
\label{eq:4.13}
\label{4.13}
\begin{eqnarray}
&&{\cal D}_\Phi\left(\tau\right) -{\cal D}_\Phi\left(0\right) 
= -\frac{1}{2}
\int_0^\infty d x e^{-x}
\ln\left[\frac{
\sin\left(i \frac{2\pi x}{E_c}+\tau\right)\pi T\ 
\sin\left(-i \frac{2\pi x}{E_c}+\tau\right)\pi T}
{\sinh^2\left(\frac{2\pi^2Tx}{E_c}\right)}\right];
\label{eq:4.13a}\\
&&{\cal D}_-\left(\tau\right) -{\cal D}_-\left(0\right) 
= 
\ln\left(\frac{\lambda}{v_F}\frac{\pi T}{|\sin \pi T\tau |}\right);
\label{eq:4.13b}\\
&&{\cal D}_+\left(0\right) = \ln\left(\frac{2\pi v_F}{\lambda E_C
e^{\bf C}}\right);
\label{eq:4.13c}\\
&&{\cal D}_+\left(\tau\right) -{\cal D}_+\left(0\right) 
=  \left[{\cal D}_-\left(\tau\right) -{\cal D}_-\left(0\right)\right]
-\left[ {\cal D}_\Phi\left(\tau\right) -{\cal D}_\Phi\left(0\right)\right] ;
\label{eq:4.13d}\\
&&{\cal D}_{\Phi+}\left(\tau\right) =
\frac{i}{2}\int_0^\infty dx e^{-x}\frac{2\pi^2T}{E_C}
\left[
\cot\left(i \frac{2\pi x}{E_c}+\tau\right)\pi T +
\cot\left(-i \frac{2\pi x}{E_c}+\tau\right)\pi T 
\right],
\label{eq:4.13e}
\end{eqnarray}
\end{mathletters}
}
where ${\bf C} \approx 0.577$ is the Euler constant.
 
To conclude this Subsection, let us proof the assumption of
Sec.~\ref{sec:2} about the Fermi liquid behavior of the system at low
energies. In order to do this, we will calculate the fermionic Green
functions, $\langle\psi_L^\dagger (\tau)\psi_L(0)\rangle$ and
$\langle\psi_R^\dagger(\tau)\psi_L(0)\rangle$, using the definitions
(\ref{eq:4.0}), (\ref{eq:4.8}), and the results
(\ref{eq:4.13}). Averaging over the bosonic fields similar to the
well-known calculation of the Debye-Waller factor, yields:
\begin{mathletters}
\label{eq:4.14}
\begin{eqnarray}
&&\langle\psi_L^\dagger(\tau)\psi_L(0)\rangle=
\frac{1}{2\pi\lambda}\exp\left\{\frac{1}{2}\left[{\cal D}_-(\tau)-{\cal
D}_-(0)\right]\right\} \times\nonumber\\
&&\exp\left\{\frac{1}{2}\left[{\cal D}_+(\tau)-{\cal
D}_+(0)\right]\right\} \exp\left\{\frac{1}{2}\left[{\cal
D}_{\Phi}(\tau)-{\cal D}_{\Phi}(0)\right]\right\}\times \nonumber\\
&&\hspace{1.2cm} \exp\left\{\frac{1}{2}\left[{\cal
D}_{\Phi+}(\tau)+{\cal D}_{\Phi+}(-\tau)\right]\right\}
\label{eq:4.14a}\\
&&\langle\psi_R^\dagger(\tau)\psi_L(0)\rangle= \frac{e^{i2\pi {\cal
N}}}{2\pi\lambda} \exp\left\{-\frac{1}{2}\left[{\cal D}_+(\tau)+ {\cal
D}_+(0)\right]\right\} \times\nonumber\\
&&\exp\left\{\frac{1}{2}\left[{\cal D}_-(\tau)-{\cal
D}_-(0)\right]\right\} \exp\left\{\frac{1}{2}\left[{\cal
D}_{\Phi}(\tau)-{\cal D}_{\Phi}(0)\right]\right\}\times \nonumber\\
&&\hspace{1.2cm} \exp\left\{\frac{1}{2}\left[{\cal
D}_{\Phi+}(\tau)-{\cal D}_{\Phi+}(-\tau)\right]\right\}.
\label{eq:4.14b}
\end{eqnarray}
Substituting Eqs.~(\ref{eq:4.13}) into Eq.~(\ref{eq:4.14}), we obtain 
\end{mathletters}
\begin{mathletters}
\label{eq:4.15}
\begin{eqnarray}
&&\langle\psi_L^\dagger(\tau)\psi_L(0)\rangle=
\frac{1}{2\pi v_F}\frac{\pi T}{\sin \pi T\tau};
\label{eq:4.15a}\\
&&\langle\psi_R^\dagger(\tau)\psi_L(0)\rangle=
\frac{e^{i2\pi {\cal N}}}{2\pi v_F}
K(\tau);\label{eq:4.15b} \\
&&K(\tau)\!=\!
\frac{\pi T}{\sin \pi T\tau}
\exp \left\{-\int_0^\infty\!\! dx e^{- \frac{E_C}{2\pi^2T}x} 
\coth\left( x\! +\! i\pi\tau T\right)
\right\}.
\nonumber
\end{eqnarray}
\end{mathletters}
The Green function (\ref{eq:4.15a}) is not affected by interactions at all.  The reason
is, that it is taken at coinciding arguments ($x_1=x_2=0$), {\sl e.g.}, outside the
interaction region. Because
$\langle\psi_L^\dagger(\tau)\psi_L(0)\rangle$ describes propagation of a chiral
particle, the information about interaction is never carried back to the observation
point $x=0$. The Green function (\ref{eq:4.15b}) acquires the free-fermion form at
$\tau > E_C^{-1}$, which corresponds to the energies below the charging energy
(note that $T\ll E_C$). In this energy range,
$\langle\psi_R^\dagger(\tau)\psi_L(0)\rangle$ corresponds to a free fermion
completely reflected from the dot. The phase factor
$\exp(i2\pi{\cal N})$ in Eq.~(\ref{eq:4.15b}) represents the scattering phase $\pi{\cal
N}$, which agrees with the Friedel sum rule (\ref{phaseshift}). Thus, our intuitive
picture of Sec.~\ref{sec:2} is proven by explicit calculation of the fermionic
propagators.

\subsection{Electrons with spin}
\label{sec:4b}
Similarly to the spinless case, we start here with the bosonization of
electron operators:
\begin{eqnarray}
\psi_{R,\alpha}(x) =\frac{\hat{\eta}_\alpha}{\sqrt{2\pi\lambda}}
\exp\left(i\frac{\hat{\varphi}^\rho_R(x)+\alpha\hat{\varphi}^\sigma_R(x)}
{\sqrt{2}}\right),
\nonumber\\
\psi_{L,\alpha}(x) =\frac{\hat{\eta}_\alpha}{\sqrt{2\pi\lambda}}
\exp\left(-i\frac{\hat{\varphi}^\rho_L(x)+\alpha\hat{\varphi}^\sigma_L(x)}
{\sqrt{2}}\right),
\label{eq:4.16}
\end{eqnarray}
where index $\alpha=\pm 1$ denotes the spin projections, and the
Majorana fermions $\eta_{\pm 1}$ satisfy the anticommutation relations
$\{\eta_{+1},\eta_{-1}\}=0$. Boson fields $\hat{\varphi}^\rho_{L,R}$
and $\hat{\varphi}^\sigma_{L,R}$ corresponding to the charge and spin
degrees of freedom respectively, satisfy the following commutation
relations:
\begin{mathletters}
\label{eq:4.17}
\begin{eqnarray}
\left[\hat{\varphi}^i_L(x), \hat{\varphi}^j_L(y)\right] &=& -i\pi\ {\rm
sgn} (x-y)\delta_{ij};
\label{eq:4.17a}\\
\left[\hat{\varphi}^i_R(x), \hat{\varphi}^j_R(y)\right] &=& i\pi\ {\rm
sgn} (x-y)\delta_{ij};
\label{eq:4.17b}\\
\left[\hat{\varphi}^i_R(x), \hat{\varphi}^j_L(y)\right] &=&
- i\pi\delta_{ij}; \quad i,j=\rho,\sigma. 
\label{eq:4.17c}
\end{eqnarray}
\end{mathletters}
Like in the case of spinless fermions, it is convenient to introduce
even and odd modes $\hat{\varphi}^{\rho,\sigma}_\pm$ for the charge
and spin sectors, and two $x$-independent fields, $\Phi^{\rho,\sigma}$
analogous to Eq.~(\ref{eq:4.8})
\begin{eqnarray}
&\displaystyle{\hat{\varphi}_L^i(x) = \frac{\hat{\varphi}_+^i(x) +
\hat{\varphi}_-^i(x) +\hat{\Phi}^i- \delta_{i\rho}\pi {\cal N}}
{\sqrt{2}};}& \quad i=\rho,\sigma \nonumber\\
&\displaystyle{\hat{\varphi}_R^i(x)= \frac{\hat{\varphi}_+^i(-x) -
\hat{\varphi}_-^i(-x) - \hat{\Phi}^i - \delta_{i\rho}\pi {\cal N}}
{\sqrt{2}}.}&
\label{eq:4.18}
\end{eqnarray}
The commutation relations for the new fields within the charge and spin
sectors coincide with Eqs.~(\ref{eq:4.7}); fields of different sectors commute
with each other. In terms of the new fields, Hamiltonian (\ref{eq:3.15})
acquires the form independent on dimensionless gate voltage ${\cal N}$:
\begin{equation}
\hat{H}_0=\frac{v_F}{4\pi}\sum_{i=\rho,\sigma}\sum_{\gamma=\pm}
\int_{-\infty}^{\infty}dx
\left(\frac{\partial \hat{\varphi}_\gamma^i}{\partial x}\right)^2 +
  \frac{E_C}{2\pi^2}
\left[\hat{\varphi}_+^\rho(0)\right]^2.
\label{eq:4.19}
\end{equation}
Backscattering Hamiltonian (\ref{reflection}) takes the form
\begin{equation}
\hat{H}_{bs} = \frac{2 |r|v_F}{\pi \lambda} \cos 
\left(\hat{\varphi}_+^\rho (0)- \pi{\cal N}\right)
\cos\hat{\varphi}_+^\sigma (0),
\label{eq:4.20}
\end{equation}
and the effective action (\ref{lraction}) can be rewritten as
\begin{eqnarray}
&&
\hat{S}=\frac{1}{\pi\lambda}\int_0^\beta d\tau_1 d\tau_2
L\left(\tau_1-\tau_2\right)
\sum_{\alpha=\pm 1}
\hat{\eta}_\alpha(\tau_1)\hat{\eta}_\alpha (\tau_2)
\times
\nonumber \\
&&\
e^{\frac{i}{2}\alpha
\left(\hat{\varphi}_-^\sigma (\tau_1)-\hat{\varphi}_-^\sigma (\tau_2)\right)}
e^{\frac{i}{2}\alpha
\left(\hat{\Phi}^\sigma(\tau_1)-\hat{\Phi}^\sigma (\tau_2)\right)}\times
\nonumber\\
&&\
e^{\frac{i}{2}
\left(\hat{\varphi}_-^\rho (\tau_1)-\hat{\varphi}_-^\rho (\tau_2)\right)}
e^{\frac{i}{2}
\left(\hat{\Phi}^\rho (\tau_1)-\hat{\Phi}^\rho (\tau_2)\right)}\times
\nonumber\\
&&\
\left\{
\cos \left[
\frac{\hat{\varphi}_+^\rho (\tau_1)+\hat{\varphi}_+^\rho (\tau_2) }{2}
+\alpha
\frac{\hat{\varphi}_+^\sigma (\tau_1)+
\hat{\varphi}_+^\sigma (\tau_2) }{2}
-\pi {\cal N}\right]
-
\right.
\nonumber\\
&&\
\left.
\sin \left[
\frac{\hat{\varphi}_+^\rho (\tau_1)-
\hat{\varphi}_+^\rho (\tau_2) }{2}
+\alpha
\frac{\hat{\varphi}_+^\sigma (\tau_1)-\hat{\varphi}_+^\sigma (\tau_2)
}
{2}
\right]
\right\}.
\label{eq:4.21}
\end{eqnarray}

\begin{mathletters}
\label{eq:4.22}
Similarly to Eq.~(\ref{eq:4.12}), we introduce the relevant bosonic
correlation functions
\begin{eqnarray}
&&{\cal D}_-^i\left(\tau\right) = \langle T_\tau
\hat{\varphi}_-(\tau)^i\hat{\varphi}_-(0)^i\rangle;
\label{eq:4.22a}\\
&&{\cal D}_+^i\left(\tau\right) = \langle T_\tau
\hat{\varphi}_+^i(\tau)\hat{\varphi}_+^i(0)\rangle;
\label{eq:4.22b}\\
&&{\cal D}_{\Phi}^i\left(\tau\right) = \langle T_\tau
\hat{\Phi}^i(\tau)\hat{\Phi}^i(0)\rangle;
\label{eq:4.22c}\\
&&{\cal D}_{\Phi +}^i\left(\tau\right) = \langle T_\tau
\hat{\Phi}^i(\tau)\hat{\varphi}^i_+(0)\rangle,
\label{eq:4.22d}
\end{eqnarray}
where index $i=\rho,\sigma$ labels charge and spin sectors
respectively, bosonic fields are taken at $x=0$, and averaging is
performed over the Hamiltonian $\hat{H}_0$ given by
Eq.~(\ref{eq:4.19}).
\end{mathletters}

The calculation of these propagators can be performed immediately by
noticing that the spin sector of the Hamiltonian (\ref{eq:4.19})
corresponds to the free bosons, and the charge sector differs from
Eq.~(\ref{eq:4.9}) only by replacement $E_C \to 2E_C$. Thus, we obtain

\wide{m}
{
\begin{mathletters}
\label{eq:4.23}
\label{4.23}
\begin{eqnarray}
&&{\cal D}_\Phi^\rho\left(\tau\right) -{\cal D}_\Phi^\rho\left(0\right) 
= -\frac{1}{2}
\int_0^\infty d x e^{-x}
\ln\left[\frac{
\sin\left(i \frac{\pi x}{E_c}+\tau\right)\pi T\ 
\sin\left(-i \frac{\pi x}{E_c}+\tau\right)\pi T}
{\sinh^2\left(\frac{\pi^2Tx}{E_c}\right)}\right];
\quad {\cal D}_\Phi^\sigma =0;
\label{eq:4.23a}\\
&&{\cal D}_-^\rho\left(\tau\right) -{\cal D}_-^\rho\left(0\right) 
=
{\cal D}_\pm^\sigma\left(\tau\right) -{\cal D}_\pm^\sigma\left(0\right) 
= 
\ln\left(\frac{\lambda}{v_F}\frac{\pi T}{|\sin \pi T\tau |}\right);
\label{eq:4.23b}\\
&&{\cal D}_+^\rho \left(0\right) = \ln\left(\frac{\pi v_F}{\lambda E_C
e^{\bf C}}\right);
\label{eq:4.23c}\\
&&{\cal D}_+^\rho \left(\tau\right) -{\cal D}_+^\rho\left(0\right) 
=  \left[{\cal D}_-^\rho\left(\tau\right) -
{\cal D}_-^\rho\left(0\right)\right]
-\left[ {\cal D}_\Phi^\rho\left(\tau\right) -
{\cal D}_\Phi^\rho \left(0\right)\right] ;
\label{eq:4.23d}\\
&&{\cal D}_{\Phi+}^\rho\left(\tau\right) =
\frac{i}{2}\int_0^\infty dx e^{-x}\frac{\pi^2T}{E_C}
\left[
\cot\left(i \frac{\pi x}{E_c}+\tau\right)\pi T +
\cot\left(-i \frac{\pi x}{E_c}+\tau\right)\pi T 
\right],\quad {\cal D}_{\Phi+}^\sigma\left(\tau\right) =
\frac{i\pi}{2}{\rm sgn}\,\tau,
\label{eq:4.23e}
\end{eqnarray}
\end{mathletters}
}
where ${\bf C} \approx 0.577$ is the Euler constant.

As we will see below, see also Ref.~\onlinecite{Matveev95}, the main
contribution to the observable quantities is associated with the time
scale $\tau \gtrsim 1/E_C$. At this time scale the effective theory
can be further simplified.  The mode $\hat{\varphi}^\rho_+$ is ``pinned'' due
to the charging energy, see Eq.~(\ref{eq:4.19}). Therefore, the
amplitude of quantum fluctuations of this mode is finite, see
Eq.~(\ref{eq:4.23c}), and the correlation function 
${\cal D}_+^\rho (\tau)$ decreases rapidly at $\tau \gtrsim 1/E_C$, as it follows
from Eqs.~(\ref{eq:4.23d}), (\ref{eq:4.23b}) and (\ref{eq:4.23a}). The
decrease of correlations means that the average of a product,
$\langle e^{i\hat{\varphi}_+^\rho(\tau_1)}
\dots e^{i\hat{\varphi}_+^\rho(\tau_n)}\rangle$, can be replaced by the
product of averages, $\langle e^{i\hat{\varphi}_+^\rho(\tau_1)}\rangle
\dots\langle e^{i\hat{\varphi}_+^\rho(\tau_n)}\rangle$, if the intervals
between the times $\tau_1,\dots,\tau_n$ exceed $1/E_C$. In other words, 
the operator functions of $\hat{\varphi}_+^\rho$ in Eq.~(\ref{eq:4.21}) can be
substituted with $c$-numbers, according the rule:
\begin{equation}
e^{i\hat{\varphi}_+^\rho} \to e^{-\frac{1}{2}{\cal D}^{\rho}_+(0)}.
\label{eq:4.24}
\end{equation}
 On the other hand, it follows from
Eqs.~(\ref{eq:4.23a}) and (\ref{eq:4.23b}),  that at $\tau > 1/E_C$
\begin{eqnarray*}
&&\langle T_\tau \!
\left(\hat{\varphi}_-^\rho (\tau) +\hat{\Phi}^\rho(\tau)
-\hat{\varphi}_-^\rho (0)-
\hat{\Phi}^\rho(0)\right)^2
\rangle\! = \\
&&
\hspace*{2cm}
2\ln\left[\frac{\pi\lambda }{v_F E_Ce^{\bf C}}
\left(\frac{\pi T}{\sin \pi T\tau}\right)^2\right],
\end{eqnarray*}
which means that such correlation function will be preserved if we
introduce another free bosonic field $\hat{\varphi}_\rho(x)$, with
commutation relation 
$\left[\hat{\varphi}_\rho (x);\hat{\varphi}_\rho (y) \right]
=-i\pi {\rm sgn}(x-y)$, and substitute
\begin{equation}
\hat{\varphi}_-^\rho (x=0 ) +\hat{\Phi}^\rho \to 
\sqrt{2}\hat{\varphi}_\rho(x=0). 
\label{eq:4.25}
\end{equation}

After substitutions (\ref{eq:4.24}) and (\ref{eq:4.25}), Hamiltonian
(\ref{eq:4.19}) becomes just a Hamiltonian of three free bosonic
fields
\begin{equation}
\hat{H}_0 = \int_{-\infty}^\infty dx \left[
\left(\frac{\partial\hat{\varphi}_\rho }{\partial x}\right)^2
+\left(\frac{\partial\hat{\varphi}_+^\sigma }{\partial x}\right)^2
+\left(\frac{\partial\hat{\varphi}_-^\sigma }{\partial x}\right)^2\right],
\label{eq:4.26}
\end{equation}
backscattering term acquires the form\cite{Matveev95}
\begin{equation}
\hat{H}_{bs} = \frac{2 |r|}{\pi }
\left(\frac{E_Ce^{\bf C}v_F}{\pi \tilde{\lambda}}\right)^{1/2} 
\cos\pi{\cal N}
\cos\hat{\varphi}_+^\sigma (0),
\label{eq:4.27}
\end{equation}
and the effective action is given by
\begin{eqnarray}
&&
\hat{S}=\frac{1}{\pi\tilde{\lambda}}\int_0^\beta d\tau_1 d\tau_2
L\left(\tau_1-\tau_2\right)
\sum_{\alpha=\pm 1}
\hat{\eta}_\alpha(\tau_1)\hat{\eta}_\alpha (\tau_2)
\times
\nonumber \\
&&\
e^{\frac{i}{2}\alpha
\left(\hat{\varphi}_-^\sigma(\tau_1)-
\hat{\varphi}_-^\sigma(\tau_2)\right)}
e^{\frac{i}{2}\alpha
\left(\hat{\Phi}^\sigma (\tau_1)-\hat{\Phi}^\sigma
(\tau_2)\right)}
e^{\frac{i}{\sqrt{2}} 
\left(\hat{\varphi}_\rho(\tau_1)-\hat{\varphi}_\rho(\tau_2)\right)}
\times
\nonumber\\
&&\
\left\{
\cos \left[\alpha
\frac{\hat{\varphi}_+^\sigma (\tau_1)+\hat{\varphi}_+^\sigma (\tau_2) }{2}
-\pi {\cal N}\right]
+
\right.
\label{eq:4.28}\\
&&\hspace*{3cm}
\left.
\cos \left[
\frac{\pi}{4}
+\alpha
\frac{\hat{\varphi}_+^\sigma (\tau_1)-\hat{\varphi}_+^\sigma (\tau_2) }{2}
\right]
\right\}.
\nonumber
\end{eqnarray}
Correlation functions of the free bosonic fields are given by
\begin{equation}
{\cal D}_\rho\left(\tau\right) -{\cal D}_\rho\left(0\right) 
=
{\cal D}_\pm^\sigma\left(\tau\right) -{\cal D}_\pm^\sigma\left(0\right) 
= 
\ln\left(\frac{\tilde{\lambda}}{v_F}\frac{\pi T}{|\sin \pi T\tau |}\right),
\label{eq:4.29}
\end{equation}
where cut-off $\tilde{\lambda}$ is of the order of $v_F/E_C$ because the
charging energy $E_C$ is the largest energy scale which can be
considered with the help of Hamiltonian (\ref{eq:4.26}).  It is easy
to check also by an explicit calculation, that at time differences larger than
$E_C^{-1}$ correlation functions of the electron operators evaluated
with the help of the Hamiltonians (\ref{eq:4.19}) and (\ref{eq:4.26})
respectively coincide.

\section{Thermodynamics of the ``open'' dot}
\label{sec:5} 
Coulomb blockade can be investigated
experimentally\cite{Molenkamp,Zhitenev} by measuring the differential
capacitance of a dot, see Eq.~(\ref{Cdif}). In the regime of a
developed blockade (weak tunneling between the dot and the electron
reservoir), $C_{\it diff}({\cal N})$ exhibits sharp peaks at half-integer
values of ${\cal N}$. In the opposite limit of no backscattering, the
differential capacitance is an ${\cal N}$--independent constant,
$C_{\it diff}({\cal N})=C$. It was shown in Ref.~\onlinecite{Matveev95}
that weak reflection from a scatterer in the channel leads to the
 capacitance oscillations with a phase depending on the exact
position of the scatterer. In this section we demonstrate that even at
$r=0$, the differential capacitance still depends on ${\cal N}$ due to
the electron backscatterings from inside the dot, which are described
by the action $\hat S$. The randomness of the backscattering events
results in the randomness of the phase of the capacitance
oscillations. We will relate the statistics of $C_{\it diff}({\cal N})$
with the one of kernel $L(\tau)$.

The starting point for the calculation of the capacitance is Eq.~(\ref{O}) for the
thermodynamic potential. In principle, Eq.~(\ref{O}) enables one to consider
the backscattering off a barrier in the channel (${\hat H}_{bs}\neq 0$), as
well as off the dot (${\hat S}\neq 0$). In the limit of weak backscattering, the
perturbation theory in ${\hat H}_{bs}$ and ${\hat S}$ can be used to calculate
$C_{\it diff}({\cal N})$. The case of ${\hat S}=0$ was considered by
Matveev\cite{Matveev95}. He has shown that for spinless fermions, a
non-vanishing result appears in the first-order perturbation theory,
whereas for the spin-$1/2$ electrons this order gives zero result. Similarly, it
is sufficient to account for the scatterings from inside the cavity in 
the first order of the perturbation theory for spinless fermions, but in the
case of spin-$1/2$ electrons we have to expand the thermodynamic potential
(\ref{O}) up to the second order in ${\hat S}$, if  ${\hat H}_{bs}=0$.
 
Backscattering in the channel leads to a finite modulation of the average
differential capacitance. The modulation amplitude can be
estimated\cite{Matveev95} by expansion of Eq.~(\ref{O}) to the second order
in ${\hat H}_{bs}$ in the spin-1/2 case. Electron scattering from inside the dot
leads to the capacitance fluctuations superimposed on this modulation. The
two contributions to the capacitance are not additive: the non-zero result
appears only in the second-order  in perturbations to the Hamiltonian
${\hat H}_{0}$, when one expands Eq.~(\ref{O}). In the domain of a
relatively strong backscattering in the channel, $|r|^2\gtrsim\Delta/E_C$,
the leading term in fluctuations is proportional to the product of ${\hat
H}_{bs}$ and
${\hat S}$. We address the capacitance fluctuations at finite $|r|$ in the
end of this section.

\subsection{Spinless fermions}
\label{sec:5a}
\subsubsection{Reflectionless contact}
\label{sec:5a1}
The first-order expansion of Eq.~(\ref{O})  in ${\hat S}$ yields:
\begin{equation} 
\delta\Omega= T\langle T_\tau \hat{S}\rangle.
\label{eq:5.1}
\end{equation}
We substitute Eq.~(\ref{eq:4.11}) into Eq.~(\ref{eq:5.1}), retain 
only ${\cal N}$-dependent terms, and obtain with the help of
Eqs.~(\ref{eq:4.13}):
\begin{equation}
\delta\Omega = \frac{1}{2\pi v_F}\int_0^\beta d\tau L(\tau)
\left[e^{i2\pi{\cal N}} K(\tau) +c.c. \right],
\label{eq:5.2}
\end{equation}
where function $K(\tau)$ is defined in Eq.~(\ref{eq:4.15b}). 
To perform the integration over $\tau$, we use the 
Lehmann representation (\ref{Lehman}) of the kernel $L$:
\begin{eqnarray}
&&\delta\Omega=\frac{1}{2\pi v_F}
\int_{-\infty}^{\infty}\!\frac{dt}{2\pi}\!\left(L^R(t)-L^A(t)\right)
\times
\nonumber\\
&&\! \int_0^{\beta}\!\! d\tau\frac{\pi T}{\sinh [\pi T(t+i\tau)]}
\left[e^{i2\pi{\cal N}} K(\tau) + c.c.\right].
\label{eq:5.3}
\end{eqnarray}
The integration over $\tau$ here can be now performed with the help of
analytic properties of function $K(\tau)$. As it follows from
Eq.~(\ref{eq:4.15b}), the function
$K(\tau) $ is analytic in the lower semiplane
${\rm Im}\,\tau<0$. To calculate the
integral  of the first term in the brackets, we deform the contour 
of integration over $\tau$ as shown in Fig.~\ref{Fig:5.1}.
\narrowtext{
\begin{figure}[h]
\vspace{0.2cm}
\epsfxsize=5.0cm
\hspace*{0.5cm}
\epsfbox{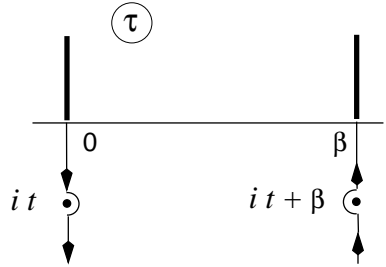}
\vspace{0.3 cm}
\caption{The integration contour used in the evaluation of
$\delta\Omega$ in the spinless case, see Eq.~(\protect\ref{eq:5.3}).
Branch cuts of $K(\tau)$ are shown by  thick  lines.
}
\label{Fig:5.1}
\end{figure} 
}
Because of the periodicity
of the integrand, the integrals over the parts of the contour running
parallel to the imaginary axis, cancel out. As the result, only the
pole contribution at $\tau =it$ remains at $t <0$. 
At $t>0$ the pole contribution disappears. Second term in
Eq.~(\ref{eq:5.3}) is integrated by using $K(\tau)^* = - K(-\tau)$ for
the real $\tau$.
With the help of the explicit expression
(\ref{eq:4.15b}) for the function $K(\tau )$, we find:
\begin{eqnarray}
\delta\Omega ({\cal N})&&=\frac{1}{2\pi i v_F}\int_0^\infty dt
\frac{\pi T}{\sinh (\pi Tt)}\times
\label{eq:5.4}\\
&&\exp\left[-\!\int_0^\infty \!\! dx e^{-x}\frac{2\pi^2 T}{E_C}
\coth\left(\frac{2\pi x}{E_C}+t\right)\pi T\right]\times
\nonumber\\
&&\left[L^R(t)e^{-i 2\pi  {\cal N}} - L^A(-t)e^{i 2\pi  {\cal N}}\right].
\nonumber
\end{eqnarray}
Finally, at low temperatures $T \ll E_C$, the ${\cal N}$-dependent
correction takes the form:
\begin{eqnarray}
\delta\Omega ({\cal N})=
&&\frac{1}{2\pi i v_F}\int_0^\infty\frac{dt}{t}
\exp\left[-\!\int_0^\infty \!\! dx \frac{e^{-x}}{x+\frac{E_C}{2\pi}t}\right]\times
\label{eq:5.5}\\
&&\left[L^R(t)e^{2\pi i {\cal N}} - L^A(-t)e^{-2\pi i {\cal N}}\right].
\nonumber
\end{eqnarray}
Equation (\ref{eq:5.5}) relates the Coulomb blockade oscillation to
the exact free-electron Green function in the dot.
The variation of $\delta\Omega$ with the gate voltage is
harmonic, however, its phase and amplitude are random quantities.
To reveal this oscillatory dependence in the average quantities,
one has to find the correlation function 
$\overline{\delta\Omega({\cal N}_1)\delta\Omega({\cal N}_2)}$.
At low temperatures, Eqs.~(\ref{eq:5.5}) and
(\ref{eq:3.25}) lead directly to the result (\ref{EHresult}) with the
dimensionless function $\Lambda_E$ given by:
\begin{eqnarray}
\Lambda_E (x)=&&\frac{1}{(2\pi)^4}\Lambda (x), 
\label{eq:5.6}\\
\Lambda (x)=&&\int_0^\infty\frac{dy}{y^2}
\exp\left[-xy 
+ 2 e^y Ei(-y)
\right],
\nonumber
\end{eqnarray}
where $Ei(x)=\int^x_{-\infty}e^tdt/t$ is the exponential integral
function\cite{Ryzhik}.  The correlation function of the differential
capacitances (\ref{Cdif}) for different values of the gate voltage and
magnetic field is given by:
\begin{equation}
\frac{\overline{\delta C_{\it diff}(1)\delta C_{\it diff}(2)}}{C^2}=
\frac{2\Delta}{E_C}\left[\Lambda\left(\!\frac{H_-^2}{H_c^2}\!\right)+
\Lambda\left(\!\frac{H_+^2}{H_c^2}\!\right)\right]\cos 2\pi n,
\label{eq:5.7}\\
\end{equation}
where we use the short hand notations $i \equiv {\cal N}_i,H_i$,
$n={\cal N}_1-{\cal N}_2$, and $H_\pm = H_1 \pm H_2$. Correlation magnetic field
$H_c$ is controlled by the charging energy and it is given by Eq.~(\ref{eq:2.25}).

The variance of the capacitance fluctuations at $H=0$ is two times larger
than in the unitary limit ($H\gg H_c$). The dimensionless crossover
function $\Lambda (x)$ is plotted in Fig.~\ref{Fig:5.2}.

\narrowtext{
\begin{figure}[h]
\epsfxsize=6.7cm
\hspace*{0.5cm}
\vspace{0.3 cm}
\epsfbox{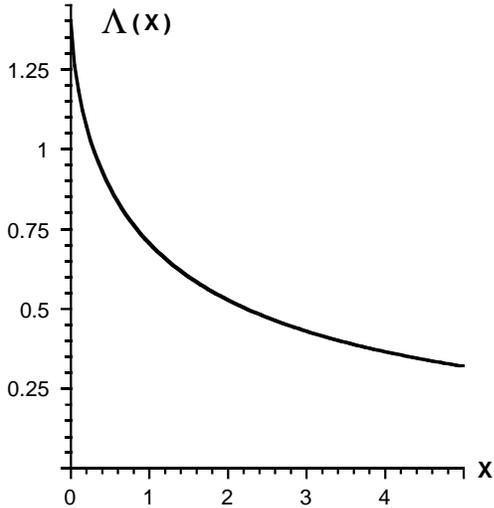}
\vspace{0.3 cm}
\caption{Function $\Lambda(x)$ determining correlation of differential
capacitances at different values of magnetic fields, see
Eqs.~(\protect\ref{eq:5.7}) and Eq.~(\protect\ref{eq:5.6}).}
\label{Fig:5.2}
\end{figure} 
} 

Correlation between the capacitances at different magnetic fields is
suppressed if $H_-$ exceeds $H_c$. In our approximation, correlation
``length'' $n_c$ in the dimensionless gate voltage is infinite. To find $n_c$,
one should take into account that varying of the gate voltage affects the
chemical potential of electrons in the dot by the level spacing, each
time when ${\cal N}$ changes by one.
If the chemical potential is shifted by $E_C$, a 
completely new set of levels determines the
kernel $L$, thus suppressing the correlations. 
It results in the estimate: $n_c\simeq E_C/\Delta$. Another effect
which leads to the decrease of the correlation function at large $n$,
is the variation of the dot shape with the gate voltage. 

\subsubsection{Finite reflection in the contact}
\label{sec:5a2}

At finite backscattering in the  contact, $r\neq 0$, Hamiltonian
(\ref{eq:4.10}) should be taken into account:
\[
\delta\Omega ({\cal N}) = \langle {\hat H_{bs}} \rangle.
\] 
Calculating this average with the help of Eq.~(\ref{eq:4.13c}), and
using Eq.~(\ref{Cdif}), we arrive to the result for the oscillating contribution of the
averaged capacitance,
\begin{equation}
\frac{\overline{\delta C ({\cal N})}}{C} = 
2 e^{\bf C} |r| \cos 2\pi {\cal N}.
\label{eq:5.70}
\end{equation}
This result was first obtained in Ref.~\onlinecite{Matveev95}.

Average of the action Eq.~(\ref{eq:5.1}) simply adds to this result.
Thus, we conclude that  finite reflection in this order of
perturbation theory does not affect the mesoscopic fluctuations of the
capacitance, see Eq.~(\ref{eq:5.7}). We will see below that this is
qualitatively different from the electrons with spin where  finite
reflection in the contact leads to the increase of the mesoscopic
fluctuations of the differential capacitance.

\subsection{Electrons with spin}
\label{sec:5b}

\subsubsection{Reflectionless contact}
In the case of spin-$1/2$ fermions, the charging energy pins only one out of
four modes. Fluctuations spin mode are not suppressed, see
Eq.~(\ref{eq:4.26}). These fluctuations average action (\ref{eq:4.28}) to zero.
To obtain a finite result for the ${\cal N}$-dependent part of the
thermodynamic potential, we have to expand $\Omega$ up to the second
order in ${\hat S}$:
\begin{equation}
\delta\Omega=-\frac{1}{2}T\langle T_{\tau}{\hat S}^2\rangle.
\label{eq:5.8}
\end{equation}
Upon the substitution of Eq.~(\ref{eq:4.28}) into Eq.~(\ref{eq:5.8}), we use
the expressions (\ref{eq:4.29}) to perform the averaging over $\hat{H}_0$.
This cumbersome albeit straightforward calculation yields
for the ${\cal N}$-dependent part of the thermodynamic potential:
\begin{eqnarray}
\delta\Omega&&=\frac{1}{(2\pi v_F)^2}\int_0^\beta d\tau_1 d\tau_2 d\tau_3
L(\tau_1)L(\tau_2)\sum_{\gamma=\pm 1}e^{-i2\pi\gamma{\cal N}}\times
\nonumber\\
&&\frac{\pi T}
{\left[\sin\pi T(\tau_1+i\gamma 0)\sin\pi T(\tau_2+i\gamma 0)\right]^{1/2}}
\times
\label{eq:5.9}\\
&&\frac{\pi T}
{\left[\sin\pi T(\tau_1+\tau_3+i\gamma 0)\sin\pi T(\tau_2-\tau_3+i\gamma
0)\right]^{1/2}}.
\nonumber
\end{eqnarray}
As in the Section~\ref{sec:5a}, it is convenient to use the Lehmann
representation (\ref{Lehman})  for the kernel $L$:
\begin{eqnarray}
\delta\Omega&&=\frac{1}{(2\pi v_F)^2}\int_{-\infty}^{\infty}\frac{dt_1}{2\pi}
\frac{dt_2}{2\pi}\times
\label{eq:5.10}\\
&&\left(L^R(t_1)-L^A(t_1)\right)\left(L^R(t_2)-L^A(t_2)\right)
\times
\nonumber\\
&&\int_0^\beta d\tau_1 d\tau_2 d\tau_3
\frac{\pi T}{\sinh \pi T(t_1+i\tau_1)}\frac{\pi T}{\sinh \pi T(t_2+i
\tau_2)}\times
\nonumber\\
&&\sum_{\gamma=\pm 1}e^{-i 2\pi \gamma{\cal N}}\frac{\pi T}
{\left[\sin\pi T(\tau_1+i\gamma 0)\sin\pi T(\tau_2+i\gamma 0)\right]^{1/2}}
\times
\nonumber\\
&&\frac{\pi T}
{\left[\sin\pi T(\tau_1+\tau_3+i\gamma 0)\sin\pi T(\tau_2-\tau_3+i\gamma
0)\right]^{1/2}}.
\nonumber
\end{eqnarray}
The evaluation of the integrals over $\tau_1$ and $\tau_2$ here is similar to the
procedure employed for the evaluation of the integral over $\tau$ in
Section~\ref{sec:5a}. Upon the integration, we find:
\begin{eqnarray}
\delta\Omega&&=\frac{1}{(2\pi v_F)^2}\int_{-\infty}^{\infty}\frac{dt_1}{2\pi}
\frac{dt_2}{2\pi}\times
\label{eq:5.11}\\
&&\left(L^R(t_1)L^R(t_2)e^{-i 2\pi {\cal N}}+L^A(t_1)L^A(t_2)e^{i2\pi{\cal N}}
\right)\times
\nonumber\\
&&\frac{\pi T}{\left[\sinh (\pi Tt_1)\sinh (\pi Tt_2)\right]^{1/2}}\times
\nonumber\\
&&\int_0^\beta d\tau_3\frac{\pi T}{\left[\sinh\pi T (t_1-i\tau_3)\sinh\pi T
(t_2+i\tau_3)\right]^{1/2}}.
\nonumber
\end{eqnarray}
Integral over $\tau_3$ can be easily evaluated with the help of the
formula
\[
\int_0^\pi\frac{d\phi}{\left[\sinh(x-i\phi)\sinh(y+i\phi )\right]^{1/2}}=
4e^{-|x+y|}{\bf K}
\left(
e^{-|x+y|}
\right),
\]
where ${\bf K}(k)$ is the complete elliptic integral of the first kind\cite{Ryzhik}.
We find for $\delta\Omega$:
\begin{eqnarray}
&&\delta\Omega=\frac{1}{(\pi
v_F)^2}\int_{0}^{\infty}\frac{dt_1dt_2\ \pi T}
{\left[\sinh (\pi Tt_1)\sinh (\pi Tt_2)\right]^{1/2}}\times
\label{eq:5.12}\\
&&\
\left[L^R(t_1)L^R(t_2)e^{-i 2\pi {\cal N}}+c.c.
\right]
e^{-\pi T (t_1+t_2)}
{\bf K}\left(
e^{-\pi T (t_1+t_2)}
\right).
\nonumber
\end{eqnarray}
In the derivation of Eq.~(\ref{eq:5.12}), we have utilized the reduced
version of action, see Eq.~(\ref{eq:4.28}), which is valid only on a
relatively long time scale larger than $1/E_C$.  Now, we average the product
of two thermodynamic potentials with the help of Eq.~(\ref{eq:3.25}). As we
will see shortly, the result of the averaging is logarithmically divergent at
large $t$. The divergence should be cut-off at $t_{1,2} \gtrsim 1/E_C$. 
Without violating the logarithmical accuracy, we can use also
asymptotic expansion of the elliptic integral 
${\bf K}(k)= -\ln\sqrt{1-k^2}$. We obtain
\begin{eqnarray}
&&\overline{\delta\Omega(1)\delta\Omega(2)}=
\frac{2\Delta^2}{(2\pi)^6}\cos 2\pi n\times
\nonumber\\
&&\int_{1/E_C}^\infty\frac{dt_1dt_2\ \pi^2T^2}{\sinh\pi Tt_1\sinh\pi Tt_2}
\left[\ln\frac{1}{T^2(t_1+t_2)^2}\right]^2\times
\nonumber\\
&&\left[e^{-t_1/\tau_D}+e^{-t_1/\tau_C}\right] 
\left[e^{-t_2/\tau_D}+e^{-t_2/\tau_C}\right].
\label{eq:5.13}
\end{eqnarray}
Here, as in Eq.~(\ref{eq:5.7}), we use the short hand notation $i= H_i,{\cal
N}_i$, and $n={\cal N}_1 - {\cal N}_2$. The diffuson and Cooperon decay times $\tau_D$
and $\tau_C$ are related  to the magnetic field values $H_1$, $H_2$
by Eq.~(\ref{tauH}).  Now we can use Eq.~(\ref{Cdif}) to find the correlation function
of the differential capacitances. In the leading logarithmic approximation we obtain 
\begin{eqnarray}
&&\frac{\overline{\delta C_{\it diff}(1)\delta C_{\it diff}(2)}}{C^2}=
\frac{8}{3\pi^2}\frac{\Delta^2}{E_C^2}\ln^4\left(\frac{E_C}{T}\right)
\cos 2\pi n \times
\nonumber\\
&&\sum_{\gamma =\pm}\left[1- 4\left(\frac{ \ln {\rm max} \left(1; 
\left[\frac{H_\gamma}{H^{<}_c}\right]^2\right)}
{ \ln \left(\frac{E_C}{T}\right)}\right)^3 + \dots \right],
\label{eq:5.14}
\end{eqnarray}
where the fields $H_\pm = H_1\pm H_2$ are assumed to be small compared to
the correlation field $H_c$ given by Eq.~(\ref{Bc}). The new temperature
dependent correlation field appearing in Eq.~(\ref{eq:5.14}) is given by
\begin{equation}
H_c^<=\frac{\Phi_0}{\cal A}\sqrt{\frac{T}{2\pi E_T}} 
\label{eq:5.140}
\end{equation}
with $\Phi_0$, ${\cal A}$, and $E_T$ being the flux quantum, the geometrical
area of the dot, and the Thouless energy respectively. Unlike the scale
$H_c$, the characteristic field $H_c^<$ is independent on the charging
energy. This smaller field scale appears due to the existence of the
``free'' excitation mode $\varphi^\sigma_+$ which is not pinned by the effect
of charging. 

With the increase of the number of channels in the dot-lead junction, the
number of free modes also increases. The role of charging (which  still
pins only one mode), and therefore of the field $H_c$, in the correlation
functions should vanish gradually. The dependence of the correlation
functions on the magnetic field at $H < H_c$ becomes power law rather than
logarithmic; however, this power law is still non-trivial, and approaches
Fermi-liquid results only in the limit of an infinite number of channels.

If $H_\pm \gtrsim H_c$, [the charging correlation field $H_c$ is given
by Eq.~(\ref{eq:2.25})], the correlation function of the fluctuation
starts decreasing much faster, $\propto 1/H_\pm^4$. 
Therefore, in order to get the representative statistics of the capacitance
fluctuations, averaging should be performed in the interval of magnetic
fields larger than the magnetic field determined by the charging energy
(\ref{eq:2.25}). Finally, in the  limit of a strong field,
$H_1=H_2\gtrsim H_c$ (unitary limit), the variance
$\overline{\delta C_{\it diff}^2}$ of the differential capacitance becomes 
four times smaller than at $H=0$. 

Our results for the correlation functions diverge logarithmically at
$T\to 0$. At lower
temperatures the pinning of the spin mode described
by action $\hat S$ should be taken into account. A variational
estimate\cite{preparation} shows that $T$ at low temperatures
should be replaced by $\Delta\ln(E_C/\Delta)$ in the above results for
${H}_c^<$ and for the correlation function. We will elaborate on
this point more in the end of the following subsection.

\subsubsection{Finite reflection in the contact
}
\label{sec:5c}

The main effect of the backscattering in the channel is that 
the Coulomb blockade appears  already in the averaged
capacitance\cite{Matveev95}. Taking into account  backscattering 
Hamiltonian (\ref{eq:4.27}) in the second order perturbation theory,
we obtain from Eq.~(\ref{Cdif}):
\begin{equation}
\delta C({\cal N}) = \frac{8 e^{\bf C}}{\pi}|r|^2 \cos 2\pi{\cal} N
\ln \left(\frac{E_C}{T}\right).
\label{eq:5.141}
\end{equation}
Due to the finite level spacing $\Delta$,  this result acquires mesoscopic
fluctuations. As we already mentioned in the introduction to this section, at $r\neq
0$ the leading term in fluctuations of the thermodynamic potential is first-order
in both $\hat{H}_{bs}$ and $\hat{S}$,
\begin{equation}
\delta\Omega=-T\langle T_{\tau}\int_0^\beta d\tau {\hat H}_{bs}(\tau){\hat
S}\rangle.
\label{eq:5.16}
\end{equation}
To calculate the average over the unperturbed state, we use the bosonized
representation of $\hat{H}_{bs}$ and $\hat{S}$ given by Eqs.~(\ref{eq:4.27})
and (\ref{eq:4.28}) respectively, and then the expressions (\ref{eq:4.29}) for
the correlation functions of the boson fields. Then, similar to the
Section~\ref{sec:5a}, we switch to the Lehmann representation
(\ref{Lehman}) for the kernel $L(\tau)$ to obtain:
\begin{eqnarray}
\delta\Omega&&=\frac{2\sqrt{e^{\bf C}}|r|\sqrt{E_C}}{\pi^2\sqrt{\pi}v_F}
\cos (\pi{\cal N})\int_{-\infty}^{\infty}\!\frac{dt}{2\pi}
\left(L^R(t)-L^A(t)\right)
\times
\nonumber\\
&&\int_0^\beta \!\frac{d\tau_1 d\tau_2 \quad (\pi T)^2} 
{\sinh [\pi Tt+i\pi T(\tau_1-\tau_2)]\left[\sin (\pi T\tau_1)
\sin (\pi T\tau_2)\right]^{1/2} }\times
\nonumber\\
&&\sum_{\gamma=\pm 1}\frac{(\pi T)^{1/2}}
{\left[\sin\pi T(\tau_2-\tau_1+i\gamma 0)\right]^{1/2}}
e^{-i\pi \gamma({\cal N}-1/4)}.
\label{eq:5.17}
\end{eqnarray}
Here only the ${\cal N}$-dependent part of the thermodynamic potential is
taken into account. Integral over $\tau_1$ in Eq.~(\ref{eq:5.17}) is
determined by the contribution of the pole at $\tau_1=\tau_2-it$, which can
be easily calculated:
\begin{eqnarray}
\delta\Omega&&=\frac{2\sqrt{e^{\bf C}}|r|\sqrt{E_C}}{\pi^2\sqrt{\pi}v_F}
\cos (\pi{\cal N})\int_{-\infty}^{\infty}\!
\frac{dt\quad (\pi T)^{1/2}}{\left[-i\sinh (\pi Tt)\right]^{1/2}}\times
\nonumber\\
&&\left[L^A(t)e^{i\pi({\cal N}-1/4)} 
+L^R(t)e^{-i\pi({\cal N}-1/4)}\right]\times
\nonumber\\
&&\int_0^\beta d\tau_2
\frac{\pi T}{\left[\sin (\pi T\tau_2) \sin (\pi T\tau_2+it)\right]^{1/2}}.
\label{eq:5.18}
\end{eqnarray}
Integration over $\tau_2$ now is completely similar to the one we performed
over the variable $\tau_3$ in Eq.~(\ref{eq:5.11}), and we find for
$\delta\Omega$:
\begin{eqnarray}
\delta\Omega&&=\frac{2\sqrt{e^{\bf C}}|r|\sqrt{E_C}}{\pi^2\sqrt{\pi}v_F}
\cos (\pi{\cal N})\int_{-\infty}^{\infty}\!\!
\frac{(\pi T)^{1/2}\ln(1/T^2t^2)dt}
{\left[-i\sinh (\pi Tt)\right]^{1/2}}\times
\nonumber\\
&&\left[L^A(t)e^{i\pi{\cal N}} + L^R(t)e^{-i\pi{\cal N}}\right].
\label{eq:5.19}
\end{eqnarray}
From Eq.~(\ref{eq:5.19}), with the help of Eq.~(\ref{eq:3.25})
and (\ref{Cdif}), we find the correlation function of mesoscopic
fluctuation of the capacitances:
\begin{eqnarray}
&&
\frac{\overline{\delta C_{\it diff}(1)\delta C_{\it diff}(2)}}{C^2}=
\frac{32 e^{\bf C}}{3 \pi^2}\frac {|r|^2\Delta}{E_C}\ln^3
\left(\frac{E_C}{T}\right)\cos 2\pi n
\times
\nonumber\\
&&\sum_{\gamma =\pm}\left[1- \left(\frac{ \ln {\rm max} \left(1; 
\left[\frac{H_\gamma}{H^{<}_c}\right]^2\right)}
{ \ln \left(\frac{E_C}{T}\right)}\right)^3 \right],
\label{eq:5.20}
\end{eqnarray}
where $i=H_i,{\cal N}_i$, $n={\cal N}_1 - {\cal N}_2$, and $H_\pm =
H_1\pm H_2$. Correlation field $H^<_c$ is defined in
Eq.~(\ref{eq:5.140}).  The amplitude of fluctuations at a partial
transmission of the channel is parametrically larger than at $r=0$, cf
Eq.~(\ref{eq:5.14}).  Furthermore, in the unitary limit the variance
of the differential capacitance is suppressed only by a half of its
zero-field value. This similarity with the case of spinless fermions is due
to the backscattering in the channel, which leads to pinning of the spin
mode.

Result (\ref{eq:5.20}) is valid at relatively high temperatures. As was shown
by Matveev\cite{Matveev95}, the divergences should be cut at energy
$\epsilon^* \simeq |r|^2E_C\cos^2\pi{\cal N}$ which corresponds to the
pinning energy of the spin mode. The higher-order corrections
in backscattering\cite{Matveev95} shows that at $T\lesssim |r|^2E_C$ the
logarithmic growth of fluctuations saturates.  Simultaneously, the
correlation functions start to depend not only on the difference
${\cal N}_1-{\cal N}_2$, but also on each of these arguments
separately. This weaker logarithmic dependence is beyond the scope of
this paper. For an estimate of the differential capacitance variance
at low temperature, one may replace $\ln (E_C/T)$ by $\ln(1/|r|^2)$ in
Eq.~(\ref{eq:5.20}). 

Finally, we elaborate on the estimate of the
characteristic energy scale which controls the low-temperature cutoff
for the reflectionless contact. Comparing Eq.~(\ref{eq:5.141}) with
Eq.~(\ref{eq:5.19}), or Eq.~(\ref{eq:5.19}) with Eq.~(\ref{eq:5.14}),
we observe that the reflectionless case results can be obtained from
formulas with finite reflection coefficient by putting $|r|^2 \mapsto
(\Delta/E_C)\ln (E_C\Delta)$. It would correspond to the energy of
pinning of the spin mode $\epsilon^* \simeq \Delta \ln
(E_C/\Delta)$ which agrees with our variational estimate\cite{preparation}.

\section{Tunneling conductance of the "open" dot}
\label{sec:6} 

In the previous Section, we considered in detail the thermodynamics of
the dot with one almost open channel, and studied mesoscopic effects
related to the discreteness of the charge. However, the majority of 
experimental work deals not with thermodynamics, but rather with 
transport through a dot. Coulomb blockade shows up as an oscillatory gate
voltage dependence of the conductance of the dot connected with two leads.

The case of small transparency of the channels connecting the dot with
leads is well studied\cite{Review,Stone,Averin90,Aleiner96}. The
conductance in the valleys can be represented as the sum of two
physically different contributions -- elastic and inelastic
cotunneling\cite{Averin90}. During the {\em elastic} cotunneling
process, an electron enters the dot, spends there a time $\simeq
\hbar/E_C$ and then leaves the dot. Without interaction, this electron
would be able to spend a time of the order of $\hbar /\Delta$.  As the
result, the conductance in the Coulomb blockade valleys is suppressed
by a factor of $\Delta/E_C$. During the {\em inelastic} cotunneling
process\cite{AverinOdintsov}, an electron enters the dot, spends there
time $\simeq \hbar/E_C$ and then another electron leaves the dot. In
this case, the final state contains an extra, in comparison with the
initial state, two-particle excitation. It means that the phase volume
of the final state is small as $T^2$; therefore, the elastic
contribution dominates at $T \leq \left(E_C\Delta\right)^{1/2}$.

\narrowtext{
\begin{figure}[h]
\vspace{0.2cm}
\epsfxsize=6.2cm
\hspace*{0.5cm}
\epsfbox{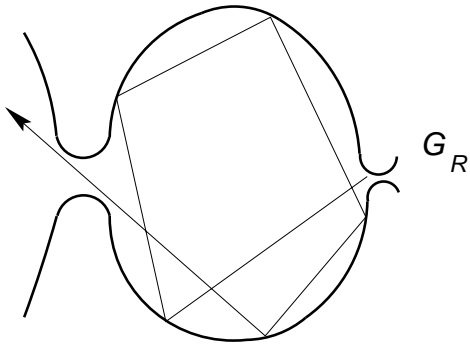}
\vspace{0.6cm}
\caption{Schematic view of the asymmetric two-terminal setup. The left
point contact has one channel almost open and the conductance of the
right point contact $G_R$ is much smaller than $e^2/2\pi\hbar$.
One of the electron trajectories contributing to the elastic cotunneling is 
also shown.
}
\label{Fig:6.1}
\end{figure} 
}

Conductance of the dot connected to each reservoir by almost perfect channel
was studied, in the limit $\Delta/E_C\to 0$, by Furusaki and
Matveev\cite{Furusaki95}. They concluded that even small reflection in any
of the channels leads to a dramatic suppression of the conductance of the
system. At any gate voltage, except the discrete points of charge
degeneracy, they obtained a $T^2$ law which closely resembles the behavior of
inelastic cotunneling in the weak tunneling regime. Pursuing the analogy
with the weak tunneling regime further, it is natural to expect that there
should be another non-vanishing at $T\to 0$ contribution from a counterpart
of the elastic cotunneling mechanism. Studying of this contribution is the
goal of this section\cite{attempt}.

Similar to Ref.~\onlinecite{Furusaki95}, two limiting cases may be
distinguished. In the first case both channels are either open or have
the same reflection coefficient. This case is technically difficult
to consider. Instead, we concentrate on the properties of the strongly
asymmetric setup; one point contact (left in Fig.~\ref{Fig:6.1}) has
the transparency close to unity and the other contact (right in
Fig.~\ref{Fig:6.1}) has a very small conductance $G_R \ll
e^2/\pi\hbar$. This case can be realized experimentally by a corresponding
adjustment of the voltages on the gates forming point
contacts\cite{MarcusPrivate}. Moreover, it follows from the scaling
arguments\cite{Furusaki95} that the asymmetry of contacts
is a relevant perturbation. In the case $\Delta = 0$, the strongly
asymmetric limit corresponds to the fixed point of a system with an
infinitesimally small initial asymmetry. Therefore we can expect
that at $\Delta/E_C\ll 1$, this limit adequately describes dots with a finite
initial degree of asymmetry.

\subsection{General formalism}
\label{sec:6a}
For calculation of such tunneling conductance, we have to modify
derivation of Secs.~\ref{sec:3a} and \ref{sec:3b} in order to take into
account the tunneling between the dot and the second lead. In comparison with
Hamiltonian (\ref{Hamiltonian}), the total Hamiltonian of the system acquires
two additional terms,
\begin{equation}
\hat{H}=\hat{H}_F + \hat{H}_C+\hat{H}_M +\hat{H}_T.
\label{eq:6.1}
\end{equation}
Here $\hat{H}_F$ describes the electron motion in the dot and in the left
lead and is given by (\ref{HF}), interaction Hamiltonian $\hat{H}_C$ is given
by Eq.~(\ref{Hc}), and $ \hat{H}_M$ is the Hamiltonian of free electrons in
the right lead
\begin{equation}
\hat{H}_M=\sum_p\xi_p \hat{a}^\dagger_p \hat{a}_p.
\label{eq:6.2}
\end{equation}
Tunneling Hamiltonian $\hat{H}_T$ describes the weak coupling
between the right lead and the dot,
\begin{equation}
\hat{H}_T=v_t \hat{\psi}^\dagger ( {\bf r}_t) 
\sum_p \hat{a}_p + h.c. \, ,
\label{eq:6.3}
\end{equation}
where ${\bf r}_t$ is the coordinate of the
tunneling contact, $\hat{\psi}^\dagger({\bf r})$ is the electron wavefunction
in the lead, and $v_t$ is the coupling constant which will be later related
to the tunneling conductance of the contact $G_R$.

Because $G_R \ll e^2/(2\pi\hbar)$, we can consider the tunneling
current $I$ as the function of applied voltage $V$ in the second order
of perturbation theory in tunneling Hamiltonian (\ref{eq:6.3}).
This gives us the standard result\cite{Mahan}
\begin{equation}
I(eV) =i\left[  
J\left(i\Omega_n\to eV+i0\right)-
J\left(i\Omega_n\to eV-i0\right) \right], 
\label{eq:6.4}
\end{equation}
where $\Omega_n=2\pi Tn$ is the bosonic Matsubara frequency, and
Matsubara current $J$ is defined as
\begin{equation}
J(i\Omega_n)=ev_t^2\nu 
\int_0^\beta d\tau e^{-i\Omega_n\tau} {\cal G}_M(\tau)\Pi (\tau).
\label{eq:6.5}
\end{equation}
Here $\nu$ is the one-electron density of states per unit area and
per one spin in the dot, ${\cal G}_M$ is the Green function of the
electrons in the leads,
\begin{equation}
{\cal G}_M \equiv -\sum_{p_1,p_2}\langle T_\tau
\hat{a}_{p_1}(\tau)\hat{a}_{p_2}^\dagger(0)\rangle
=\nu_M \frac{\pi T}{\sin\pi T\tau },
\label{eq:6.6}
\end{equation}
with $\nu_M$ being the one-electron density of states per one spin in
the lead, and function $\Pi(\tau)$ is given by
\begin{equation}
\Pi(\tau) = \nu^{-1}
\langle T_\tau \bar{\psi}\left(\tau;{\bf r}_t\right)
{\psi}\left(0; {\bf r}_t\right)
\rangle. 
\label{eq:6.7}
\end{equation}
  Averages in Eqs.~(\ref{eq:6.6}) and
(\ref{eq:6.7}) are performed with respect to the equilibrium distribution
of the system without tunneling. We choose to introduce $\nu$ into
Eq.~(\ref{eq:6.4}) and into definition (\ref{eq:6.6}) to make function
$\Pi (\tau )$ dimensionless.

In the absence of the interaction, $E_C =0$, propagator $\Pi(\tau)$ is
nothing but the Green function of non-interacting system; its ensemble
average has the form analogous to Eq.~(\ref{eq:6.6}),
\begin{equation}
\left.\overline{\Pi(\tau)}\right|_{E_C=0} 
= \frac{\pi T}{\sin\pi T\tau }.
\label{eq:6.8}
\end{equation}
Then, substitution of Eqs.~(\ref{eq:6.6}) and (\ref{eq:6.8}) into
Eq.~(\ref{eq:6.5}) and analytic continuation (\ref{eq:6.4}) gives the
tunneling current $I=s G_RV$ ($s$ is the spin degeneracy), where the
tunneling conductance of the contact per one spin is
\begin{equation}
G_R = \frac{2\pi e^2}{\hbar}v_t^2\nu_M\nu .
\label{eq:6.9}
\end{equation}
With the help of Eq.~(\ref{eq:6.9}) we can rewrite Eq.~(\ref{eq:6.5}) in terms
of the bare conductance of the point contact:
\begin{equation}
J(i\Omega_n)=\frac{G_R}{2\pi e}
\int_0^\beta d\tau \frac{\pi T e^{-i\Omega_n\tau} }
{\sin\pi T\tau }
\Pi (\tau).
\label{eq:6.10}
\end{equation}

As we will see below, function $\Pi(\tau)$ can be analytically continued
from the real axis to the complex plane, so that the result is analytic in a
strip $0 < {\rm Re }\,\tau < \beta$, and has branch cuts along ${\rm
Re}\,\tau=0,\beta$ lines. It allows one to deform the contour of
integration as shown in Fig.~\ref{Fig:6.10}, and to obtain
\begin{eqnarray}
J(i\Omega_n)&=&\frac{G_R T}{2 e}
\int_{-\infty}^\infty \!dt e^{\Omega_nt}
\left[\theta (-\Omega_n) \theta(t)-
\theta (\Omega_n) \theta(-t)  
\right]\times
\nonumber\\
&&
\left(\frac{\Pi (it+0) }
{\sinh[\pi T(t-i0)]}-
\frac{\Pi (it-0) }
{\sinh[\pi T(t+i0)]
}
\right).
\label{eq:6.100}
\end{eqnarray}
Now the analytic continuation (\ref{eq:6.4}) can be  performed,
because the periodicity of the Matsubara Green functions was already
taken into account. This gives
\begin{eqnarray}
I(eV)&=&i\frac{G_R T}{2 e}
\int_{-\infty}^\infty \!dt e^{-ieVt}
\times
\nonumber\\
&&
\left[
\frac{\Pi (it+0) }
{\sinh[\pi T(t-i0)]}
-\frac{\Pi (it-0) }
{\sinh[\pi T(t+i0)]}
\right].
\label{eq:6.101}
\end{eqnarray}
Next, we  use the analyticity of $\Pi(\tau)$ in the strip $0<{\rm
Re}\,\tau <\beta$, and shift the integration variable $t \to t-i\beta/2$
in the first term in brackets in Eq.~(\ref{eq:6.101}), and $t \to
t+i\beta/2$ in the second term. Bearing in mind that
$\Pi(\tau)=-\Pi(\tau +\beta )$, we find
\begin{equation}
I=\left(T \sinh \frac{eV}{2T}\right) G_R \int_{-\infty}^\infty dt
e^{-ieVt}
\frac{\Pi \left(it+\frac{\beta}{2}\right)}{\cosh \pi T t}.
\label{eq:6.102}
\end{equation}
Linear conductance $G$ is therefore given by
\begin{equation}
G= G_R \int_{-\infty}^\infty dt
\frac{\Pi \left(it+\frac{\beta}{2}\right)}{2\cosh \pi T t}.
\label{eq:6.103}
\end{equation}

\narrowtext{
\begin{figure}[h]
\vspace{-0.5cm}
\epsfxsize=6.7cm
\hspace*{0.5cm}
\epsfbox{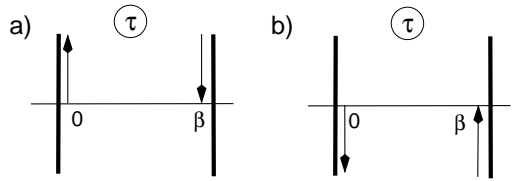}
\vspace{0.3 cm}
\caption{The integration contour used in the evaluation of
the conductance, see Eq.~(\protect\ref{eq:6.10}) for (a) $\Omega_n <0$,
and (b) for $\Omega_n > 0$. Branch cuts of the analytic continuation
of $\Pi (\tau )$ are shown by  thick lines.}
\label{Fig:6.10}
\end{figure} 
}

Let us turn now to the actual calculation of the function $\Pi(\tau)$.
It was shown in Ref.~\onlinecite{Furusaki95} that the interaction
drastically affects the form of the function (\ref{eq:6.7}), however,
some contributions were not taken into account. Our purpose is to
construct an effective action theory, similar to that of Sec.~\ref{sec:3},
for calculation of $\Pi (\tau)$. Once again, we wish to get rid of the
fermionic degrees of freedom of the dot.  Similar to Eq.~(\ref{Q1}),
it is convenient to rewrite charge operator in terms of the variables
of the channel. However, here we have to keep in mind the fact that the
tunneling events described by operators ${\psi}^\dagger\left({\bf
r}_t\right)$ and ${\psi}\left({\bf r}_t\right)$ change the charge in
the system by $+e$ and $-e$. It can be taken 
into account\cite{Furusaki95} by
introducing  three additional operators: Hermitian operator  $\hat{n}$, and
unitary operators  $\hat{F}, \hat{F}^\dagger$  with the following
commutation relations:
\begin{equation}
\left[\hat{n},\hat{F}^\dagger\right]=\hat{F}^\dagger.
\label{eq:6.11}
\end{equation}
We can definitely choose Hilbert subspace in a way such that operator
$\hat{n}$ has integer eigenvalues.  Finally, these operators commute
with all the fermionic degrees of freedom.  
Then, we can change the definition of the charge operator [cf. Eq.~(\ref{Q1})]
to
\begin{equation}
\frac{\hat{Q}}{e}= - 
\int_{\mbox{channel}} d{\bf r}\psi^\dagger\psi+\hat{n},
\label{eq:6.12}
\end{equation}
and rewrite Eq.~(\ref{eq:6.7}) as 
\begin{equation}
\Pi(\tau) = \nu^{-1}
\langle T_\tau \hat{\bar{F}}(\tau )\bar{\psi}\left(\tau;{\bf r}_t\right)
\hat{{F}}(0){\psi}\left(0; {\bf r}_t\right)
\rangle. 
\label{eq:6.13}
\end{equation}
It is easy to see from Eqs.~(\ref{eq:6.12}) and
(\ref{eq:6.11}) that operators $\hat{F}^\dagger, \hat{F}$ in
Eq.~(\ref{eq:6.13}) change the charge by $+e$ and $-e$ respectively, in
accordance with the initial definition of charge.

After this manipulation, the Hamiltonian of the system and correlation
function (\ref{eq:6.13}) become quadratic in the fermionic operators of
the dot, so that part of the system can be integrated out. We use the
identity similar to Eq.~(\ref{transformation}):
\wide{m}{
\begin{eqnarray}
{\rm Tr}_2&&\left\{e^{-\beta \hat{H}_F}T_\tau
\bar{\psi}_2(\tau)\psi_2(0)\right\}= e^{-\beta \hat{H}_1 }{\rm Tr}_2
\left[ e^{-\beta \hat{H}_2} T_\tau \bar{\psi}_2(\tau)\psi_2(0)
e^{-\int_0^\beta d\tau \hat{H}_{12}(\tau)} \right]=
\label{eq:6.14}\\
&&e^{-\beta \Omega_2} \langle T_\tau
\bar{\psi}_2(\tau)\psi_2(0)\rangle_2 e^{-\beta \hat{H}_1 } T_\tau e^{
\frac{1}{2}\int_0^\beta \!\! d\tau _1\int_0^\beta\!\! d\tau _2
\langle\hat{H}_{12}(\tau_1)\hat{H}_{12}(\tau_2)\rangle_2}+\nonumber \\
&&
e^{-\beta \Omega_2}
e^{-\beta \hat{H}_1 }\!\int_0^\beta
d\tau _3\int_0^\beta\! d\tau _4 T_\tau\
\langle T_\tau
\bar{\psi}_2(\tau)\hat{H}_{12}(\tau_3)\rangle_2
\langle T_\tau
{\psi}_2(0) \hat{H}_{12}(\tau_3))\rangle_2
 e^{ \frac{1}{2}\int_0^\beta
\!\! d\tau _1\int_0^\beta\!\! d\tau _2
\langle\hat{H}_{12}(\tau_1)\hat{H}_{12}(\tau_2)\rangle_2},
\nonumber
\end{eqnarray}
}
\noindent where all $\psi^\dagger_2, \psi_2$ are the fermionic operators of
the dot,  and   the rest of the notation is the same as that in
Eq.~(\ref{transformation}).

The calculation of the product
$\langle\hat{H}_{12}(\tau_1)\hat{H}_{12}(\tau_2)\rangle_2$ was performed
in Sec.~\ref{sec:3a}, see Eq.~(\ref{A1}), and all the steps 
leading to the derivation of the effective action (\ref{action}) can
be  repeated here. Calculation of the remaining operator products
can be performed along the lines of Appendix~\ref{ap:1}. This yields
\begin{eqnarray}
\langle T_\tau\psi (\tau_1; {\bf r}_t) \hat{H}_{12}(\tau_2)\rangle_2
= -\psi (\tau_2,0)R^*(\tau_1-\tau_2),
\label{eq:6.15}\\
\langle T_\tau\bar{\psi} (\tau_1; {\bf r}_t) \hat{H}_{12}(\tau_2)\rangle_2
= -\bar{\psi} (\tau_2,0)R(\tau_2-\tau_1),
\nonumber
\end{eqnarray}
where, similar to Eq.~(\ref{A1}),  
$\psi(\tau; x)=e^{\tau \hat{H}_1}\psi( x)e^{-\tau \hat{H}_1}$
are the one dimensional fermionic operators of the channel in the
interaction representation, $\bar{\psi}(\tau)=\psi^\dagger (-\tau )$.
Kernel $R$ describes the motion of an electron from the tunnel
contact to the entrance of the single  mode channel, and it is given by
\begin{equation}
R(\tau)=
\frac{1}{2m}
\int dy
\phi({y})
\partial_{x}
{{\cal G}}(\tau; {\bf r},{\bf r}_t).
\label{eq:6.16}
\end{equation}
Here, ${\cal G}$ is the exact Matsubara Green function of the closed dot
subjected to the zero boundary condition. The wave function $\phi(y)$
describes the transverse motion in the single-mode channel, and the
coordinates $x$ in the derivative of the Green function ${\cal G}$ is set to
$+0$.

Kernel $R(\tau)$ is the random quantity with the zero
averages. In the universal regime, products of retarded $R^R(t)$ and
advanced $R^A(t)$ counterparts of $R(\tau)$ entering into the Lehmann
representation (\ref{Lehman}) have the following non-vanishing averages:
\begin{mathletters}
\label{eq:6.17}
\begin{eqnarray}
&&\frac{1}{\nu}\langle
 R^R_{H_1}(t_1)R^A_{H_2}(t_2)\rangle=\Delta {v_F}\delta
(t_1+t_2)\theta (t_1) e^{-\frac{t_1}{\tau_H^C}}, \label{eq:6.17a} \\
&&\frac{1}{\nu}\langle
 R^R_{H_1}(t_1)\left[R^R_{H_2}(t_2)\right]^*\rangle=
\nonumber\\
&&
\quad\quad
\Delta{v_F}
\delta
(t_1-t_2)\theta (t_1) e^{-\frac{t_1}{\tau_H^D}} \label{eq:6.17b},
\end{eqnarray}
where the decay times $\tau_H^{C,D}$ associated with applied
magnetic fields $H_{1,2}$ are given by Eq.~(\ref{tauH}). All the higher
momenta can be found by using the Wick theorem\cite{Footnote1}. Deriving
Eq.~(\ref{eq:6.17b}) we use Eqs.~(\ref{eq:3.21}), (\ref{universal}), and the
identity ${\cal G}^R(t;r_1,r_2) = \left[{\cal G}^A(-t;r_2,r_1)\right]^*$.
\end{mathletters}

To complete the derivation of the effective theory, we use
Eqs.~(\ref{eq:6.14}) and (\ref{eq:6.15}), introduce left and right
moving fermions similarly to Sec.~(\ref{sec:3a}), and thus obtain the
effective action representation for $\Pi(\tau)$ from
Eq.~(\ref{eq:6.13}):
\begin{mathletters}
\label{eq:6.18}
\begin{eqnarray}
&&\Pi (\tau) = \Pi_{in}(\tau) +  \Pi_{el}(\tau);
\label{eq:6.18a} \\
&&\Pi_{in}=-\frac{{\cal G}(- \tau; {\bf r}_t, {\bf r}_t ) }
{\nu \langle T_\tau e^{-\hat{S}}\rangle}
\langle T_\tau e^{-\hat{S}}
\hat{\bar{F}}(\tau)\hat{F}(0)
\rangle;\label{eq:6.18b}\\
&&\Pi_{el}\!=\frac{1}
{\nu\langle T_\tau e^{-\hat{S}}\rangle}\int_0^\beta
\!\! d\tau_1d\tau_2 
R(\tau_1-\tau) R^*(-\tau_2)\times \label{eq:6.18c}\\
&&\langle T_\tau e^{-\hat{S}}
\hat{\bar{F}}(\tau)\hat{F}(0)
\left[\bar{\psi}_L(\tau_1)+\bar{\psi}_R(\tau_1)\right]
\left[\psi_L (\tau_2) + \psi_R (\tau_2)\right]
\rangle.
\nonumber
\end{eqnarray}
Here the averaging is performed with respect to the Hamiltonian
\end{mathletters}
\begin{eqnarray}
\hat{H}_0 &=& 
iv_F\int_{-\infty}^\infty dx \left\{\psi_L^\dagger\partial_x\psi_L -
\psi_R^\dagger\partial_x\psi_R
\right\} \label{eq:6.19} \\
&&+ \frac{E_C}{2}\left(\int_{-\infty}^0dx
:\psi_L^\dagger\psi_L +\psi^\dagger_R\psi_R:+{\cal N}-\hat{n}\right)^2,
\nonumber
\end{eqnarray}
and action $\hat{S}$ is given by Eq.~(\ref{lraction}).  The difference
between the Eq.~(\ref{eq:6.19}) and Eq.~(\ref{eq:3.15}), is
caused by the different definitions of the charge operator in
Eqs.~(\ref{Q1}) and (\ref{eq:6.12}). 

Two contributions can be distinguished in the correlation function
(\ref{eq:6.18}). Inelastic contribution (\ref{eq:6.18b}) was considered in
Ref.~\onlinecite{Furusaki95} in the approximation corresponding to $\hat{S}=
0$, and with the Green function of the dot ${\cal G}$ replaced by its
averaged value $\overline{\cal G}$. The obtained results vanish at low
temperatures. The reason for the vanishing is that this term does not allow
the introduced electron to leave the dot; charge of the dot at the moment of
tunneling suddenly changes by $+e$ and all the other electrons have to
redistribute themselves to accommodate this charge. The logarithmical
divergence of the imaginary time action corresponding to such evolution
(orthogonality catastrophe) completely suppresses this contribution at $T\to
0$. Conversely, the second contribution, $\Pi_{el}$
from Eq.~(\ref{eq:6.18c}), contains the kernel $R(\tau)$ which promotes an
electron from the tunneling contact to the channel. Because the very same
tunneling electron is introduced to and then removed from the dot, there is
no need in the redistribution of other electrons, so no orthogonality
catastrophe occurs. As the result, the elastic contribution survives at
$T\to 0$, analogously to the elastic cotunneling contribution for the weak
coupling regime.

In what follows, we will be interested in the low temperature behavior
of the system, so we will retain elastic contribution (\ref{eq:6.18c})
only. Similarly to the Sec.~\ref{sec:4} results for electrons with
spin and spinless electrons differ significantly, and we will consider
those two cases separately.

\subsection{Spinless electrons}

We follow the lines of Sec.~\ref{sec:4a} in the bosonization of the
chiral fermionic fields. In order to account for the appearance of the
operator ${\hat n}$ in the Hamiltonian [compare Eqs.~(\ref{eq:3.15})
with Eq.~(\ref{eq:6.19})], we change slightly the  transformation
(\ref{eq:4.8}): 
\begin{eqnarray}
\hat{\varphi}_L(x)=
\frac{\hat{\varphi}_+(x) + \hat{\varphi}_-(x) +\hat{\Phi}}
{\sqrt{2}}
 - \pi {\cal N}+\pi\hat{n};
\nonumber\\
\hat{\varphi}_R(x)=
\frac{\hat{\varphi}_+(-x) - \hat{\varphi}_-(-x) -\hat{\Phi}}
{\sqrt{2}}
 - \pi {\cal N}+\pi\hat{n},
\label{eq:6.20}
\end{eqnarray}
where operator $\hat{n}$ commutes with the bosonic fields
$\hat{\varphi}_{\pm}, \hat{\Phi}$. In order to preserve the
commutation relation
$\left[\hat{F}^\dagger,\hat{\varphi}_{L,R}\right]=0$, we change the
operator $\hat{F}^\dagger$ as
\begin{equation}
\hat{F}^\dagger \mapsto \hat{F}^\dagger e^{-i\sqrt{2}\hat{\Phi}},
\hat{F} \mapsto \hat{F} e^{i\sqrt{2}\hat{\Phi}}, 
\label{eq:6.21}
\end{equation} 
The fact that $\hat{F}^\dagger$ commutes with bosonic fields
$\hat{\varphi}_{L,R}$ is obvious from Eqs.~(\ref{eq:6.11}) and
Eq.~(\ref{eq:4.7d}).

Substitution of Eqs.~(\ref{eq:4.0}), (\ref{eq:6.20}) and 
(\ref{eq:6.21}) into Eqs.~(\ref{eq:6.18b}) yields
\begin{mathletters}
\label{eq:6.22}
\begin{equation}
\Pi_{in}(\tau)=-\frac{{\cal G}(-\tau; {\bf r}_t, {\bf r}_t ) }
{\nu \langle T_\tau e^{-\hat{S}}\rangle}
\langle T_\tau e^{-\hat{S}}
\hat{\bar{F}}(\tau)\hat{F}(0)
e^{i\sqrt{2}\left[\hat{\Phi}(0)-\hat{\Phi}(\tau )\right]}
\rangle
\label{eq:6.22a}
\end{equation}
for the inelastic part of the cotunneling, see also
Ref.~\onlinecite{Furusaki95}. For the elastic contribution, we find
\begin{eqnarray}
&&\Pi_{el}(\tau)\!=\frac{2}
{\pi\nu\lambda\langle T_\tau e^{-\hat{S}}\rangle}
\int_0^\beta\!\!d\tau_1d\tau_2 R(\tau_1-\tau )R^*(-\tau_2)\times 
\nonumber\\
&&\langle T_\tau e^{-\hat{S}}
\hat{\bar{F}}(\tau)\hat{F}(0)
e^{\frac{i}{\sqrt{2}}\left[2\hat{\Phi}(0)-2\hat{\Phi}(\tau )
+\hat{\Phi}(\tau_1)
-\hat{\Phi}(\tau_2)
\right]}\times\nonumber\\
&&\hat{\eta}(\tau_1)\hat{\eta}(\tau_2)
(-1)^{\hat{n}(\tau_1)+\hat{n}(\tau_2)}
\exp\left[i\frac{\hat{\varphi}_-(\tau_1)-
\hat{\varphi}_-(\tau_2)}{\sqrt{2}}\right]\times
\nonumber \\
&&\prod_{i=1}^2\cos\left[\frac{\hat{\varphi}_+(\tau_i)}
{\sqrt{2}}+\frac{\pi}{4}-\pi {\cal
N}\right]
\rangle.
\label{eq:6.22b}
\end{eqnarray}
Averaging in Eqs.~(\ref{eq:6.22}) is performed over the Hamiltonian
given by Eq.~(\ref{eq:4.9}), finite backscattering is described by
(\ref{eq:4.10}), and action (\ref{eq:4.11}) is modified as
\end{mathletters}
\begin{eqnarray*}
\hat{S}\!=\!&&\frac{1}{2\pi\lambda}\int_0^\beta\!\! d\tau_1 d\tau_2
L\left(\tau_1\!-\!\tau_2\right)\hat{\eta}(\tau_1)\hat{\eta}(\tau_2)
(-1)^{\hat{n}(\tau_1)+\hat{n}(\tau_2)}
\times
\\
&&\exp\left[i\frac{\hat{\varphi}_-(\tau_1)-
\hat{\varphi}_-(\tau_2)}{\sqrt{2}}\right]
\exp\left[i\frac{\hat{\Phi}(\tau_1)-
\hat{\Phi}(\tau_2)}{\sqrt{2}}\right]\times
\nonumber\\
&&\cos\left[\frac{\hat{\varphi}_+(\tau_1)}{\sqrt{2}}
+\frac{\pi}{4}-\pi{\cal
N}\right]
\cos\left[\frac{\hat{\varphi}_+(\tau_2)}{\sqrt{2}}+
\frac{\pi}{4}-\pi{\cal
N}\right].  
\end{eqnarray*}

We will consider only the elastic contribution (\ref{eq:6.22b}) 
because it does not vanish at low temperatures.

\subsubsection{Reflectionless contact}

In the lowest in $\Delta/E_C$ approximation we can neglect action
$\hat{S}$ in Eq.~(\ref{eq:6.22b}) at all. Then averaging over bosonic
fields can be performed with the help of Eqs.~(\ref{eq:4.12}) and
Eq.~(\ref{eq:4.13}), average of the product Majorana fermions
operator is given by Eq.~(\ref{eq:4.6}), and the relevant correlation 
function of the operators $\hat{n}, \hat{F}, \hat{F}^\dagger$ is given
by
\begin{equation}
\langle T_\tau \hat{\bar{F}}(\tau) \hat{F}(0)
(-1)^{\hat{n}(\tau_1) +\hat{n}(\tau_2) }\rangle = 
{\rm sgn} (\tau -\tau_1){\rm sgn} (\tau -\tau_2).
\label{eq:6.23}
\end{equation}
Equation (\ref{eq:6.23}) follows from Eq.~(\ref{eq:6.11}) and from the
fact that operators $\hat{n}, \hat{F}$ commute with Hamiltonian
(\ref{eq:4.10}) and thus do not have their own dynamics. We obtain
\begin{eqnarray}
\Pi_{el}(\tau)&=&\frac{2\pi\left|K(\tau)\right|^2}
{\nu v_F E_C^2 e^{2 {\bf C}}}
\int_0^\beta\!\!d\tau_1d\tau_2 R(\tau_1-\tau )R^*(-\tau_2)\times 
\nonumber\\
&&\left[
\frac{\pi T}{\sin \pi T (\tau_1-\tau_2)}
\frac{K(\tau_2)K(\tau_1-\tau)}{K(\tau_1)K(\tau_2-\tau)} +
\right. \label{eq:6.24}\\
&&\left.e^{i2\pi{\cal N}}
K(\tau_1-\tau_2)
\frac{K(-\tau_2)K(\tau_1-\tau)}{K(\tau_1)K(\tau-\tau_2)}+ c.c.
\right],
\nonumber
\end{eqnarray}
where function $K(\tau)$ is defined by Eq.~(\ref{eq:4.15b}), and
${\bf C}$ is the Euler constant.

\narrowtext{
\begin{figure}[h]
\vspace{0.2cm}
\epsfxsize=6.7cm
\hspace*{0.5cm}
\epsfbox{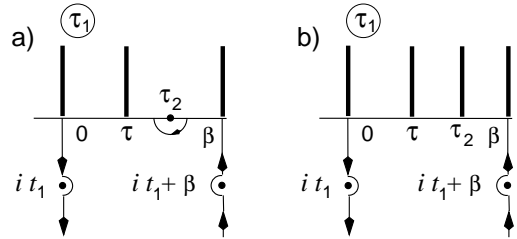}
\vspace{0.3 cm}
\caption{The integration contour used in the evaluation of
the conductance in the spinless case, for the first (a) and the second
term (b) in Eq.~(\protect\ref{eq:6.25}). Branch cuts of the integrand are shown by
thick lines.
Contribution of the semi-pole at $\tau_1 =\tau_2$ in the first term
in brackets in Eq.~(\protect\ref{eq:6.25}) is canceled by the term complex
conjugated to it.
}
\label{Fig:6.2}
\end{figure} 
} 

Before performing the analytic continuation, see Eq.~(\ref{eq:6.103}), we
have to transform integrals over imaginary times in Eq.~(\ref{eq:6.24}) to
the integrals over real time. In order to do so, we use Lehmann
representation (\ref{Lehman}) for the kernel $R(\tau)$:
\begin{eqnarray}
\Pi_{el}&(&\tau)=
\frac{2\pi\left|K(\tau)\right|^2}
{\nu v_F E_C^2 e^{2 {\bf C}}}
\int\frac{dt_1}{2\pi}\int\frac{dt_2}{2\pi}\times
\label{eq:6.25}\\
&&
\left(R^R(t_1)-R^A(t_1)\right)
\left(R^R(t_2)-R^A(t_2)\right)^*\times
\nonumber\\
&&
\int_0^\beta\!\!d\tau_1d\tau_2 
\frac{\pi^2 T^2}{
\sinh [\pi T(t_1 +i\tau_1 -i\tau)]
\sinh [\pi T(t_2+i\tau_2)]
}
\times 
\nonumber\\
&&\left[\frac{\pi T}{\sin \pi T (\tau_1-\tau_2)}
\frac{K(\tau_2)K(\tau_1-\tau)}{K(\tau_1)K(\tau_2-\tau)} +
\right. \nonumber\\
&&\left.e^{i2\pi{\cal N}}
K(\tau_1-\tau_2)
\frac{K(-\tau_2)K(\tau_1-\tau)}{K(\tau_1)K(\tau-\tau_2)}+ c.c.
\right].
\nonumber
\end{eqnarray}
Integration can be now performed in a manner similar to
Sec.\ref{sec:5a}. Using the fact that function $K(\tau_1)$ is 
analytical within the lower complex semiplane ${\rm Im}\,\tau_1 <0,$ we deform
the contour of integration as shown in Fig.~\ref{Fig:6.2}.

Because of the periodicity of the integrand, the integrals over the
parts of the contour running parallel to the imaginary axis, cancel
out. As the result, only the pole contribution at $\tau_1 =it_1+\tau$ 
remains at $t_1 <0$.  At $t_1>0$ the pole contribution disappears. 
Analogously, the complex conjugated terms are contributed by pole
$\tau_1 =it_1+\tau$ at $t_1 >0$.
As the result, we obtain from Eq.~(\ref{eq:6.25})
\wide{m}{
\begin{eqnarray}
\Pi_{el}(\tau)&=&
\frac{2\pi\left|K(\tau)\right|^2}
{\nu v_F E_C^2 e^{2 {\bf C}}}
\int\frac{dt_1dt_2}{2\pi}
\left(R^R(t_2)-R^A(t_2)\right)^*\int_0^\beta\!\!d\tau_2 
\frac{\pi T}{
\sinh [\pi T(t_2-i\tau_2)]
}\times
\label{eq:6.26}\\
&&
\left[\frac{\pi T}{\sin \pi T (\tau-\tau_2+it_1)}
\left(
\frac{R^A(t_1)K(\tau_2)K(it_1)}{K(it_1+\tau)K(\tau_2-\tau)} +
\frac{R^R(t_1)K(-\tau_2)K(-it_1)}{K(-it_1-\tau)K(\tau-\tau_2)}
\right)+
\right. \nonumber\\
&&\left.
e^{i2\pi{\cal N}}
K(\tau-\tau_2+ it_1)
\frac{R^A(t_1)K(-\tau_2)K(it_1)}{K(it_1+\tau)K(\tau-\tau_2)}- 
e^{-i2\pi{\cal N}}
K(\tau_2-\tau-it_1)
\frac{R^R(t_1)K(\tau_2)K(-it_1)}{K(-it_1-\tau)
K(\tau_2-\tau)}\right],
\nonumber
\end{eqnarray}
where we wrote explicit expressions for all the terms. 
Integration over $\tau_2$ can be now easily performed by deformation
of the integration contours shown in Fig.~\ref{Fig:6.3}, and we obtain
\begin{eqnarray}
&&\Pi_{el}(\tau)=
\frac{2\pi\left|K(\tau)\right|^2}
{\nu v_F E_C^2 e^{2{\bf C}}}
\int {dt_1dt_2}
\left\{R^R(t_1)[R^R(t_2)]^*+
R^A(t_1)[R^A(t_2)]^*\right\}
\frac{\pi T}{
\sin [\pi T(it_2-it_1-\tau)]
}-
\label{eq:6.27}\\
&&
\frac{2\pi\left|K(\tau)\right|^2}
{\nu v_F E_C^2 e^{2 {\bf C}}}
\int {dt_1dt_2}
\left[\frac{\pi T}{\sin \pi T (\tau-it_2+it_1)}
\left(
\frac{R^A(t_1)[R^A(t_2)]^*K(it_2)K(it_1)}{K(it_1+\tau)K(it_2-\tau)} +
\frac{R^R(t_1)[R^R(t_2)]^*K(-it_2)K(-it_1)}{K(-it_1-\tau)K(\tau-it_2)}
\right)+
\right. \nonumber\\
&&\left.
e^{i2\pi{\cal N}}
K(\tau-it_2+ it_1)
\frac{R^A(t_1)[R^R(t_2)]^*K(-it_2)K(it_1)}{K(it_1+\tau)K(\tau-it_2)}- 
e^{-i2\pi{\cal N}}
K(it_2-it_1-\tau)
\frac{R^R(t_1)[R^A(t_2)]^*K(it_2)K(-it_1)}{K(-it_1-\tau)
K(it_2-\tau)}\right].
\nonumber
\end{eqnarray}
}

\narrowtext{
\begin{figure}[h]
\vspace{-0.5cm}
\epsfxsize=6.7cm
\hspace*{0.5cm}
\epsfbox{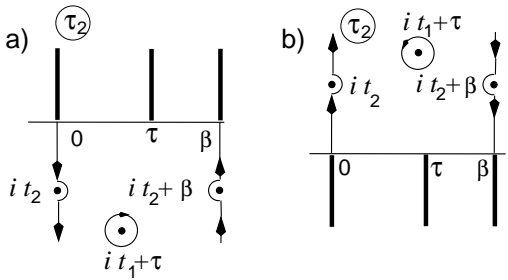}
\vspace{0.3 cm}
\caption{The integration contours used in the evaluation of
the integral over $\tau_2$ in Eq.~(\protect\ref{eq:6.26}) for  (a) the first
and third terms in brackets, and (b) for the second and fourth term.
Only first and second term have poles at $\tau_2 =\tau+it_1$. 
}
\label{Fig:6.3}
\end{figure} 
} 

\noindent One can see from Eq.~(\ref{eq:6.27}) that function $\Pi_{el}(\tau)$
is indeed analytic in the strip $0<{\rm Re}\,\tau < \beta$, which
justifies the steps leading to Eq.~(\ref{eq:6.103}).

Finally, we substitute Eq.~(\ref{eq:6.27}) into Eq.~(\ref{eq:6.103}).
For small temperatures $T\ll E_C$, we have from Eq.~(\ref{eq:4.15b})
$K({\beta}/{2}+it)=\pi T/\cosh\pi Tt$. As the result, the first term
in Eq.~(\ref{eq:6.27}) produces a contribution $\propto T^2$ and can be
neglected. The remainder can be recast into the  formula
\begin{equation}
G=\frac{2\pi G_R}{\nu v_F E_{C}^2 e^{2 {\bf
C}}}\left|\int_{-\infty}^{0}\!\! dt
K(it)\left[R^A(t)e^{i2\pi{\cal N}}+R^{R}(-t)\right]\right|^2,
\label{eq:6.28}
\end{equation}
which gives non-averaged conductance of the dot. Here we used the fact that
the characteristic scale of the integration over $t_1, t_2 \simeq 1/E_C$ is
much smaller than $\beta$. Equation (\ref{eq:6.28}) is reminiscent of the
Landauer formula. However, the form-factor $K(it)$ entering into this formula
indicates that a large number of states in the dot participate in the
transport, unlike the case of non-interacting electrons.

Now, we are prepared to study the statistics of the conductance. Using the
explicit expression (\ref{eq:4.15b}) for function $K$ and formula
(\ref{eq:6.17b}), we find the average conductance
\begin{equation}
\overline{G} = G_R\frac{2 \Delta}{E_C}e^{-2{\bf C}}\Lambda(0),
\label{eq:6.29}
\end{equation}
where ${\bf C} \approx 0.577$ is the Euler constant, and
$\Lambda(0)\approx 1.398$
is given by Eq.~(\ref{eq:5.6}). This expression is analogous to the
elastic cotunneling for the case of weak
coupling\cite{Averin90,Aleiner96}.

For the correlation function of the mesoscopic fluctuations of the
conductance, we find with the help of Eqs.~(\ref{eq:6.17})
\begin{equation}
\frac{\overline{\delta G(1)\delta G(2)}}{\overline{G}^2} = 
\left(\frac{\cos \pi n}{\Lambda (0)}\right)^2
\left[\Lambda^2\left(\!\frac{H_-^2}{H_c^2}\!\right)+
\Lambda^2\left(\!\frac{H_+^2}{H_c^2}\!\right)\right],
\label{eq:6.30}
\end{equation}
where we use again the short hand notations $i \equiv {\cal N}_i,H_i$,
$n={\cal N}_1-{\cal N}_2$, and $H_\pm = H_1 \pm H_2$.  Correlation
magnetic field $H_c$ is defined in Eq.~(\ref{eq:2.25}), and the dimensionless
function $\Lambda (x)$ is given by Eq.~(\ref{eq:5.6}) and is
plotted in Fig.~\ref{Fig:5.2}.  Once again, we see that even though
the averaged conductance does not any longer oscillate with the gate
voltage, the discreteness of charge manifests itself in the
oscillatory behavior of the conductance correlation function.  It is also
noteworthy that the mesoscopic conductance fluctuations are of the order of
the average, similarly to the weak coupling regime\cite{Aleiner96}.

\subsubsection{Finite reflection in the contact.} 
\label{sec:6a1}

So far, we have shown that the conductance in the tunneling setup is
non-vanishing at $T \to 0$, which is analogous to the elastic
cotunneling in the weak tunneling regime. However, the oscillatory
dependence of the conductance showed up not in the average conductance
but rather in the correlation function of mesoscopic fluctuations. On
the other hand, as we saw in Sec.~\ref{sec:5} finite backscattering
leads to the oscillatory dependence in the averaged quantities. The
purpose of this subsection is to study how the finite reflection in
the contact affects the elastic cotunneling, and to demonstrate that it
indeed leads to the oscillatory dependence of the averaged
conductance on the gate voltage.

To treat the finite reflection in the contact, we have to expand the
denominator and numerator in Eq.~(\ref{eq:6.22b}) up to the first order
in the backscattering Hamiltonian (\ref{eq:4.10}). Performing
averaging over the bosonic fields, we obtain with the help of
Eqs.~(\ref{eq:4.13}), (\ref{eq:4.7}), (\ref{eq:6.23}), and
(\ref{eq:4.15b}):
\wide{m}
{
\begin{eqnarray}
\Pi_{el}(\tau)&=& - |r|\frac{\left|K(\tau)\right|^2}
{2\pi\nu v_F E_C e^{ {\bf C}}}
\int_0^\beta\!\!d\tau_1d\tau_2d\tau_3 
 R(\tau_1-\tau )R^*(-\tau_2)\frac{\pi T}{\sin \pi T (\tau_1-\tau_2)}
\times\label{eq:6.31}\\
&&
\left[
\frac{K(\tau_2)K(\tau_1-\tau)}{K(\tau_1)K(\tau_2-\tau)}
\left\{
\frac{ e^{-i2\pi {\cal N}}
K(\tau_3) K(\tau-\tau_3 )}{K(-\tau_3)K(\tau_3-\tau)}
\frac{\sin \pi T(\tau_2-\tau_3) K(\tau_2-\tau_3)}
{\sin \pi T(\tau_1-\tau_3) K(\tau_1-\tau_3)}
+\right. \right.\nonumber\\
&&\left. \left.
\frac{ e^{i2\pi {\cal N}}
K(-\tau_3) K(\tau_3-\tau )}{K(\tau_3)K(\tau-\tau_3)}
\frac{\sin \pi T(\tau_1-\tau_3) K(\tau_1-\tau_3)}
{\sin \pi T(\tau_2-\tau_3) K(\tau_2-\tau_3)}
-  2\cos 2\pi {\cal N} \right\}
+ c.c.\right].
\nonumber
\end{eqnarray}
Here we retained only the terms which do not vanish after ensemble
averaging. Then, we can use the Lehmann representation for the kernel
$R(\tau)$ and perform the integration over $\tau_{1,2}$ in the manner of
the previous subsection. It yields
\begin{eqnarray}
\Pi_{el}(\tau)&=&- |r|\frac{\left|K(\tau)\right|^2}
{2\pi\nu v_F E_C e^{ {\bf C}}}
\int\!\!dt_1dt_2\int_0^\beta d\tau_3 
\frac{\pi T}{\sin \pi T (\tau - it_2 +it_1)}
\left[
\frac{R^A(t_1)[R^A(t_2)]^*K(it_2)K(it_1)}{K(it_1+\tau)K(it_2-\tau)}
\times \right.
\label{eq:6.32}\\
&&
\quad\quad
\left\{
\frac{ e^{-i2\pi {\cal N}}
K(\tau_3) K(\tau-\tau_3 )}{K(-\tau_3)K(\tau_3-\tau)}
\frac{\sin \pi T(it_2-\tau_3) K(it_2-\tau_3)}
{\sin \pi T(\tau+it_1 -\tau_3) K(\tau + it_1 -\tau_3)}
+\right.\nonumber\\
&&\quad\quad
\left.\left.
\frac{ e^{i2\pi {\cal N}}
K(-\tau_3) K(\tau_3-\tau )}{K(\tau_3)K(\tau-\tau_3)}
\frac{\sin \pi T(\tau+i t_1-\tau_3) K(\tau+it_1-\tau_3)}
{\sin \pi T(it_2-\tau_3) K(i t_2 - \tau_3)}
- 2 \cos 2\pi {\cal N}\right\} +c.c.
\right].
\end{eqnarray}
}

We deform the integration contour over $\tau_3$, as shown in
Fig.~\ref{Fig:6.4}, and substitute the result into Eq.~(\ref{eq:6.103}).
Then, we average the product of the Green functions $R^A[R^A]^*$ 
using Eq.~(\ref{eq:6.17b}). As the result, we find the
oscillating part of the ensemble averaged conductance
\begin{equation}
G= \alpha_1 G_R \left(\frac{|r|\Delta}{E_C}\right)\cos 2\pi{\cal N}.
\label{eq:6.33}
\end{equation}
Here $\alpha_1$ is the numerical coefficient given by
\begin{eqnarray*}
&&\alpha_1=\frac{4 e^{-{\bf C}}}{\pi}
\int_0^{\infty}\frac{dx dy}{x^2} e^{2e^x Ei(-x)}
\sin (\pi e^{-y})\times\\ 
&&\sinh\left[
e^{-y} Ei(y) + e^{x+y} Ei(-x-y)
 - e^y Ei(-y)
\right]\approx 1.458,
\end{eqnarray*}
with $Ei(x)$ being the exponential-integral function\cite{Ryzhik}.

Equation~(\ref{eq:6.33}) confirms our expectation that the finite
backscattering leads to the oscillatory dependence of the averaged
conductance on the gate voltage. Although the amplitude of
oscillations (\ref{eq:6.33}) is small compared to the average value of the
conductance (\ref{eq:6.29}), it still exceeds at low temperatures the
contribution of the inelastic cotunneling\cite{Furusaki95} to the conductance
oscillations.

\narrowtext{
\begin{figure}[h]
\vspace{0.2cm}
\epsfxsize=5cm
\hspace*{0.5cm}
\epsfbox{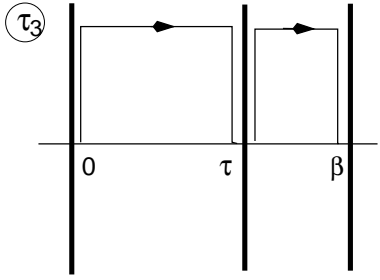}
\vspace{0.3 cm}
\caption{The integration contour used in the evaluation of
the integral over $\tau_3$ in Eq.~(\protect\ref{eq:6.32}). Integral along the
part of the contour parallel to the real axis cancels out the $2\cos 2\pi
{\cal N}$ term. 
}
\label{Fig:6.4}
\end{figure} 
}

\subsection{Electrons with spin.}
\label{sec:6b}
We use formulas of Sec.~\ref{sec:4b} to bosonize the
chiral fermionic fields. In order to account for the appearance of the
operator ${\hat n}$ in the Hamiltonian [compare Eqs.~(\ref{eq:3.15})
with Eq.~(\ref{eq:6.19})], we change the  transformation
(\ref{eq:4.18}): 
\begin{eqnarray}
&\displaystyle{\hat{\varphi}_L^i(x) = \frac{\hat{\varphi}_+^i(x) +
\hat{\varphi}_-^i(x) +\hat{\Phi}^i- \delta_{i\rho}\pi ({\cal N}-\hat{n})}
{\sqrt{2}};}& \nonumber\\
&\displaystyle{\hat{\varphi}_R^i(x)= \frac{\hat{\varphi}_+^i(-x) -
\hat{\varphi}_-^i(-x) - \hat{\Phi}^i - \delta_{i\rho}\pi ({\cal N} -\hat{n})}
{\sqrt{2}},}&
\label{eq:6.34}
\end{eqnarray}
where $i=\rho,\sigma$, and operator $\hat{n}$ commutes with the bosonic fields
$\hat{\varphi}_{\pm}^{\rho,\sigma}, \hat{\Phi}^{\rho,\sigma}$. 
To preserve the commutation relation
$\left[\hat{F}^\dagger,\hat{\varphi}_{L,R}^{\rho,\sigma}\right]=0$, we 
change the operators $\hat{F}^\dagger, \hat{F}$ similarly to Eq.~(\ref{eq:6.21}) 
\begin{equation}
\hat{F}^\dagger \mapsto \hat{F}^\dagger e^{-i\hat{\Phi}^\rho},
\hat{F} \mapsto \hat{F} e^{i\hat{\Phi}^\rho}.
\label{eq:6.35}
\end{equation} 
Substitution of Eqs.~(\ref{eq:4.16}), (\ref{eq:6.34}) and 
(\ref{eq:6.35}) into Eqs.~(\ref{eq:6.18b}) results in the formulas similar
to Eq.~(\ref{eq:6.22}):
\wide{m}
{
\begin{mathletters}
\label{eq:6.36}
\begin{eqnarray}
&&\Pi_{in}(\tau)=-\frac{2{\cal G}(-\tau; {\bf r}_t, {\bf r}_t ) }
{\nu \langle T_\tau e^{-\hat{S}}\rangle}
\langle T_\tau e^{-\hat{S}}
\hat{\bar{F}}(\tau)\hat{F}(0)
e^{i\left[\hat{\Phi}(0)-\hat{\Phi}(\tau )\right]}
\rangle,
\label{eq:6.36a}\\
&&\Pi_{el}(\tau)\!=\frac{1}
{\pi\nu\lambda\langle T_\tau e^{-\hat{S}}\rangle}
\int_0^\beta\!\!d\tau_1d\tau_2 R(\tau_1-\tau )R^*(-\tau_2) 
\sum_{\alpha=\pm 1}\langle T_\tau e^{-\hat{S}}
\hat{\bar{F}}(\tau)\hat{F}(0)\hat{\eta}_\alpha (\tau_1)
\hat{\eta}_\alpha(\tau_2)
e^{{i}\left[\hat{\Phi}^\rho (0)-\hat{\Phi}^\rho (\tau )]
\right]}
\times\nonumber\\
&&\quad
e^{\frac{i}{2}\alpha
\left(\hat{\varphi}_-^\sigma (\tau_1)-\hat{\varphi}_-^\sigma(\tau_2)\right)}
e^{\frac{i}{2}\alpha
\left(\hat{\Phi}^\sigma (\tau_1)-\hat{\Phi}^\sigma (\tau_2)\right)}
e^{\frac{i}{2}
\left(\hat{\varphi}_-^\rho(\tau_1)-\hat{\varphi}_-^\rho (\tau_2)\right)}
e^{\frac{i}{2}
\left(\hat{\Phi}^\rho (\tau_1)-\hat{\Phi}^\rho (\tau_2)\right)}
\times
\nonumber\\
&&\quad
\left\{
\sin \left[
\frac{\hat{\varphi}_+^\rho (\tau_1)-
\hat{\varphi}_+^\rho (\tau_2) }{2}
+\alpha
\frac{\hat{\varphi}_+^\sigma (\tau_1)-\hat{\varphi}_+^\sigma (\tau_2)
}
{2}+ \frac{\pi}{2}\left(\hat{n}(\tau_1)- \hat{n}(\tau_2)\right)
\right]
-
\right.
\nonumber\\
&&\quad \quad
\left.
\cos \left[
\frac{\hat{\varphi}_+^\rho (\tau_1)+\hat{\varphi}_+^\rho (\tau_2) }{2}
+\alpha
\frac{\hat{\varphi}_+^\sigma (\tau_1)+
\hat{\varphi}_+^\sigma (\tau_2) }{2} +
\frac{\pi}{2}\left(\hat{n}(\tau_1)+ \hat{n}(\tau_2)\right)
-\pi {\cal N}\right]
\right\}\rangle,
\label{eq:6.36b}
\end{eqnarray}
\end{mathletters}
}
 for the inelastic and the elastic cotunneling respectively.
However, for the calculation it is more convenient to proceed directly
to the low energy effective theory (\ref{eq:4.27}), because the main
contribution to the conductance comes from the energy scale smaller
than the charging energy $E_C$. First, we integrate out the symmetric
charge mode $\hat{\varphi}_+^\rho$. Then, we wish to use the substitution
(\ref{eq:4.25}). The important difference brought into the problem by
accounting for the second junction, is that Eqs.~(\ref{eq:6.36}) contain the
charged field ${\hat \Phi}^\rho$ itself, and not only the combination ${\hat
\Phi}^\rho - \phi_-^\rho(x=0)$, as we had before. Fortunately, the
corresponding change can be simply accounted for by the introduction of one
more chiral field $\hat{{\phi}}$, so that we have 
\begin{eqnarray}
\hat{\varphi}_-^\rho (x=0 ) +\hat{\Phi}^\rho \to 
\sqrt{2}\hat{\varphi}_\rho(x=0), \nonumber\\
 \hat{\Phi}^\rho - \hat{\varphi}_-^\rho (x=0 ) \to 
\sqrt{2}\hat{\phi}(x=0)
\label{eq:6.37}
\end{eqnarray}
instead of Eq.~(\ref{eq:4.25}). The field $\hat{{\phi}}$ enters neither the
effective action, nor the backscattering Hamiltonian, so it can be
immediately integrated out, and we find the low energy representation  
\begin{eqnarray}
&&e^{i[\hat{\Phi}(\tau)-\hat{\Phi} (0)]} \mapsto
\langle
e^{i[\hat{\phi} (\tau)-\hat{\phi} (0)]/\sqrt{2}} 
\rangle 
e^{i[\hat{\varphi}^\rho (\tau)-\hat{\varphi}^\rho (0)]/\sqrt{2}}
=\nonumber\\
&&\quad \frac{\pi}{E_C e^{\bf C}}\sqrt{\frac{v_F}{\tilde{\lambda}}}
\left(\frac{\pi T}{\sin \pi T\tau}\right)^{1/2}
e^{i[\hat{\varphi}^\rho (\tau)-\hat{\varphi}^\rho (0)]/\sqrt{2}}.
\label{eq:6.38}
\end{eqnarray}
The prefactor in the last formula can be found by requiring the
averages calculated with the help of the effective theory and the
initial theory to coincide.  Using Eqs.~(\ref{eq:6.37}) and
(\ref{eq:6.38}), we obtain from Eq.~(\ref{eq:6.36})
\wide{m}
{
\begin{mathletters}
\label{eq:6.39}
\begin{eqnarray}
\Pi_{in}(\tau)&=&-\frac{2{\cal G}(-\tau; {\bf r}_t, {\bf r}_t ) }
{\nu \langle T_\tau e^{-\hat{S}}\rangle}
\frac{\pi}{E_C e^{\bf C}}\sqrt{\frac{v_F}{\tilde{\lambda}}}
\left(\frac{\pi T}{\sin \pi T\tau}\right)^{1/2}
\langle T_\tau e^{-\hat{S}}
\hat{\bar{F}}(\tau)\hat{F}(0)
e^{i[\hat{\varphi}^\rho (\tau)-\hat{\varphi}^\rho (0)]/\sqrt{2}}
\rangle,
 \label{eq:6.39a} \\
\Pi_{el}(\tau)&=&\frac{1}
{\pi\nu\langle T_\tau e^{-\hat{S}}\rangle}
\frac{\pi}{E_C e^{\bf C}}\sqrt{\frac{v_F}{\tilde{\lambda}^3}}
\left(\frac{\pi T}{\sin \pi T\tau}\right)^{1/2}
\int_0^\beta\!\!d\tau_1d\tau_2 
R(\tau_1-\tau )R^*(-\tau_2) 
\sum_{\alpha=\pm 1}\langle T_\tau e^{-\hat{S}}
\hat{\bar{F}}(\tau)\hat{F}(0)\times
\label{eq:6.39b}
\\
&&
\hat{\eta}_\alpha (\tau_1)\hat{\eta}_\alpha(\tau_2)
e^{i[\hat{\varphi}^\rho (\tau)-\hat{\varphi}^\rho (0)]/\sqrt{2}}
e^{\frac{i}{2}\alpha
\left(\hat{\varphi}_-^\sigma(\tau_1)-
\hat{\varphi}_-^\sigma(\tau_2)\right)}
e^{\frac{i}{2}\alpha
\left(\hat{\Phi}^\sigma (\tau_1)-\hat{\Phi}^\sigma
(\tau_2)\right)}
\times
\nonumber\\
&&
\left\{
\cos \left[\alpha
\frac{\hat{\varphi}_+^\sigma (\tau_1)+\hat{\varphi}_+^\sigma (\tau_2) }{2}
+\frac{\pi}{2}\left(\hat{n}(\tau_1)+ \hat{n}(\tau_2)\right)-
\pi {\cal N}\right]
+
\cos \left[
\frac{\pi}{4}
+\alpha
\frac{\hat{\varphi}_+^\sigma (\tau_1)-\hat{\varphi}_+^\sigma (\tau_2) }{2}
+\frac{\pi}{2}\left(\hat{n}(\tau_1)- \hat{n}(\tau_2)\right)
\right]
\right\}\rangle,
\nonumber
\end{eqnarray}
where the averaging is performed with respect to the Hamiltonian
(\ref{eq:4.26}).  The effective action (\ref{eq:4.28}) acquires the
form
\end{mathletters}
\begin{eqnarray}
\hat{S}&=&\frac{1}{\pi\tilde{\lambda}}\int_0^\beta\!\! d\tau_1 d\tau_2
L\left(\tau_1-\tau_2\right)
\sum_{\alpha=\pm 1}
\hat{\eta}_\alpha(\tau_1)\hat{\eta}_\alpha (\tau_2)
\times
e^{\frac{i}{2}\alpha
\left(\hat{\varphi}_-^\sigma(\tau_1)-
\hat{\varphi}_-^\sigma(\tau_2)\right)}
e^{\frac{i}{2}\alpha
\left(\hat{\Phi}^\sigma (\tau_1)-\hat{\Phi}^\sigma
(\tau_2)\right)}
e^{\frac{i}{\sqrt{2}} 
\left(\hat{\varphi}_\rho(\tau_1)-\hat{\varphi}_\rho(\tau_2)\right)}
\times
\label{eq:6.40}\\
&&
\left\{
\cos \left[\alpha
\frac{\hat{\varphi}_+^\sigma (\tau_1)+\hat{\varphi}_+^\sigma (\tau_2) }{2}
+\frac{\pi}{2}\left(\hat{n}(\tau_1)+ \hat{n}(\tau_2)\right)-
\pi {\cal N}\right]
+
\cos \left[
\frac{\pi}{4}
+\alpha
\frac{\hat{\varphi}_+^\sigma (\tau_1)-\hat{\varphi}_+^\sigma (\tau_2) }{2}
+\frac{\pi}{2}\left(\hat{n}(\tau_1)- \hat{n}(\tau_2)\right)
\right]
\right\},
\nonumber
\end{eqnarray}
}
and the backscattering Hamiltonian $\hat{H}_{bs}$ is given
by\cite{Furusaki95}
\begin{equation}
\hat{H}_{bs} = \frac{2 |r|}{\pi }
\left(\frac{E_Ce^{\bf C}v_F}{\pi \tilde{\lambda}}\right)^{1/2} 
(-1)^{\hat n}\cos\pi{\cal N}
\cos\hat{\varphi}_+^\sigma (0),
\label{eq:6.41}
\end{equation}
cf. Eq.~(\ref{eq:4.27}). Cut-off $\tilde{\lambda}$ in Eqs.~(\ref{eq:6.38}) -
(\ref{eq:6.41}) should be of the order of $v_F/E_C$, because the
charging energy $E_C$ is the largest energy scale which can be
considered with the help of Hamiltonian (\ref{eq:4.26}).

In the following subsections we will apply effective description
(\ref{eq:6.39}) - (\ref{eq:6.41}) and (\ref{eq:6.103}) to find the
tunneling conductance of the dot in the asymmetric setup.

\subsubsection{Reflectionless contact.}
Let us consider first the elastic contribution (\ref{eq:6.39b}) (we
will see below that inelastic contribution should be taken into
account in order to obtain the correct temperature dependence).  In
the lowest-order approximation we neglect the action $\hat{S}$, and obtain
with the help of Eqs.~(\ref{eq:4.29}):
\begin{eqnarray}
\Pi_{el}^{(0)}(\tau)&=&\frac{1}
{\nu v_F E_C e^{ {\bf C}}}\frac{\pi T}{\sin \pi T\tau}
\int_0^\beta\!\!\!d\tau_1d\tau_2 \times\label{eq:6.42}\\
&&R(\tau_1-\tau )R^*(-\tau_2)
\frac{\pi T}{\sin \pi T (\tau_1-\tau_2)}\times \nonumber\\
&&\sum_{\gamma = \pm 1}
\left(
\frac{\sin \pi T\tau_1
\sin \pi T(\tau_2-\tau+ i\gamma 0)}
{\sin \pi T\tau_2 \sin \pi T(\tau_1-\tau
+ i\gamma 0)}
\right)^{1/2}. 
\nonumber
\end{eqnarray} 
When deriving Eq.~(\ref{eq:6.42}), we used the expression similar to
Eq.~(\ref{eq:6.23}),
\begin{eqnarray}
&&\langle T_\tau \hat{\bar{F}}(\tau) \hat{F}(0)
\cos \left[\frac{\pi}{2}
\left({\hat{n}(\tau_1) - \hat{n}(\tau_2) }\right)
\right]
\rangle =\label{eq:6.43} \\
&&\quad
\cos \left[\frac{\pi}{2}\left(\theta (\tau-\tau_1) -
\theta (\tau-\tau_2)
\right)\right].
\nonumber
\end{eqnarray}
All the further manipulations with Eq.~(\ref{eq:6.42}) are absolutely
analogous to the steps of Sec.~\ref{sec:6a} in the derivation of
Eq.~(\ref{eq:6.28}) from Eq.~(\ref{eq:6.24}). Instead of Eq.~(\ref{eq:6.28}),
here we find 
\begin{eqnarray}
G&=&\frac{G_R}{\nu v_F E_{C} e^{ {\bf
C}}}
\int_{E_C^{-1}}^{\infty}\!\! dt_1 dt_2 \frac{\pi T}
{\left( \sinh \pi Tt_1\sinh \pi Tt_1\right)^{1/2} }\times
\nonumber\\
&&\left[
R^A(-t_1)[R^A(-t_2)]^*
+R^R(t_1)[R^R(t_2)]^*\right],
\label{eq:6.44}
\end{eqnarray}
where the divergences (as we will see, logarithmical) should be cut off at times of the
order of $1/E_C$. Notice that there is no ${\cal N}$ dependence of the non-averaged
conductance. The reason is similar to the absence of the oscillations of the
capacitance in the first order in level spacing $\Delta$ -- the oscillations are washed
out by the quantum fluctuations of the spin mode, which is not pinned. We will see
later that the oscillatory  term in the conductance is smaller than the leading
non-oscillatory contribution to the conductance by a factor $\simeq
(\Delta/T)^{1/2}$.

Let us now proceed with the statistics of the elastic cotunneling
conductance (\ref{eq:6.44}). Ensemble averaging performed with the
help of Eq.~(\ref{eq:6.17a}) gives for the average conductance
\begin{equation}
\overline{G} = G_R \frac{2 \Delta e^{- {\bf C}}}{E_C}
\ln\left(\frac{E_C}{T}\right),
\label{eq:6.45}
\end{equation}
where ${\bf C}\approx 0.577$ is the Euler constant. At very low
temperatures, $T$ should be substituted by $\Delta \ln(E_C/\Delta)$, see
discussion in Sec.~\ref{sec:5b}.

Equation (\ref{eq:6.45}) deserves some discussion. Firstly, we notice
the presence of the large logarithmic factor in comparison with
Eq.~(\ref{eq:6.29}). It can be understood using the arguments of the
orthogonality catastrophe\cite{Anderson67} similar to those applied in
Ref.~\onlinecite{Furusaki95} for the inelastic cotunneling.  Consider
the elastic cotunneling process where an electron is introduced at
time $t=0$, and then taken away at time $t \gtrsim E_C^{-1}$. Because the
introduction of an electron costs extra energy $\simeq E_C$, all the
other electrons tend to redistribute themselves by moving one electron
charge through the left point contact, Fig~\ref{Fig:6.1}. One can
describe such a tendency as sudden change of the boundary condition in
each of the spin channels. Because all the spin channels are
symmetric, each spin mode should transfer charge $e/2$. According to
the Friedel sum rule, it corresponds to the additional scattering phase
shift $\delta =\pm \pi/2$ in each spin mode. It is
known\cite{Nozieres69} that such a sudden change causes creation of a large
number of electron-hole excitations, so that the resulting state is
orthogonal to the ground state. The probability for the system to
retain its initial state during time $t$ is $P(t) \simeq
1/(tE_C)^\chi$, where the index $\chi$ is related to the phase shift
in all channels, $\chi = \sum (\delta/\pi)^2$. In our case we have
four spin modes (two in the dot and two in the reservoir), therefore
$\chi=1$. In order to find the total conductance, we have to sum over
all possible times that electron spends in the dot: $G \propto 
 \int_{E_C^{-1}}^\beta dt/(E_Ct)$, which results in the logarithmic
temperature dependence (\ref{eq:6.45}). The similar argument for the
spinless electrons gives the phase shift $\delta = \pi$ in each of the two
channels. Thus,  $\chi =2$, all the relevant dynamics occurs during the times
smaller than $1/E_C$, and the divergent factor disappears.

Secondly, even though the elastic contribution is dominant in the
value of the conductance, in order to find the temperature dependence,
one has to take into account the inelastic contribution
(\ref{eq:6.39a}) which yields\cite{coefficient}
\begin{equation}
G_{in} = G_R\frac{\pi^3 Te^{-{\bf C}}}{2E_C }.
\label{eq:6.46}
\end{equation}
We see that this term has stronger temperature dependence than
(\ref{eq:6.45}) and therefore, the resulting conductance
$\overline{G}+G_{in}$ always grows  as the temperature increases.  

Mesoscopic fluctuations of the contribution (\ref{eq:6.44}) to the
conductance can be obtained with the help of Eq.~(\ref{eq:6.17}),
\begin{equation}
\frac{\overline{\delta G(H_1)\delta G(H_2)} }
{ \overline{G}^2} =\frac{1}{2}\sum_{\gamma =\pm}
\left(
1-\frac{ \ln {\rm max} \left(1; 
\left[\frac{H_\gamma}{H^{<}_c}\right]^2\right)}
{ \ln \left(\frac{E_C}{T}\right)} 
\right)^2.
\label{eq:6.47}
\end{equation}
Here the correlation field $H^{<}_c$ is defined by Eq.~(\ref{eq:5.140}),
and the field combinations $H_\pm = H_1 \pm H_2$ are assumed to be much
smaller than $H_c$ [the charging correlation field $H_c$ is given by
Eq.~(\ref{eq:2.25})]. The correlation function of the conductance
fluctuations starts to decrease fast, as $1/H_\pm^4$, only at fields 
$H_\pm\gtrsim H_c$. Similarly to the case of capacitance fluctuations
discussed in the previous Section, to obtain a representative statistics of
the conductance, the magnetic field should be varied in a range wider than
$H_c$.

Equation~(\ref{eq:6.47}) shows that the amplitude of conductance fluctuations are
of the order of its average value, like in the case of the spinless
fermions. However, unlike Eq.~(\ref{eq:6.30}), the correlation function
(\ref{eq:6.47}) does not reveal any oscillations with the gate voltage
${\cal N}$. 

In order to reveal this oscillatory dependence on ${\cal N}$, one has to expand
$\Pi_{el}(\tau)$ from Eq.~(\ref{eq:6.39b}) up to the first order in action
(\ref{eq:6.40}). The procedure of averaging over all the relevant operators
is absolutely similar to the derivation of Eq.~(\ref{eq:6.42}), and we obtain
\wide{m}
{
\begin{eqnarray}
\Pi_{el}^{(1)}(\tau)&=&-\frac{1}
{\nu v_F^2 E_C e^{ {\bf C}}}\frac{\pi T}{\sin \pi T\tau}
\int_0^\beta\!\!\!d\tau_1\dots d\tau_4
R(\tau_1-\tau )R^*(-\tau_2)L(\tau_3 -\tau_4)
\times  \label{eq:6.48}\\
&&
\sum_{\gamma = \pm 1}
e^{i2\pi\gamma{\cal N}}
\left(
\frac{\sin \pi T \tau_1
 \sin \pi T(\tau_2-\tau- i\gamma 0)}
{\sin \pi T\tau_2\sin \pi T(\tau_1-\tau
+ i\gamma 0)}
\frac{\sin \pi T(\tau_4 - \tau - i\gamma 0)\sin \pi T\tau_3 }
{\sin \pi
T\tau_4
\sin \pi T(\tau_3 -\tau + i\gamma 0)}
\right)^{1/2}
\times 
\nonumber\\
&&
\frac{\pi^2T^2 }
{\left[\sin \pi
T(\tau_1-\tau_2 + i\gamma 0)
\sin \pi T(\tau_3 -\tau_4 + i\gamma 0)
\sin \pi T(\tau_3 -\tau_2 + i\gamma 0)
\sin \pi T(\tau_1 -\tau_4 + i\gamma 0)
\right]^{1/2}}.
\nonumber
\end{eqnarray}
} Integration over imaginary times in Eq.~(\ref{eq:6.48}) is rather
straightforward, and technically very close to that in the
Sec.~\ref{sec:6a1}. Unlike Eq.~(\ref{eq:6.44}), here the result for the
non-averaged conductance is ${\cal N}$-dependent. We obtain for the
oscillating contribution: 
\begin{eqnarray}
G_{\rm osc}\!&=&\!\frac{G_R}{ \nu v_F^2 E_{C} e^{ {\bf
C}}}\frac{\alpha_2}{(2\pi^2 T)^{1/2}} \int_{E_C^{-1}}^\infty\!
\left[\prod_{i=1}^3 \!dt_i 
 \left(\frac{\pi
T}{\sinh \pi Tt_i }\right)^{1/2}\right] \nonumber\\
 &&
\times
\left[i
e^{-i2\pi{\cal N}} R^R(t_1)\left[R^A(t_2)\right]^*L^R(t_3)
+c.c.\right],
\label{eq:6.49}
\end{eqnarray}
where $\alpha_2$ is a numerical coefficient:
\begin{equation}
\alpha_2 = \int_{-\infty}^\infty \frac{dxdy}{|\sinh y|^{1/2}(\cosh
x)^{3/2}(\cosh (y-x))^{1/2}}\approx 11.31.
\label{eq:6.50}
\end{equation}
Ensemble averaging of Eq.~(\ref{eq:6.49}) is then performed  with the
help of Eqs.~(\ref{eq:3.25}) and (\ref{eq:6.17}), and the final result is
\begin{eqnarray}
&&\frac{\overline{\delta G_{\rm osc}(1)\delta G_{\rm osc}(2)} }
{ {G}^2_R} = \alpha_3 \frac{\Delta}{ T}\left(\frac{\Delta}{ E_C}\right)^2 
\ln^3\left( \frac{E_C}{ T}\right)\cos 2\pi n\times
\nonumber\\
&&
\quad
\left(\Lambda_+ + \Lambda_-\right)
\left(\Lambda_+^2 + \Lambda_-^2\right),
\label{eq:6.51}\\
&&
\Lambda_\pm =
1-\frac{ \ln {\rm max} \left(1; 
\left[\frac{H_\pm}{H^{<}_c}\right]^2\right)}
{ \ln \left(\frac{E_C}{T}\right)},
\nonumber
\end{eqnarray}
where the short hand notation for the conductance arguments is $i= {\cal
N}_i, H_i$, $H_\pm = H_1 \pm H_2$, $n= {\cal N}_1 - {\cal N}_2$, correlation
magnetic field $H_c^<$ is given by Eq.~(\ref{eq:5.140}), and the numerical
coefficient $\alpha_3$ is given by
$\alpha_3 = \alpha_2^2e^{-2{\bf C}} /(2\pi^4)\approx 0.207$. 

The variance of the conductance fluctuations (\ref{eq:6.51}) in the unitary
limit ($H\gtrsim H_c$) is suppressed  by a factor of four compared to its
zero-field value.

\subsubsection{Finite reflection in the contact.}
For the spinless electrons, finite reflection leads to the oscillations in
the averaged conductance already in the first order of perturbation
theory in $r\ll 1$. On the contrary, for the electrons with spin,
backscattering leads only to the enhancement of the oscillating part
of the correlation function of the mesoscopic fluctuations. In order to demonstrate
this, we expand Eq.~(\ref{eq:6.39b}) up to the first order in the backscattering
Hamiltonian (\ref{eq:6.41}). We obtain 
\wide{m}
{
\begin{eqnarray}
\Pi_{el}^{(b)}(\tau)&=&
-\frac
{1}
{\nu v_F E_C e^{ {\bf C}}}
\sqrt\frac{E_Ce^{\bf C}}{\pi}
\frac{\pi T}{\sin \pi T\tau}
\int_0^\beta\!\!\!d\tau_1\dots d\tau_3
R(\tau_1-\tau )R^*(-\tau_2)\sum_{\gamma = \pm 1}
e^{i2\pi\gamma{\cal N}}
\times
\label{eq:6.52}\\
&&
\left(
\frac{\sin \pi T \tau_1
 \sin \pi T(\tau_2-\tau- i\gamma 0)}
{\sin \pi T\tau_2\sin \pi T(\tau_1-\tau + i\gamma 0)}
\right)^{1/2}
\frac{(\pi T)^{3/2}}
{\left[
\sin \pi T(\tau_1-\tau_2 + i\gamma 0)
\sin \pi T(\tau_3 -\tau_2 + i\gamma 0)
\sin \pi T(\tau_1 -\tau_3 + i\gamma 0)
\right]^{1/2}}.
\nonumber
\end{eqnarray}
}
Performing the contour deformation for the integration over $\tau_{1,2}$,
as we did before, and the analytic continuation (\ref{eq:6.103}), we find
\begin{eqnarray}
G&=&\frac{G_R}{\pi \nu v_F E_{C} e^{ {\bf C}}}\frac{\alpha_2 |r|
\sqrt{E_C e^{\bf C} }}{(2\pi^2
T)^{1/2}}\times 
\label{eq:6.53}
\\ 
&&\int_{E_C^{-1}}^\infty\!dt_1dt_2 
\frac{\pi T}{
\left(
\sinh \pi Tt_1\sinh \pi Tt_2
\right)^{1/2}}
\times \nonumber\\ 
&&
\quad
\left\{i
e^{-i2\pi{\cal N}} R^R(t_1)\left[R^A(t_2)\right]^*
+c.c.\right\},
\nonumber
\end{eqnarray}
where the numerical coefficient $\alpha_2$ is given by
Eq.~(\ref{eq:6.50}). Average of Eq.~(\ref{eq:6.53}) obviously vanishes, and
for the mesoscopic fluctuations we obtain with the help of
Eqs.~(\ref{eq:6.17})
\begin{equation}
\frac{\overline{\delta G(1)\delta G(2)}}{G_R^2} = 
\alpha_4 \frac{|r|^2\Delta^2}{E_C T}
\ln^2\left( \frac{E_C}{ T}\right)
 \left(\Lambda_+^2 + \Lambda_-^2\right)
\cos 2\pi n,
\label{eq:6.54}
\end{equation}
where we use the same short-hand notation as in Eq.~(\ref{eq:6.51}).
The numerical coefficient $\alpha_4$ in Eq.~(\ref{eq:6.54}) is given by
$\alpha_4= \alpha_2^2 e^{-\bf C}/\pi^4\approx 0.737$.

Calculation of the contribution of the backscattering into the average
conductance requires  accounting of Eq.~(\ref{eq:6.41}) in the
second order perturbation theory. On the dimensional grounds we expect
this contribution to be 
\[
\overline{G}_{osc}({\cal N}) \simeq G_R \frac{|r|^2 \Delta}{T}\cos
2\pi{\cal N}.
\]
The low-temperature power law divergence in this formula and in
Eq.~(\ref{eq:6.53}) should be cut off at the energy\cite{Furusaki95}
$\epsilon^*\simeq |r|^2 E_C\cos^2\pi {\cal N}$. Calculation of the precise
behavior of the conductance at $T\lesssim \epsilon^*$ which can be performed
with the help of refermionization technique of Ref.~\onlinecite{Furusaki95} is
beyond the scope of the present paper. However, the perturbation
theory results indicate that the modulation of the conductance
$\bar{G}_{osc}({\cal N})$ at low temperature should be of the order of the average
conductance $\bar{G}$.
   
\section{Conclusion}
In this paper, we considered mesoscopic effects in the Coulomb blockade (CB) regime.
The emphasis was put on the case when the quantum dot is connected to a lead by
a perfectly transparent single-mode channel. We have demonstrated that
the earlier conclusion that the CB vanishes  under this condition \cite{Matveev95} is
only an approximation, which resulted from neglecting the electron trajectories
returning to the channel after traversing the dot. We have shown that the CB
persists, and its period is still determined by a single electron charge. However, CB
oscillations in all the observable quantities acquire a random phase and therefore it
is revealed in the correlation functions  of mesoscopic fluctuations.    We
constructed an analytic, well-controlled theory describing those fluctuations.

Our results are substantially different from the known results
in noninteracting models of mesoscopic systems. For instance, the correlated ground
state involves all the one-electron wave functions in the energy strip of the order
of the charging energy $E_C$. The number of states in this strip is $n\sim
E_C/\Delta\gg 1$ (here $\Delta$ is the level spacing).  The correlation magnetic flux
for the mesoscopic fluctuations $\Phi=\Phi_0\sqrt{E_T/E_C}$ is
controlled by the energy scale $E_C$. The large number of states in
the relevant energy strip leads to the robustness of the oscillatory
dependence over about $E_C/\Delta$ peaks.

We obtained the closed analytic expression for some experimentally relevant
characteristics. Final results are summarized in Tables ~\ref{Table:2} for the
thermodynamic quantities (differential capacitance of an almost open dot), and in
Table~\ref{Table:1} for the transport quantities (tunneling conductance of an almost
open dot).
\wide{m}{
\begin{table}[h]
\caption{
The ensemble-averaged differential capacitance
\protect\cite{Matveev95} and the correlation function of the mesoscopic
fluctuations in the Coulomb blockade regime; only the contributions oscillating with
the gate voltage ${\cal N}$ are presented in this Table. Here $|r|^2 \ll 1$ is the
reflection coefficient in an almost open channel, and $\beta =1,2$ for the  orthogonal
and unitary ensembles respectively. For more details see Eqs.~(\protect\ref{eq:5.7}),
(\protect\ref{eq:5.8}), (\protect\ref{eq:5.141}), and (\protect\ref{eq:5.20}). 
 }
\vspace*{0.3cm}
\begin{tabular}{| c| c | c |}
& 
\raisebox{0cm}[0.4cm][0.3cm]{
$\displaystyle{\overline{\delta C_{\it diff}({\cal N}) }/C
}$
}
&
$\displaystyle{
\overline{\delta C_{\it diff} ({\cal N}_1)\delta C_{\it diff} ({\cal N}_2)}/C^2
}$
\\
\hline
$\displaystyle{
s=0
}$
&
\raisebox{0cm}[0.5cm][0.5cm]
{
$\displaystyle{
3.56 |r| \cos 2\pi {\cal N}
}$
}
&
\raisebox{0cm}[0.5cm][0.5cm]
{
$\displaystyle{
\frac{5.59 }{\beta} \left(\frac{ \Delta}{E_C}\right)
\cos 2\pi \left({\cal N}_1-{\cal N}_2\right)
}$
}
 \\
\hline
$\displaystyle{
s={1}/{2}
}$
&
\raisebox{0cm}[0.5cm][0.5cm]
{
$\displaystyle{
4.53  |r|^2 \cos 2\pi {\cal N}
\ln \left(\frac{E_C}{T}\right)}
$ 
}
&
\raisebox{0cm}[0.8cm][0.8cm]
{
$\displaystyle{
\frac{ 0.54 }{\beta}
 \left(\frac{ \Delta }{E_C }\right)
\ln^3\left(\frac{E_C}{T}\right)
\left[
\left(\frac{ \Delta}{E_C }\right)
\ln\left(\frac{E_C}{T}\right)
+
7. 12  |r|^2  
\right]
\cos 2\pi \left({\cal N}_1-{\cal N}_2\right)
}$
} 
\end{tabular}
\label{Table:2}
\end{table} 

\begin{table}[h]
\caption{
Conductance of the quantum  dot in a strongly asymmetric setup, see
Fig.~\protect\ref{Fig:6.1},  and the correlation function of its mesoscopic fluctuations,
oscillating with gate voltage ${\cal N}$. Tunneling conductance $G_R$ is much smaller
than $e^2/(2\pi\hbar)$. For the detailed results, see Eqs.~ (\protect\ref{eq:6.29}),
(\protect\ref{eq:6.30}), (\protect\ref{eq:6.33}),  (\protect\ref{eq:6.45}) -
(\protect\ref{eq:6.47}), (\protect\ref{eq:6.51}) and (\protect\ref{eq:6.54}).
 }
\vspace*{0.3cm}
\begin{tabular}{| c| c | c |}
& 
\raisebox{0cm}[0.4cm][0.3cm]{
$\displaystyle{\overline{G ({\cal N}) }/G_R
}$
}
&
$\displaystyle{
\overline{\delta G ({\cal N}_1)\delta G ({\cal N}_2)}/G_R^2
}$
\\
\hline
$\displaystyle{
s=0
}$
&
\raisebox{0cm}[0.5cm][0.5cm]
{
$\displaystyle{
\frac{ \Delta}{E_C}\left( 0.63 + 1.46 |r| \cos 2\pi {\cal N}\right)
}$
}
&
\raisebox{0cm}[0.5cm][0.5cm]
{
$\displaystyle{
\frac{0.78 }{\beta} \left(\frac{ \Delta}{E_C}\right)^2
\cos 2\pi \left({\cal N}_1-{\cal N}_2\right)
}$
}
 \\
\hline
$\displaystyle{
s={1}/{2}
}$
&
\raisebox{0cm}[0.5cm][0.5cm]
{
$\displaystyle{
\frac{1.12 \Delta}{E_C}\left[\ln \left(\frac{E_C}{T}\right)+ 7.75 \frac{T}{\Delta}+ {\cal
O}(|r|^2)\right] }$ 
}
&
\raisebox{0cm}[0.8cm][0.8cm]
{
$\displaystyle{
 \left(\frac{ \Delta^2 }{E_CT }\right)
\ln^2\left(\frac{E_C}{T}\right)
\left[\frac{0.83}{\beta^2}
\left(\frac{ \Delta}{E_C }\right)
\ln\left(\frac{E_C}{T}\right)
+
\frac{1.48 |r|^2 }{\beta } 
\right]
\cos 2\pi \left({\cal N}_1-{\cal N}_2\right)
}$
} 
\end{tabular}
\label{Table:1}
\end{table} 
} 

Up to now, only one experiment studying the effect of the opening of
the channel on the Coulomb blockade was published\cite{Leo}. It was
found that at the quantized value of the channel conductance, the
Coulomb blockade oscillations disappeared completely, in disagreement
with our predictions. We attribute this finding to a relatively simple
geometry of the dot\cite{Leo}, allowing for an adiabatic (rather than 
chaotic) propagation of an electron through the entire confined region.

\acknowledgements 
We are grateful to B.L. Altshuler, P.W. Brouwer,
A.I. Larkin and K.A. Matveev for useful discussions and to
S.M. Cronenwett and C.M. Marcus for valuable information. Hospitality
of the Aspen Center for Physics and of the International Centre for
Theoretical Physics in Trieste is also acknowledged.  The work at the 
University of
Minnesota is supported by NSF Grant DMR-9423244.

\appendix
\section{Derivation of Eq.~(\ref{A1})}
\label{ap:1} 
Let us discretize the space in the direction along the channel axis. The
fermionic Hamiltonian $\hat{H}_F$ acquires the form
\begin{equation}
\hat{H}_F\!\! =\! \int\! d{\bf r}_\perp\! \sum_n\! 
\frac{
\left(\psi^\dagger_n\!-\!\psi^\dagger_{n+1}\right)
\left(\psi_n\!-\!\psi_{n+1}\right)
}
{2ma}
 + a\psi^\dagger_n \hat{H}_\perp \psi_n,
\label{eq:a1}
\end{equation}
where the transverse part of the motion is described by the operator
\begin{equation}
\hat{H}_\perp=-\frac{\nabla_{\bf r}^2}{2m} + U_n({\bf r}) -\mu,
\label{eq:a2}
\end{equation}
and $a$ is the discretization step. Fermionic operators in Eq.~(\ref{eq:a1})
satisfy the anticommutation relation 
$\left\{\psi^\dagger_n({\bf r})
\psi_{n^\prime}({\bf r}^\prime)\right\} 
= a^{-1}\delta_{nn^\prime}\delta\left({\bf
r}_\perp -{\bf r}_\perp^\prime \right)$.
The continuous limit of $a \to 0$, which will be taken in the end of
the calculation corresponds to the usual Schr\"{o}dinger equation.
Let us separate the space into two regions; region ``1'' includes all
the lattice sites with $n \leq 0$ and region ``2'' includes sites with
$n > 0$. The terms entering into decomposition of the Hamiltonian,
see also Eq.~(\ref{transformation}), $\hat{H}_F =
\hat{H}_1+\hat{H}_2+\hat{H}_{12}$, have the form
\begin{mathletters}
\label{eq:a3}
\begin{eqnarray}
&&\displaystyle{
\hat{H}_1\!\! =\! \int\! d{\bf r}_\perp\! \sum_{n<0}\! 
\left[\frac{
\left(\psi^\dagger_n\!-\!\psi^\dagger_{n+1}\right)
\left(\psi_n\!-\!\psi_{n+1}\right)
}
{2ma}
 +\right.}\nonumber\\
&&\hspace*{2.5cm}\displaystyle{
\left. a
\psi^\dagger_n
\left(\hat{H}_\perp+\frac{\delta_{0,n}}{2ma}\right) 
\psi_n
\right],}\label{eq:a3a}\\
&&\displaystyle{
\hat{H}_2\!\! =\! \int\! d{\bf r}_\perp\! \sum_{n\geq 1}\! 
\left[\frac{
\left(\psi^\dagger_n\!-\!\psi^\dagger_{n+1}\right)
\left(\psi_n\!-\!\psi_{n+1}\right)
}
{2ma}
 +\right.}\nonumber\\
&&\hspace*{2.5cm}\displaystyle{
\left. a
\psi^\dagger_n
\left(\hat{H}_\perp+\frac{\delta_{1,n}}{2ma}\right) 
\psi_n
\right],}\label{eq:a3b}\\
&&\displaystyle{
\hat{H}_{12}\! =- \int\! d{\bf r}_\perp 
\frac{\psi^\dagger_0\psi_{1} +
\psi^\dagger_{1}\psi_0
}
{2ma}}.
\label{eq:a3c}
\end{eqnarray}
\end{mathletters}
We obtain the average in Eq.~(\ref{transformation}) using Eq.~(\ref{eq:a3c}):
\begin{eqnarray}
&&\displaystyle{\frac{1}{2}\langle
T_\tau\hat{H}_{12}(\tau_1)\hat{H}(\tau_2) \rangle_2=} \label{eq:a4}\\
&&\ \ \
\displaystyle{\int
\frac{d{\bf r}_\perp d{\bf r}_\perp^\prime} 
{4m^2a^2} 
\langle
T_\tau {\psi}_0({\bf r} ;\tau_1)\bar {\psi}_0({\bf r}^\prime, \tau_2)
\rangle T_\tau {\psi}_1({\bf r}^\prime,\tau_2) \bar{\psi}_1 ({\bf
r},\tau_1)}.
\nonumber
\end{eqnarray}
Using the definition of the Matsubara Green function 
\begin{equation}
-\langle\psi_n(\tau)\bar{\psi}_m(0)\rangle_2 = {\cal G}_{nm}(\tau)
=\left[\frac{1}{\partial_\tau - \hat{H}_2}\right]_{nm}
\label{eq:a5}
\end{equation}
and low energy representation for the fermionic operator (only one
transverse mode $\phi ({\bf r}_\perp)$ in the channel)
\[
\psi_0({\bf r}_\perp) = \psi_0 \phi ({\bf r}_\perp),
\]
we rewrite Eq.~(\ref{eq:a4}) in the form
\begin{eqnarray}
&&\displaystyle{\frac{1}{2}\langle
T_\tau\hat{H}_{12}(\tau_1)\hat{H}(\tau_2) \rangle_2= - T_\tau
{\psi}_0(\tau_1)\bar {\psi}_0(\tau_2)
\times} \label{eq:a6}\\
&&\ \ \
\displaystyle{ \frac{1}{4m^2a^2} \int
{d{\bf r}_\perp d{\bf r}_\perp^\prime} 
\phi ({\bf r}_\perp)\phi ({\bf r}_\perp^\prime)
{\cal G}_{11}
\left(\tau_1-\tau_2; {\bf r}_\perp, {\bf r}_\perp^\prime\right)}.
\nonumber
\end{eqnarray}
As it follows from Eqs.~(\ref{eq:a3}) and (\ref{eq:a5}), Green function
satisfies the equation
\begin{mathletters}
\begin{eqnarray}
&&\displaystyle{
\left\{\left(\partial_\tau - \hat{H}_\perp\right)\delta_{nn^\prime} -
\frac{2\delta_{nn^\prime}-\delta_{nn^\prime+1}-\delta_{nn^\prime-1}
}
{2ma^2}
\right\} 
{\cal G}_{n^\prime m}=}\nonumber\\
&&\hspace*{2.4cm}\displaystyle{a^{-1}\delta_{nm}
\delta\left({\bf r}_1-{\bf r}_2 \right), \quad n > 1};
\label{eq:a7a}\\
&&\displaystyle{
\left\{\left(\partial_\tau - \hat{H}_\perp\right)\delta_{nn^\prime} -
\frac{2\delta_{nn^\prime}-\delta_{nn^\prime-1}
}
{2ma^2}
\right\} 
{\cal G}_{n^\prime m}=}\nonumber\\
&&\hspace*{2.4cm}\displaystyle{a^{-1}\delta_{nm}
\delta\left({\bf r}_1-{\bf r}_2 \right), \quad n = 1;}
\label{eq:a7b}
\end{eqnarray}
We see that the difference between Eq.~(\ref{eq:a7a}) and
Eq.~(\ref{eq:a7b}) can be described by the boundary condition
\end{mathletters}
\begin{equation}
{\cal G}_{0n}=0; \quad {\cal G}_{n0}=0.
\label{eq:a8}
\end{equation}
Now, we introduce coordinate $x=an$, and take the
continuous limit $a \to 0$. Substituting
${\cal G}_{11} =a^2\partial^2_{xx^\prime}{\cal
G}(x,x^\prime)|_{x,x^{\prime} = +0}$ into Eq.~(\ref{eq:a6}),  
we obtain Eq.~(\ref{A1}). 

\section{Derivation of Eqs.~(\ref{4.13})}
\label{ap:2}

We introduce more general than Eq.~(\ref{eq:4.12}) 
correlation functions
\begin{mathletters}
\label{eq:b1}
\begin{eqnarray}
&&{\cal D}_-\left(\tau; x_1,x_2\right) = \langle T_\tau
\hat{\varphi}_-(\tau;x_1)\hat{\varphi}_-(0;x_2)\rangle;
\label{eq:b1a}\\
&&{\cal D}_+\left(\tau;x_1,x_2\right) = \langle T_\tau
\hat{\varphi}_+(\tau;x_1)\hat{\varphi}_+(0;x_2)\rangle;
\label{eq:b1b}\\
&&{\cal D}_{\Phi}\left(\tau\right) = \langle T_\tau
\hat{\Phi}(\tau)\hat{\Phi}(0)\rangle;
\label{eq:b1c}\\
&&{\cal D}_{\Phi +}\left(\tau;x\right) = \langle T_\tau
\hat{\Phi}(\tau)\hat{\varphi}_+(0,x)\rangle.
\label{eq:b1d}
\end{eqnarray}
Correlation functions (\ref{eq:4.12}) are related to those from
Eq.~(\ref{eq:b1}) by
\end{mathletters}  
\begin{equation}
{\cal D}_{\pm}(\tau)=
\int_{-\infty}^\infty \frac{dx}{\pi }\frac{\lambda {\cal D}_{\pm}(\tau;0,x)}
{\lambda^2+x^2}, \quad 
{\cal D}_{\Phi +}\left(\tau\right)={\cal D}_{\Phi +}\left(\tau;0\right),
\label{eq:b2}
\end{equation}
where the high momenta cut-off is introduced consistently with
Eqs.~(\ref{eq:4.0}). Equations of motion for propagators (\ref{eq:b1})
follow from Eqs.~(\ref{eq:4.7}) and (\ref{eq:4.9}) and they are given
by
\begin{mathletters}
\label{eq:b3}
\begin{eqnarray}
&&
\left(\partial_\tau + iv_F\partial_x\right)
{\cal D}_-(\tau; x,y) = -i\pi \delta(\tau){\rm sgn}(x-y);
\label{eq:b3a}\\
&&\left(\partial_\tau + iv_F\partial_x\right)
{\cal D}_+(\tau; x,y)=
i\ {\rm sgn}\,x \ \frac{E_C}{2\pi}{\cal D}_+(\tau;0,y)\label{eq:b3b}\\
&&\hspace*{5cm} -i\pi \delta(\tau){\rm sgn}(x-y);
\nonumber\\
&&\partial_\tau{\cal D}_{\Phi}\left(\tau\right) = 
i\frac{E_C}{2\pi}{\cal D}_{\Phi +}\left(\tau;0\right);
\label{eq:b3c}\\
&&\partial_\tau {\cal D}_{\Phi +}\left(\tau;x\right) =
 i\pi \delta \left( \tau \right) -
i\frac{E_C}{2\pi}{\cal D}_{ +}\left(\tau;0,x\right).
\label{eq:b3d}
\end{eqnarray}
Performing imaginary time Fourier transform ${\cal D}(\tau) =T
\sum_{\Omega_n} e^{-i\Omega_n\tau }{\cal D}(\Omega_n )$, (here
$\Omega_n=2\pi nT$ is the bosonic Matsubara frequency) we find the
solution of Eq.~(\ref{eq:b3b})
\end{mathletters}
\begin{eqnarray}
&&{\cal D}_+(\Omega_n;x,y)\!=\!
\frac{\pi }{\Omega_n}\!
\left\{\!
 {\rm sgn}(x-y) + 2\theta\left[\Omega_n(y-x)\right]
e^{\frac{\Omega_n}{v_F}(x-y)}\!
\right\} \nonumber \\ 
&&\ 
-\frac{E_C}{2\pi}{\cal D}_+(\Omega_n;0,y)
\left\{
 {\rm sgn}\, x + 2\theta\left(-\Omega_nx\right)
e^{\frac{\Omega_n}{v_F}x}
\right\}.
\label{eq:b4}
\end{eqnarray}
Substituting $x=0$ in the both sides of Eq.~(\ref{eq:b4}), we 
find ${\cal D}_+(\Omega_n;0,y)$ and, then Eq.~(\ref{eq:b2}) yields:
\begin{mathletters}\label{eq:b5}
\begin{equation}
{\cal D}_+(\Omega_n)=
\frac{\pi}{|\Omega_n|+\frac{E_C}{2\pi}}
f\left(\frac{|\Omega_n|\lambda}{v_F}\right).
\label{eq:b5a}
\end{equation}
Propagator ${\cal D}_-$ is found by putting $E_C=0$ in  Eq.~(\ref{eq:b5a}),
\begin{equation}
{\cal D}_-(\Omega_n)=
\frac{\pi}{|\Omega_n|}
f\left(\frac{|\Omega_n|\lambda}{v_F}\right),
\label{eq:b5b}
\end{equation}
and remaining propagators are found from the time Fourier transform of
Eqs.~(\ref{eq:b3c}) -(\ref{eq:b3d}) and Eq.~(\ref{eq:b5a}):
\begin{eqnarray}
{\cal D}_{\Phi }(\Omega_n)= \frac{E_C}{2\pi |\Omega_n|}
\frac{\pi }{|\Omega_n|+\frac{E_C}{2\pi}}.
\label{eq:b5c}\\
{\cal D}_{\Phi+}(\Omega_n)=
\frac{\pi\ {\rm sgn}\, \Omega_n}{|\Omega_n|+\frac{E_C}{2\pi}}.
\label{eq:b5d}
\end{eqnarray}
In Eqs.~(\ref{eq:b5}), the cut-off function $f(x)$ is defined as
\end{mathletters}
\begin{equation}
f(x)=\int_0^\infty \frac{2dy}{\pi}\frac{e^{-xy}}{1+y^2}.
\label{eq:b6}
\end{equation}

Inverse Fourier transform of Eqs.~(\ref{eq:b5}) gives Eqs.~(\ref{eq:4.13}).
Let us write here for completeness the result for the propagator 
${\cal D}_+(0)$ at arbitrary temperatures:
\begin{equation}
{\cal D}_+\left(0\right) = \ln\left(\frac{2\pi v_F}{\lambda E_C
e^{\bf C}}\right) + \int_0^\infty dx\left[\cot x - \frac{1}{x}\right]
e^{-x\frac{E_C}{2\pi^2T}}.
\label{eq:b7}
\end{equation}
At small temperatures $T\ll E_C$, Eq.~(\ref{eq:b7}) reduces to
Eq.~(\ref{eq:4.13c}). At large temperatures, $T \gg E_C$, we obtain
\[
{\cal D}_+\left(0\right) = \ln\left(\frac{ v_F}{2\pi T\lambda}
\right) + \frac{2\pi^2 T}{E_C}.
\]
All the results associated with the Coulomb blockade, see {\em e.g.}
Eq.~(\ref{eq:4.14b}) contain exponential terms of the form $e^{-{\cal D}_+(0)}$. 
This gives the suppression of the charge quantization at high temperature 
by a factor of $e^{-2\pi^2 T/E_C}$. This clearly indicates, that the effects
considered in this paper and in Refs.~\onlinecite{Matveev95,Furusaki95}, can
not be obtained in any order of perturbation theory in charging energy
$E_C$.

\end{multicols}


\begin{references}
\bibitem{Brouwer1}This subsection was motivated by discussions with Piet
Brouwer.
\bibitem{attempt} In a recent work\cite{ButtikerPreprint}, an attempt
has been made to describe the Coulomb blockade for an open channel by
replacing the averaging over a grand canonical ensemble by the
averaging over a canonical ensemble. We do not see any grounds for
such an approach because the quantum fluctuations of the number of
electrons in the dot are parametrically large, $\langle(\delta\hat{
Q}/e)^2\rangle \simeq \ln(\epsilon_F/E_C)$.
\bibitem{Footnote1}The
same formulas can be obtained also starting from the random matrix
formulation of Sec.~\ref{sec:3b} and correlation function
(\ref{eq:6.7}) in the form $\Pi(\tau)= -\Delta M
\langle T_\tau\psi_\alpha(\tau)\bar{\psi}_\alpha(0)\rangle$ where
$\alpha$ is an arbitrary state of the dot.
\bibitem{coefficient}Numerical coefficient in Eq.~(\ref{eq:6.46})
differs from Ref.~\onlinecite{Furusaki95} because we define $G_R$ as
the conductance per one spin and our definition of the charging
energy $E_C$ also differs by a factor of 2.
\end{references}
\end{document}